\documentclass[12pt,letterpaper,twoside,singlespace]{thesis} 

\usepackage{graphicx}
\usepackage{times}
\usepackage{psfrag}

\title{\bf Search for relic neutralinos with Milagro}
\thesisauthor{Lazar Fleysher}
\degree{Doctor of Philosophy}
\department{Physics}
\thesisadviser{Allen I. Mincer}   
\submitmonth{May}                 
\submityear{2003}                 

\begin{document}
    \pagenumbering{roman}
    \maketitle
    \copyrightpage        
    \thispagestyle{empty}

\epigraph{George  Orwell, ``1984''}{%
Apart from very short notes, it was usual to dictate everything into
the speakwrite, which was of course impossible for his present purpose.
He dipped the pen into the ink and then faltered for just a second. A
tremor had gone through his bowels. To mark the paper was the decisive
act. In small clumsy letters he wrote:
}

\begin{center}
{\bf Search for relic neutralinos with Milagro}
\end{center}
    \begin{preface}{Dedication}

\begin{center}
I dedicate this to my family.
\end{center}
\end{preface}

\begin{preface}{Acknowledgements}
\epigraph{George  Orwell, ``1984''}{%
"What happens to you here is forever", O'Brien had said. That was a true
word.
}

I consider myself lucky because I can boast that I have not one, not
two, but three thesis advisors. Allen Mincer is the main one and I thank
you for taking risk in allowing me to proceed with the dark matter
search project. I also thank you for taking responsibility for me while
I was a graduate student at NYU. There are countless reasons why I am
grateful to you so I will just thank the fate who had given you to me.

I would like to thank Todd Haines from Los Alamos National Laboratory
who had been my advisor from the moment my foot landed on the steps of
the Milagro building at the laboratory. Not only am I obliged for your
help and support of my work on Milagro detector itself, but also for
your guidance in scientific research. I am thankful to you for proposing
the dark matter search as the main topic of my thesis.

I also mention Peter Nemethy who has been my advisor as well. While your
official duties were to guide my brother through his research, you payed
considerable attention to me as well. I thank you for introducing me to
Milagro in the first place.

During my work on Milagro I met many people whom I consider as founts of
knowledge and I am thankful to them for allowing me to glean the
knowledge. These people are Gaurang Yodh, Cy Hoffman, Gus Sinnis, Don
Coyne, Jim Ryan, Jordan Goodman, David Berley, David Williams, Tony 
Shoup and Bob Ellsworth. Thank you.

I also cherish my experience with the many post docs and graduate
students who made my life in Los Alamos more enjoyable. These are Rob
Atkins, Wystan Benbow, Diane Evans, Isabel Leonor, Joe McCullough,
Julie McEnery, Richard Miller, Frank Samuelson, Andy Smith, Kelin Wang,
Morgan Wascko and Stefan Westerhoff. Thank you again.

I am also in debt to Gerard Jungman and Salman Habib from the T-division
at the Los Alamos Laboratory for their contribution to this project.

To the names listed above I would like to add my twin-brother Roman. 
Roman has always been and always will be a special person in my life. 
The person who understands me from ``half a word'', my ally, my critic,
my friend and my co-dreamer. Without him many dreams would still be
dreams.

There are people whose help and support is often taken for granted. 
These are my parents and grandparents. They are the people who made me,
who taught me what ``right'' and what ``wrong'' is, who nourished and
diversified my interests. These are the people who taught me that an
educated person can do a job better regardless of whether it is
conducting esoteric research, moving furniture, weeding vegetable fields
or climbing telephone poles. Thank you for giving me all that and much
more.

I would also like to thank my wife Asya Shpiro and our daughter Sonya 
for their understanding and support during the final stages of my thesis
research.

I am also grateful to all my friends who supported and inspired me
during the ``dark'' phases of my life and shared joy and happiness
during the ``bright'' ones.

I was honored by the presence of professors Allen Mincer, Georgi Dvali, 
Todd Haines, Peter Nemethy and Engelbert Schucking at my dissertation 
defense.

\vspace{0.5in}
\ \\
Lazar Fleysher\\
New York University\\
April 14, 2003

\end{preface}
    \begin{preface}{Abstract}

The neutralino, the lightest stable supersymmetric particle, is a strong
theoretical candidate for the missing astronomical "dark matter".
Depending on their annihilation cross section, relic neutralinos from
early formation of the Universe trapped in orbits around massive objects
may currently be annihilating at measurable rates.  The Minimal
Supersymmetric extension of the Standard Model predicts that the gamma
rays emerging from one of the annihilation modes will give a distinct
monochromatic signal with energy between 100GeV and 10TeV, depending on
the neutralino mass. An additional "continuum" spectrum signal of
photons will be produced by the decay of secondaries produced in the
non-photonic annihilation modes.

Milagro is an air shower array which uses the water Cherenkov technique
and is capable of detecting TeV gamma rays from the direction of the Sun
with an angular resolution of less than a degree. It is the first
instrument capable of establishing a limit on the gamma-ray flux from
neutralino annihilations near the Sun.

In this report results of a search for neutralino to photon annihilation
with the Milagro gamma-ray observatory are presented. Results of a Monte
Carlo computer simulation of the neutralino annihilation density in the
Solar System suggest that a large portion of neutralino annihilations
($40-50\%$) happens outside the Sun which may give rise to a detectable
gamma-ray signal from the solar region. No significant gamma-ray signal
was observed from the Sun resulting in an upper limit on the sought for
photon flux. The upper limit can be translated to a neutralino-mass
dependent limit on the product of the neutralino-proton scattering
crossection $\sigma_{p\chi}$, the integrated photon yield per neutralino
in neutralino-neutralino annihilation $b_{\gamma}$ and the local
galactic halo dark matter density $\rho_{0}$. For example, assuming a
1TeV neutralino and ignoring the continuum contribution to the signal
gives an upper limit of $\frac{\rho_{0}}{0.3\
(GeVcm^{-3})}\frac{\sigma_{p\chi}}{10^{-41}\ 
(cm^{2})}b_{\gamma}^{\delta}<2.3$.

\end{preface}

    \tableofcontents          
    \listoffigures            
    \listoftables             
    \listofappendices        


\begin{thesisbody}


\chapter{Introduction}

\epigraph{George  Orwell, ``1984''}{%
Chapter 1, like Chapter 3, had not actually told him anything that he did
not know; it had merely systematized the knowledge that he possessed
already.}

\section{The dark matter problem.}

Perhaps, there is no problem of greater importance to cosmology and
astrophysics than that of the ``dark matter''. It is centered around the
notion that there may exist an enormous amount of non-luminous matter in
the Universe. The presence of the matter, which does not radiate and can
not be seen directly, can only be inferred by observing the effects it has
on other directly observable astronomical objects. 

It has always been known that there is matter in the sky which does not
emit any kind of radiation. For instance, the planets do not shine, but
their contribution to the mass of the solar system is negligible, so
worrying about non-luminous matter was not of a great concern. 

The first evidence that there is a significant amount of dark matter came
from Zwicky \cite{Zwicky}, in the thirties, from investigations of
clusters of galaxies. It was found that velocities of the galaxies in a
cluster were about 10 times larger than expected, indicating that there is
invisible gravitating matter in a cluster, holding the galaxies together.

Somewhat more reliable evidence was found in the 1970s by
Rubin~\cite{Rubin} by studying the rotation curves of spiral galaxies.
Kepler's law states that the rotational velocity around a gravitational
center depends only on the distance from the center and on the total mass
contained within the orbit. Thus, one expects: 
\[ GM(r)\cong v^{2}r \]
where $v$ -- is the circular velocity at a distance $r$ from the center of
the galaxy, $M(r)$ is the mass enclosed in the sphere $r$ and spherical
symmetry is assumed.

If the mass were associated with light (luminous matter) $v$ would
decrease as $r^{-1/2}$ beyond the point where the light cuts off. However,
it was found (See, for instance~\cite{rot_curve_book, Ostriker}, or
fig~\ref{rot_curve}) that $v\approx const$, corresponding to $M(r)\propto
r$, implying existence of dark halos around spiral galaxies.  The halos
could be made of brown dwarf stars, jupiters, planets and $100M_{\odot}$
black holes. Collectively, these objects are called MACHOs\footnote{MaCHO
is Massive Compact Halo Object} and are the main baryonic dark matter
candidates.

\begin{figure}
\centering
\includegraphics[width=3.9in]{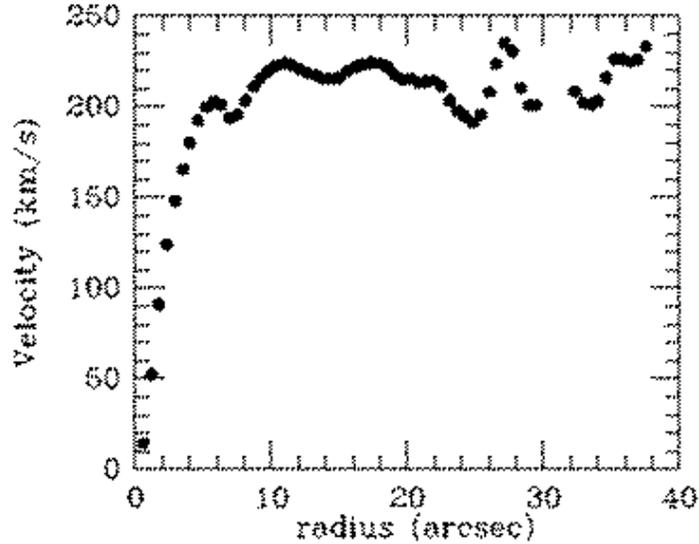}
\caption{A ``typical'' rotation curve of a ``typical'' galaxy, determined 
from 21cm observations.\cite{rot_curve_book}}
\label{rot_curve}
\end{figure}

Dark matter has important consequences for the evolution of the
Universe.  The standard, Hot Big Bang cosmology is remarkably
successful: it provides a reliable and tested account of the history of
the Universe from at least $t\sim 10^{-2}s$ until today ($t\sim 14\
Gyr$). At present, there is no strong experimental evidence
contradicting the theory. According to the theory, the Universe must
conform to one of three possible types with negative, positive or zero 
curvature. The value of the cosmological density parameter,%
\footnote{$\Omega =\rho/\rho_{c}$, $\rho$ -- energy density of the 
Universe, $\rho_{c}$ -- critical parameter;
\[ \Omega\left\{\begin{array}{cl}
<1 & \mbox{negative curvature, the Universe will expand forever} \\
=1 & \mbox{zero curvature Universe} \\
>1 & \mbox{positive curvature, the Universe will recollapse, eventually}
\end{array}\right. \]}
$\Omega_{total}$, determines which of the three possibilities applies to
our world. There is, however, a somewhat philosophical or even
aesthetical argument that makes $\Omega_{total}=1$ attractive. The point
is that as the Universe evolves, the value of $\Omega_{total}$ changes.
In fact, the value of $\Omega_{total}=1$ is unstable. If the Universe is
open $\Omega_{total}<1$, it will expand forever, until it is totally
empty $\Omega\stackrel{t\rightarrow\infty}{\longrightarrow}0$. On
contrary, if it is closed $\Omega_{total}>1$, it will recollapse to a
state with extremely high density
$\Omega\stackrel{t\rightarrow\infty}{\longrightarrow}\infty$. The
inflationary cosmology~\cite{Kolb}, which provides the most compelling
explanation for the smoothness of the Cosmic Microwave Background
Radiation (CMBR), predicts that the early Universe was extremely close
to flat $|\Omega_{total}-1|<{\cal O} (10^{-60})$, leading to the belief
that $\Omega_{total}$ is exactly one.

In fact, the most recent results from the studies of the CMBR with the
WMAP~\cite{WMAP} observatory yield $\Omega_{total}=1.02\pm 0.02$. The
same study implies that the matter component of the total energy density
is $\Omega_{m}=0.27\pm 0.04$ while ordinary baryonic component
constitutes only about $15\%$ of all matter in the Universe
$\Omega_{b}=0.044\pm 0.004$. The rest of the energy density
$\Omega_{\Lambda}=(\Omega_{total}-\Omega_{b})=0.73\pm 0.04$ is an
unknown from of energy (so-called ``dark energy'').

In any event, the abundance of baryons is not likely to account for all
matter even if $\Omega_{total}$ turns out to be slightly less than unity
and non-baryonic dark matter is almost required to dominate the
Universe. The particles or fields which comprise non­baryonic dark
matter must have survived from the Big Bang, and therefore, must either
be stable or have lifetimes in excess of the current age of the
Universe.  Among the non-baryonic dark matter candidates there are
massive neutrinos, axions~\cite{Turner} and stable supersymmetric
particles.

\section{Supersymmetry.}

The main goal of the elementary particle physics is to devise a model
which combines all particles and their interactions into one theory. 
The hope is that the development of supersymmetric theories (See, for
example,~\cite{Sohnius}) is a step towards the stated goal.  In these
theories, bosonic and fermionic fields are allowed to transform into one
another, and each particle is described by a multiplet containing bosons
and fermions. In such models, loops, divergent in quantum field
theories, cancel. Theoretical strong points of supersymmetry have
motivated many accelerator searches for supersymmetric particles. Most
of these have been guided by the Minimal Supersymmetric extension of the
Standard Model (MSSM) and are based on a missing-energy signature caused
by the escape of the lightest supersymmetric particles. In the MSSM, the
convergence of the renormalized gauge couplings at the grand unification
scale requires all masses of supersymmetric particles to appear between
$100\ (GeV)$ and $10\ (TeV)$~\cite{Amaldi}. Laboratory searches have set
lower mass limits, requiring lightest supersymmetric particles in MSSM
to possess masses greater than $20-30\ (GeV)$~\cite{PartDataG}.  Even
though no convincing evidence for existence of supersymmetric particles
has been found, they all have been given names. Bosonic ordinary
particles have fermionic superpartners with the same name except with
the suffix ``ino'' added, while fermionic ordinary particles have
bosonic superpartners with prefix ``s'' added. For example, Higgsino is
a superpartner for Higgs boson and selectron is a superpartner for
electron. There are several superpartners which have the same quantum
numbers and so can mix together in linear combinations. Since those do
not necessarily correspond to any ordinary particle, they are given
different names. For instance, the photino, Higgsino and Zino can mix
into arbitrary combinations called the neutralinos. The lightest
neutralino is a stable supersymmetric particle and makes the ``best''
candidate for a solution of the ``dark matter problem''(first suggested
in~\cite{PagelsandPrimack}, also see~\cite{Jungman} for an extensive
review).



\section{Detection Methods.}

There are several ways to test the hypothesis that stable neutralinos
exist and contribute to the dark matter. They include direct searches
with extremely sensitive devices which can detect energy deposited by an
elastically scattered neutralino off a nucleus and indirect searches
which look for products of neutralino-neutralino annihilations.

\subsection{Direct detection.}

\begin{figure}
\centering
\includegraphics[width=0.7\textwidth]{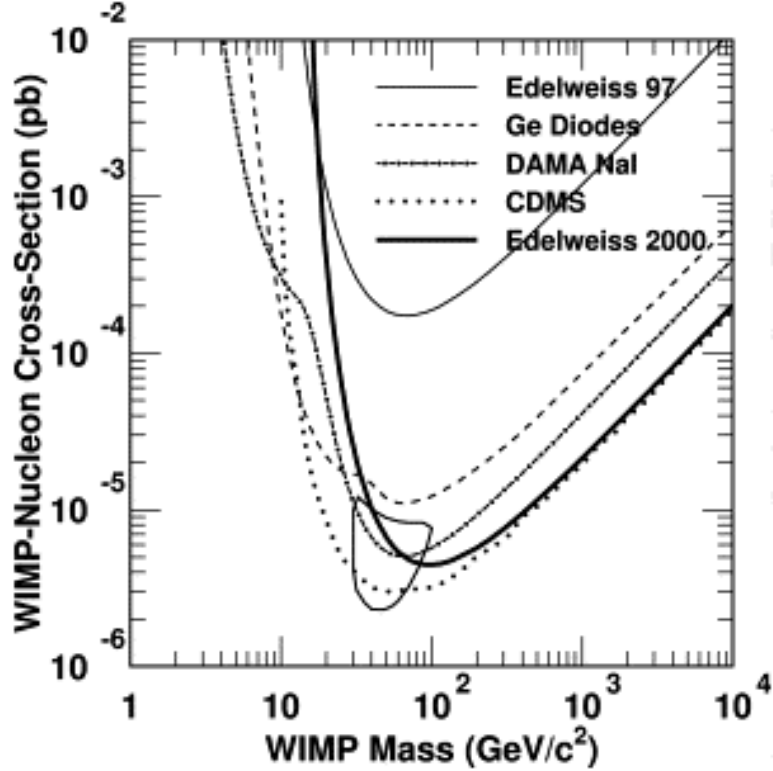}
\caption{Exclusion regions for neutralino-nucleon crossections obtained
         from different direct-search experiments. Closed contour is 
         allowed region at $3\sigma$ confidence level from the DAMA 
         experiment. The plot is adopted from~\cite{Edelweiss}.}
\label{fig:intro:edelweiss}
\end{figure}

The Italian/Chinese collaboration (DAMA) has reported an annual
modulation in the total count rate over 4 years. They interpret this as
consistent with the annual modulation predicted for WIMPs~\cite{DAMA}.
This, however, is not a widely accepted result because of some possible
modulating systematic errors. The CDMS experiment has obtained data that
appear to exclude the DAMA result~\cite{CDMS}. They reach a
spin-independent WIMP-nucleon scrossection limit around $2\cdot
10^{-42}\ (cm^{2})$ in the mass range $20-100\ (GeV)$. Edelweiss has
also released results that significantly cut into the DAMA allowed
region~\cite{Edelweiss}. The summary of the limits of direct searches is
shown on the figure~\ref{fig:intro:edelweiss}.

\subsection{Indirect Detection.}

Indirect searches also have received considerable attention from
experimenters. For example, the Kamiokande and SuperKamiokande
underground neutrino detectors have set limits on solar and terrestrial
neutralino-induced muon fluxes~\cite{Mori, SuperK2001}.

Another possible method for detecting dark matter particles is from
their annihilation into $\gamma$-rays. One of the many possible
gamma-producing channels is production of monochromatic gamma-rays:

\[ \chi\chi \ \ \ \rightarrow \ \ \ \gamma\gamma , \ \ \ 
   \chi\chi \ \ \ \rightarrow \ \ \ Z\gamma \]

Even though it is difficult to estimate the rates of these processes
because of uncertainties in the supersymmetric parameters, cross
sections and the neutralino distribution, since the annihilating
neutralinos move at galactic velocities $v/c\sim 10^{-3}$ the outgoing
photons will give very distinct monochromatic signals\footnote{If these
two lines can be resolved, the relative strength of the two could give a
handle on the composition of the neutralino. This is because despite the
fact that the two processes are closely related, there are some
differences which depend on the composition.} in each annihilation mode:
\[ E_{\gamma}=M_{\chi}, \ \ \ \ E_{\gamma}=
       M_{\chi}\left(1-\left(\frac{m_{Z}}{2M_{\chi}}\right)^{2}\right) \]
which has no conceivable origin from any known astrophysical sources. 

As was mentioned earlier, neutralinos, if they are to be the dark matter,
should have non-zero relic abundance today, but their number density is so
small that almost no annihilations happen. An observation of such an event
from some random point in the Universe is not feasible. However, since the
density of neutralinos in the vicinity of a gravitational center will be
larger than in other parts of the Universe and because the annihilation
rate is proportional to the square of the neutralino density, there will
be an enhanced flux of high energy $\gamma$-rays from such regions.
Therefore, it is tempting to look at signals from well studied gravitating
objects, such as nearby galaxies and the Milky Way Galaxy, and examine the
energy spectrum for a monoenergetic signal.

The present high energy gamma-ray experiments, such as EGRET and the
Whipple atmospheric Cherenkov Telescope, lack the sensitivity to detect
annihilation line fluxes predicted for most of the allowed supersymmetric
models and halo profiles. However, the next generation ground-based and
satellite gamma-ray experiments, such as GRANITE-III, VERITAS and GLAST,
will allow exploration of large portions of the MSSM parameter space,
assuming that the dark matter density is peaked at the galactic
center.~\cite{Bergstrom3}

The Sun is also a large gravitating object and one could study the solar
spectrum for the neutralino annihilation signal. Of course, that is
possible only with a non-optical high resolution instrument, capable of
monitoring the Sun at energies between $100\ (GeV)$ and $10\ (TeV)$.
Several semi-analytical estimates for the detection rates for several
ground-based and satellite experiments are available in the literature.
However, a more careful computer simulation following the decaying 3-D
neutralino orbits with detailed elastic scattering and planetary
perturbations accompanied by simulations of the solar magnetic field
smearing and shadowing of the galactic cosmic rays by the Sun will
provide a more definitive prediction on the neutralino annihilation
rate.

The structure of this work is the following:
chapter~\ref{chapter:air_showers} discusses how high energy cosmic
particles can be detected. This is followed by a brief description of
the Milagro detector, capable of monitoring the overhead sky at energies
near $1\ (TeV)$, in chapter~\ref{chapter:detector}. A presentation of
the data analysis techniques employed in the current work is given in
chapter~\ref{chapter:analysis_techniques}. 
Chapter~\ref{chapter:simulations} discusses the computer simulations
which are used to predict the gamma ray flux from the near-solar
neutralino annihilations. Chapter~\ref{chapter:results} discusses the
results of the search for the relic neutralinos and is followed by a
summary in chapter~\ref{chapter:conclusion}.

\chapter{Extensive Air Showers}
\label{chapter:air_showers}

\epigraph{George  Orwell ``1984''}{%
All one knew was that every quarter astronomical numbers of boots were 
produced on paper, while perhaps half the population of Oceania  went 
barefoot.

}

There are several main reasons which govern the choice of a detector type
to be used in high energy photon search from the Sun. First of all, it
should be a non-optical device capable of monitoring the solar region.
Because, the Earth's atmosphere is opaque to gamma rays, satellite-based
detectors need to be constructed to detect gamma rays. Indeed, small
detectors sensitive to gamma rays at energies below a few $GeV$ have been
constructed and used successfully.\footnote{Future satellite detectors
such as GLAST should register particles with energies as high as $300\
(GeV)$~\cite{glast}.} These detectors employ techniques developed for
accelerator experiments where an incoming photon's direction is determined
by $e^{+}e^{-}$ tracking detectors and the photon's energy is usually
measured by a total-absorption calorimeter. However, the expected low and
rapidly decreasing with photon energy $\gamma$-ray flux requires detectors
with rather large collection areas and long exposure periods. Such
detectors can be built on the surface of the Earth only.

Even though direct detection of $\gamma$-rays is not possible by
ground-based instruments, at energies above several $GeV$ indirect
gamma-ray detection is possible. Such very high energy photons initiate
extensive air shower (EAS) cascades of secondary particles which are
detectable by ground-based detectors. Knowledge of the EAS structure is 
required to infer information about the primary photon.

\section{Development of EAS.}

Although an extrapolation from known particle physics might be necessary
to describe the initial phase of the shower development, it is believed
that the structure of the EAS is well understood. A high-energy primary
photon interacts with electromagnetic fields of air molecules in the upper
atmosphere producing an electron-positron pair which in turn produces
high-energy photons via bremsstrahlung. The resulting electro-magnetic
cascade grows geometrically as it propagates through the atmosphere. The
shower growth stops when the mean energy of electrons and positrons falls
below the critical energy ($E_{c}\sim 85\ (MeV)$) where the ionization
energy-loss mechanism, which does not produce additional shower particles,
becomes dominant. After this point (called the shower maximum) the energy
of particles and their number in the shower start to decrease as the
shower continues its propagation towards the ground level. Nevertheless, a
large number of shower particles may reach the ground and may be detected.

Moreover, because the secondary particles are ultra-relativistic, they
retain the directionality of the incident gamma ray and the cascade
arrives to the ground as a thin front perpendicular to the direction of
the primary photon. The density of shower particles in the front will
decrease with distance from the extrapolated incident gamma-ray
trajectory. This trajectory is called the core of the shower.

The shower development is a stochastic process and while some analytical
calculations have been performed, computer simulations are generally
employed to study the properties of the air shower cascades.

\subsection{Longitudinal Development of Extensive Air Showers.}

\begin{table}
\centering
\begin{tabular}{|c|c|c|c|c|c|c|} \hline
$E_{th}$,  & \multicolumn{3}{c|}{k=electron} 
                 & \multicolumn{3}{c|}{k=photon} \\ \cline{2-7}
MeV &  $A$ &  $a$ &  $b$  &  $A$ &  $a$  & $b$   \\ \hline
 1  & 0.92 & 0.00 &  0.45 & 4.80 & -0.88 &  0.83 \\ \hline
 5  & 0.75 & 0.19 & -1.22 & 2.98 & -0.69 & -1.49 \\ \hline
 10 & 0.63 & 0.35 & -2.57 & 2.13 & -0.57 & -3.45 \\ \hline
 20 & 0.50 & 0.53 & -4.22 & 1.45 & -0.36 & -5.51 \\ \hline
\end{tabular}
\caption{Values of parameters $A$, $a$ and $b$ for modified Greisen 
and NKG formulae.}
\label{table:form_parameters}
\end{table}

\begin{figure}
\centering
\includegraphics[width=0.4\textwidth]{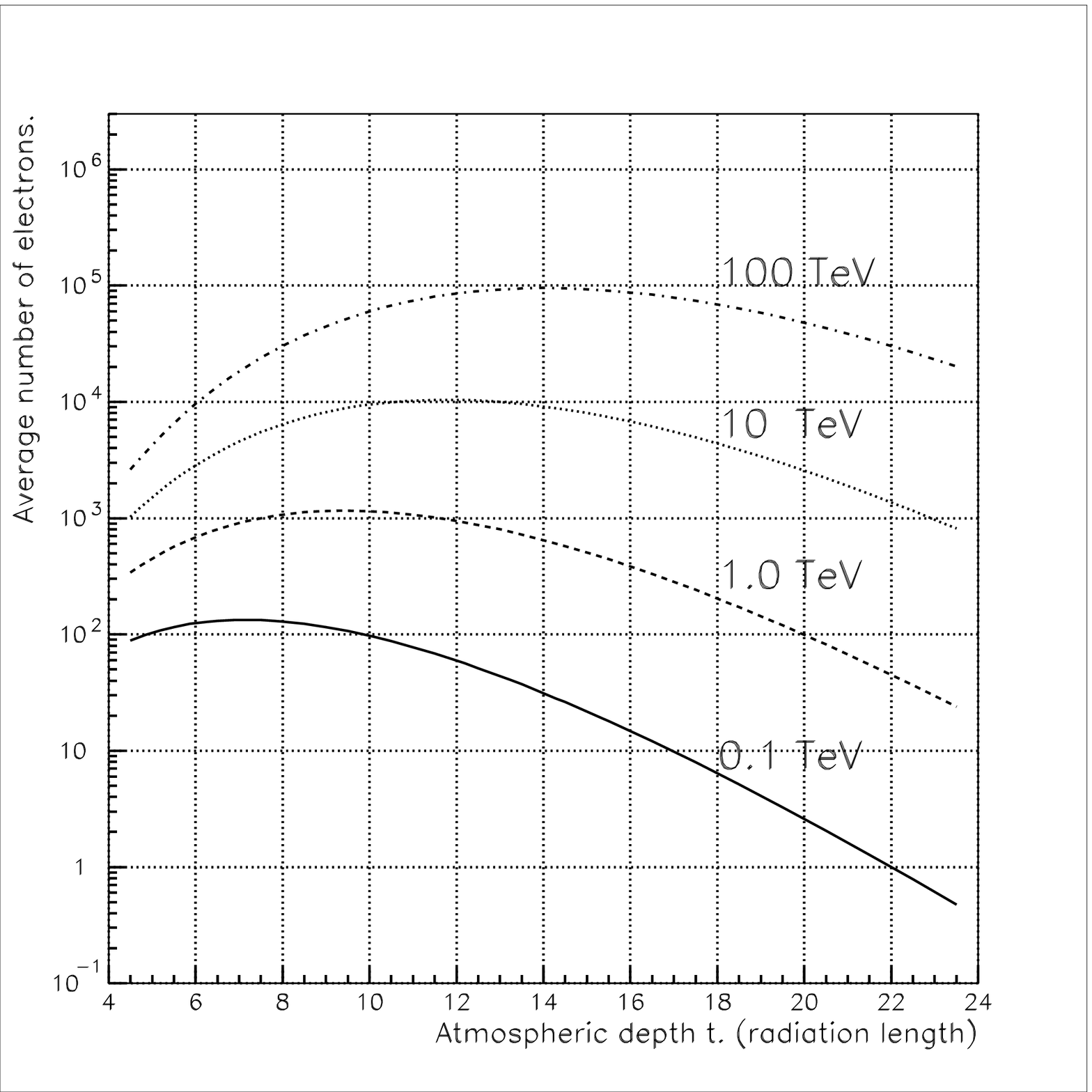} 
\hspace{0.02\textwidth}
\includegraphics[width=0.4\textwidth]{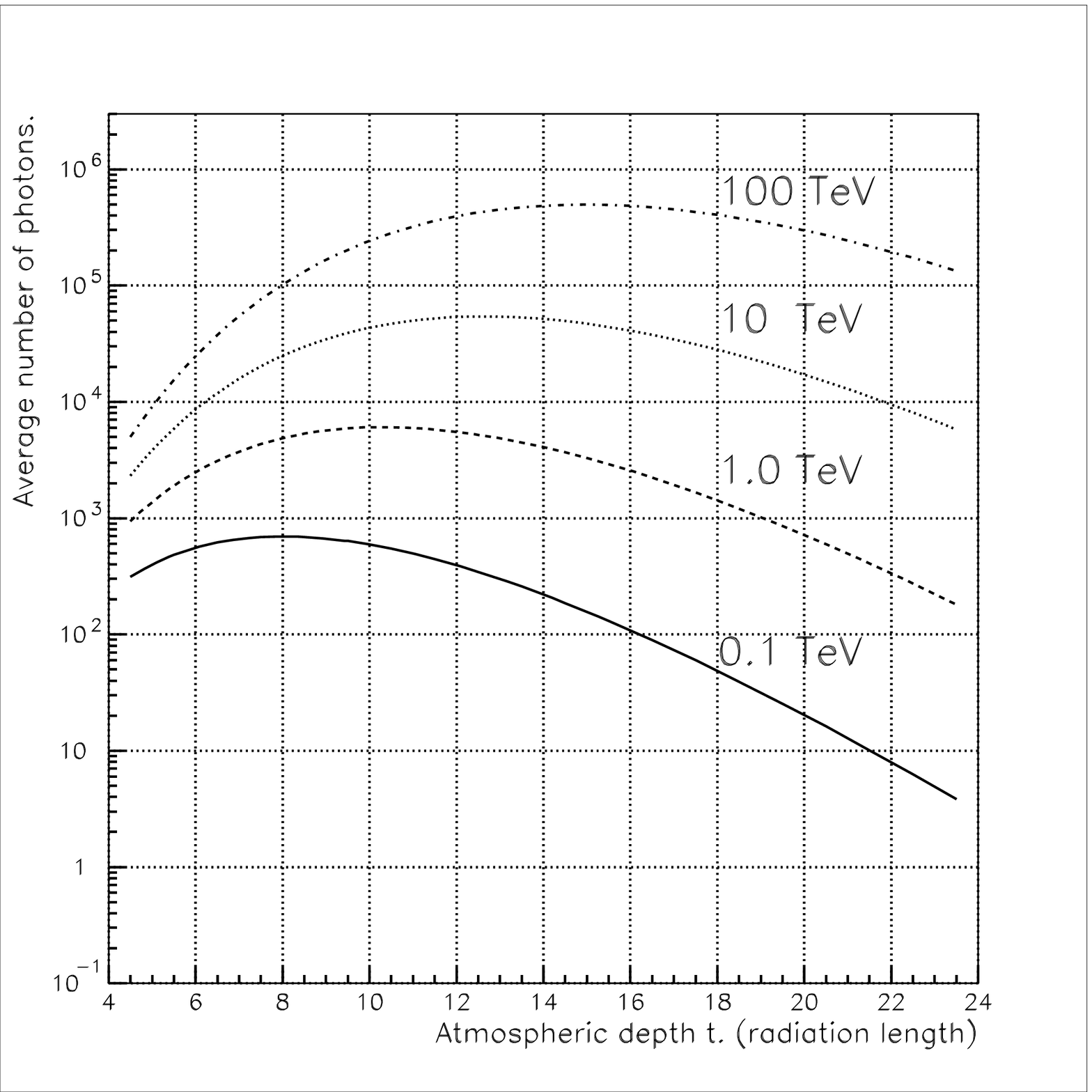}
\caption{Longitudinal development of electron (left) and photon (right) 
         compontents of gamma-ray showers with $E_{th}=1\ (MeV)$ 
         particle detection threshold.}
\label{fig:long_dev}
\end{figure}

The average number of electrons $N_{e}$ and photons $N_{\gamma}$ in an
electromagnetic shower can only depend on the primary energy $E_{0}$ and
the thickness of the traversed matter $t$. Moreover, if $E_{0}$ is
expressed in units of critical energy $E_{c}$ and $t$ is in units of
radiation lengths $X_{0}$, the number of electrons and photons is almost
independent of the specific shower propagation medium. Usually, however,
detectors can register particles with energies above some $E_{th}$,
thus, often, it is desired to know the number of particles in a shower
with energies greater than $E_{th}$. According to~\cite{sciascio} the
average number of particles $N_{k}(E_{0},E_{th},t)$ of type $k$ with
energy above $E_{th}$ at atmospheric depth $t$ in a shower initiated by
a photon with energy $E_{0}$ can be described by a modified Greisen
formula:

\[ N_{k}(E_{0},E_{th},t)=A_{k}(E_{th})\frac{0.31}{\sqrt{y}}
                                           e^{t_{k}(1-1.5\ln s_{k})} \]
\[ y=\ln\frac{E_{0}}{E_{c}}, \ \ \ t_{k}=t+a_{k}(E_{th}), \ \ \
                                       s_{k}=\frac{3t_{k}}{t_{k}+2y} \]

The parameterization is valid for $4<t<24$ and $0.1<E_{0}<10^{3}\ (TeV)$. 
The coefficients $A_{k}(E_{th})$ and $a_{k}(E_{th})$ are given in the
table~\ref{table:form_parameters} The graphical illustration of the
number of particles in a shower is presented in
figure~\ref{fig:long_dev}.

\subsection{Lateral Development of Extensive Air Showers.}

\begin{figure}
\centering
\includegraphics[width=0.4\textwidth]{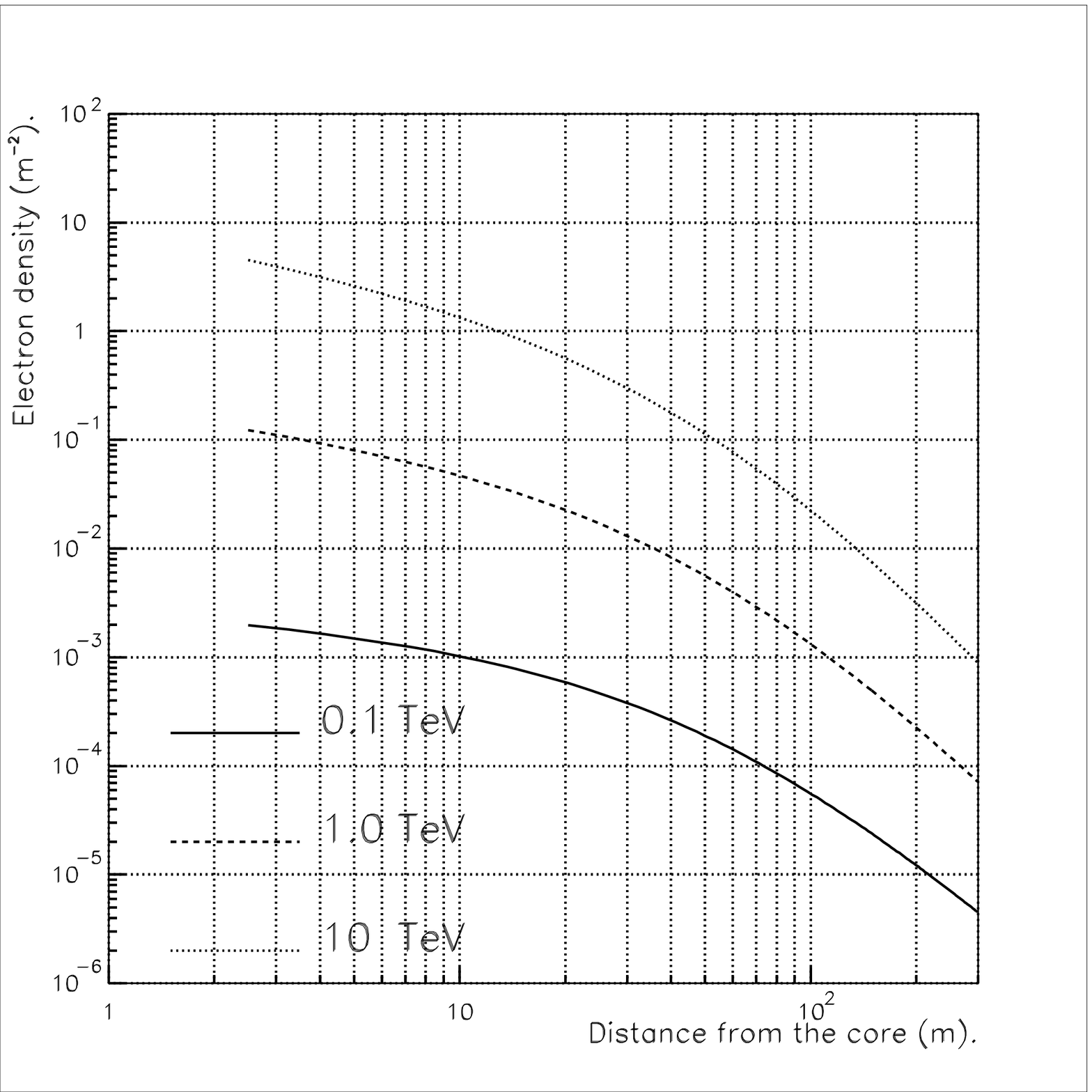} 
\hspace{0.02\textwidth}
\includegraphics[width=0.4\textwidth]{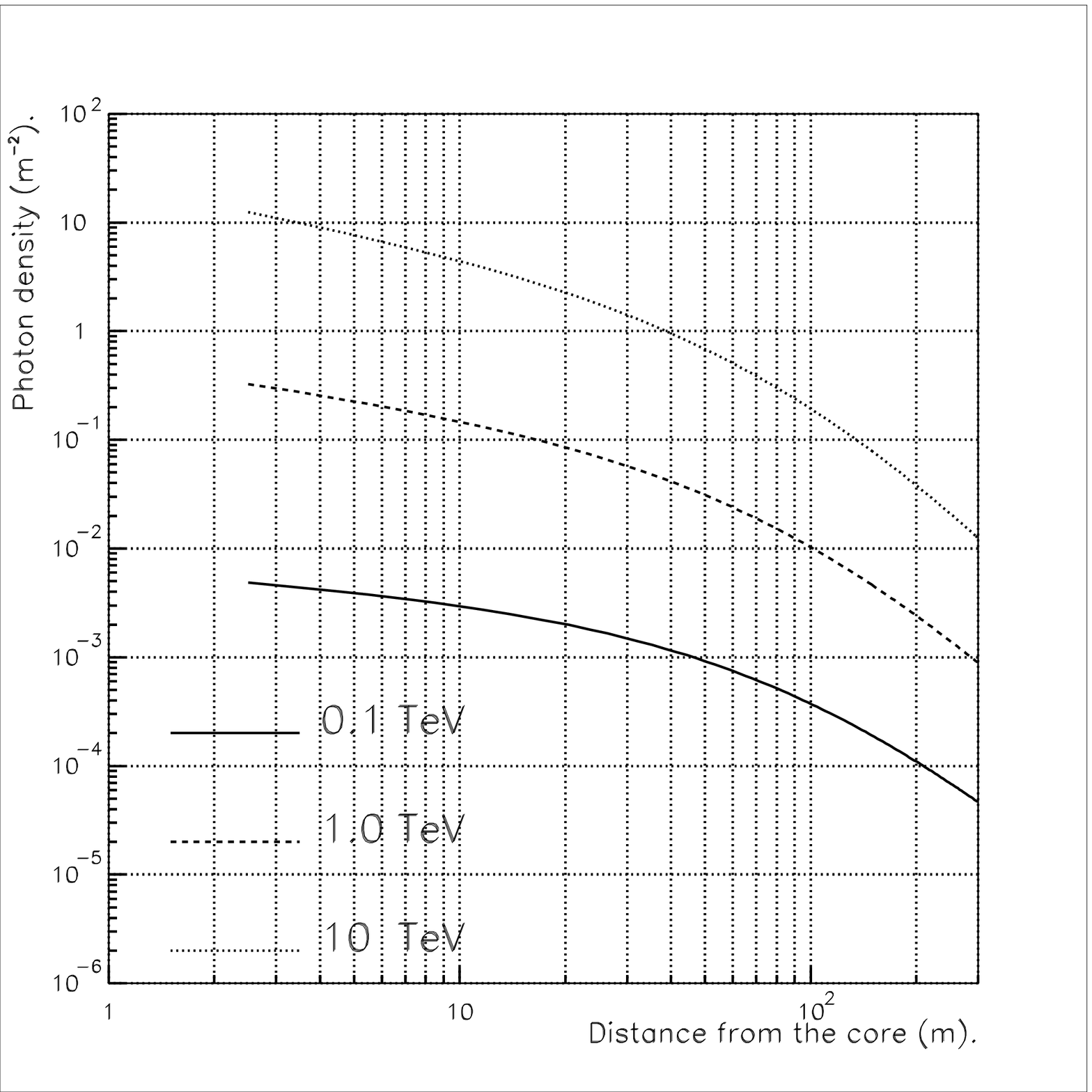}
\caption{Density of electrons (left) and photons (right) in a gamma-ray 
         shower at atmospheric depth of 20 radiation lengths as a 
         function of the core distance with $E_{th}=1\ (MeV)$. Curves 
         are normalized to the total number of respective particles.}
\label{fig:lat_dev}
\end{figure}

The average surface density $\rho_{k}(E_{0},E_{th},t,r)$ of particles of
type $k$ in the shower front with energies greater than $E_{th}$ at a
distance $r$ from the shower axis and at the atmospheric depth $t$ can
be described by a modified Nishimura-Kamata-Greisen (NKG)
function~\cite{sciascio}

\[ \rho_{k}(E_{0},E_{th},t,r)=
     \frac{N_{k}(E_{0},E_{th},t)}{R_{k}^{2}} f(r/R_{k},\tilde{s}_{k}) \]

\[ \tilde{s}_{k}=\frac{3(t+b_{k}(E_{th}))}{t+b_{k}(E_{th})+2y}, \ \ \ \
 f(x,z) =\frac{1}{2\pi}\cdot\frac{1}{B(z,4.5-z)} x^{z-2}(1+x)^{z-4.5} \]

where $B(z,w)$ is the beta-function so that 
$2\pi\int_{0}^{\infty}f(x,z)xdx=1$. The characteristic scattering length 
for photons $R_{\gamma}=\frac{m_{e}c^{2}\sqrt{4\pi/\alpha}}{E_{c}}X_{0}$ 
is the Moli\`ere scattering unit and for electrons --- 
$R_{e}=R_{\gamma}/2$.

The values of parameters $b(E_{th})$ are given in the
table~\ref{table:form_parameters} and the average density of photons
and electrons per unit area is shown of figure~\ref{fig:lat_dev} as a
function of distance from the shower axis.

\subsection{Temporal Distribution of Extensive Air Shower Particles.}

\begin{figure}
\centering
\includegraphics[width=0.56\textwidth]{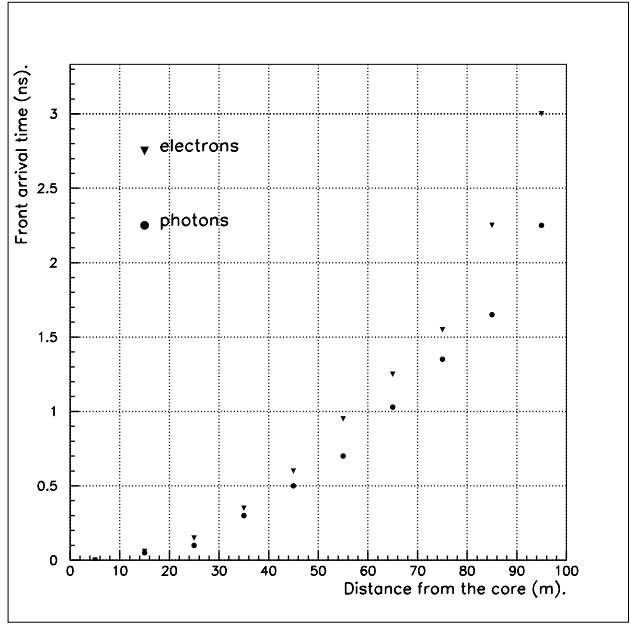}
\caption{Average arrival time of the shower front as a function of core 
distance (illustration).}
\label{fig:air_showers:temporial}
\end{figure}

Because the air shower detectors determine the primary particle direction
using particle arrival times, knowledge of the shape of the shower front
is important for achieving the best possible angular resolution. Results
of Monte Carlo simulations~\cite{sciascio} of shower front are shown on
figure~\ref{fig:air_showers:temporial}. The shower front appears to have a
parabolic shape as a function of distance from the core. At large
atmospheric depths air shower photons travel faster than air shower
electrons thus the photonic front is curved less than the electronic one.
The thickness of the shower is defined by the distribution of the shower
particle arrival times at distance $r$ from the shower core and increases
with the core distance $r$. At small core distances the fluctuations of
arrival time around its average appear to be smaller for the photon
component than for the electron one and, consequently, photonic thickness
is smaller than the electronic one. At large core distances, however, the
electronic contribution to the shower is quite small due to electron
ionization losses in the atmosphere compared to the photonic component.

\section{Cosmic rays.}

Among the particles which enter the Earth's atmosphere gamma rays present
a very small fraction. Most of the particles are so-called cosmic rays
consisting of protons, helium nuclei and the nuclei of the heavier
elements such as carbon, oxygen and iron. Just as gamma rays, cosmic rays
initiate cascades in the atmosphere. Heavier cosmic rays may interact with
air nuclei and produce high energy nucleons. High energy protons
interacting with air nuclei may produce neutral and charged pions. Neutral
pions have a rather short lifetime and decay, dominantly into photons
which may, in turn, produce electromagnetic cascades.  Charged pions have
longer lifetime and may decay into muons and neutrinos or may interact
with the air nuclei creating secondary high energy hadrons and replenish
the cascade. Muons, produced in the cascade, may also survive to the
ground level. The shower stops its growth when secondary high energy
hadrons and photons can not be produced.

Because cosmic rays are charged particles they interact with the
interstellar and interplanetary magnetic fields and do not provide
directional information about their sources. Thus, cosmic rays may
constitute an unwanted background for a gamma-ray telescope. Presence of
muons in hadronic cascades is often exploited to differentiate
cosmic-ray cascades from the gamma-ray ones.

\section{Air shower detection methods.}

The detectors used in high-energy astrophysical experiments are based on
those developed for laboratory ones. Since the showers extend over large
areas large detectors are necessary to sample the shower. Cloud/bubble
chambers are not suitable for electronic data recording and gas-filled
discharge tubes are not practical for large area detectors. Because
charged particles constitute a large fraction of the shower particles
scintillation and Cherenkov radiation detection techniques are employed in
modern air shower detectors. Cherenkov detectors detect radiation produced
when a charged particle moves through a dielectric medium at velocity
greater than that of light in the medium. Scintillation counters detect
light (luminescence) produced as a result of recombination of the
electron-hole pairs created by ionizing particles traversing the
scintillation medium.

If the Earth's atmosphere is used as the detection medium this results
in the air-Cherenkov and ``fluorescence'' detectors. Air-Cherenkov
detectors typically have energy thresholds of about several hundreds of
$GeV$, while fluorescence ones can detect high energy cosmic rays with
energies above $100\ (PeV)$. Such detectors typically have good angular 
resolution but are very narrow field-of-view devices and can operate 
only on cloudless, moonless nights.

Scintillation arrays have also been built and, due to their sparseness,
have rather high energy thresholds (typically above several tens of
$TeV$). Such detectors have worse angular resolution but can observe the
entire overhead sky 24 hours a day regardless of weather conditions.

The goal of the Milagro project is to built a detector sensitive to
cosmic gamma rays at energies around 1~TeV and capable of continuously
monitoring the overhead sky with angular resolution of less than 1
degree.

\chapter{The Milagro Detector}
\label{chapter:detector}

\epigraph{George  Orwell ``1984''}{%
In a sense it told him nothing that was new, but that was part of the
attraction. It said what he would have said, if it had been possible for
him to set his scattered thoughts in order.}


Milagro employs the water Cherenkov detection technique which is widely
used in particle physics experiments but is new to air shower detection.
The use of water as a detection medium has several advantages: it is
possible to construct a large instrument that can detect nearly every
relativistic charged shower particle falling within its area by
observing the Cherenkov radiation the particle produced.  At a typical
detector altitude, there are 4-5 times more photons in an extensive air
shower than charged particles. In a conventional EAS array these photons
are undetected.  When these photons enter the water, they convert to
electron-positron pairs or Compton scatter electrons which, in turn,
produce Cherenkov radiation that can be detected.  Consequently, Milagro
has a very low energy threshold for an EAS array.\footnote{Tibet is a
conventional EAS array with energy threshold of several TeV. Such a low
threshold could be achieved only due to its high altitude of 4300 m
above sea level~\cite{Tibet}.}

This chapter presents the Milagro detector with its physical and
electronic components, event reconstruction methodology and performance
characteristics. For a more detailed description see 
references~\cite{milagritoNIM} and~\cite{milagroNIM}.

\section{General description.}

\begin{figure}
\centering
\includegraphics[width=5.0in]{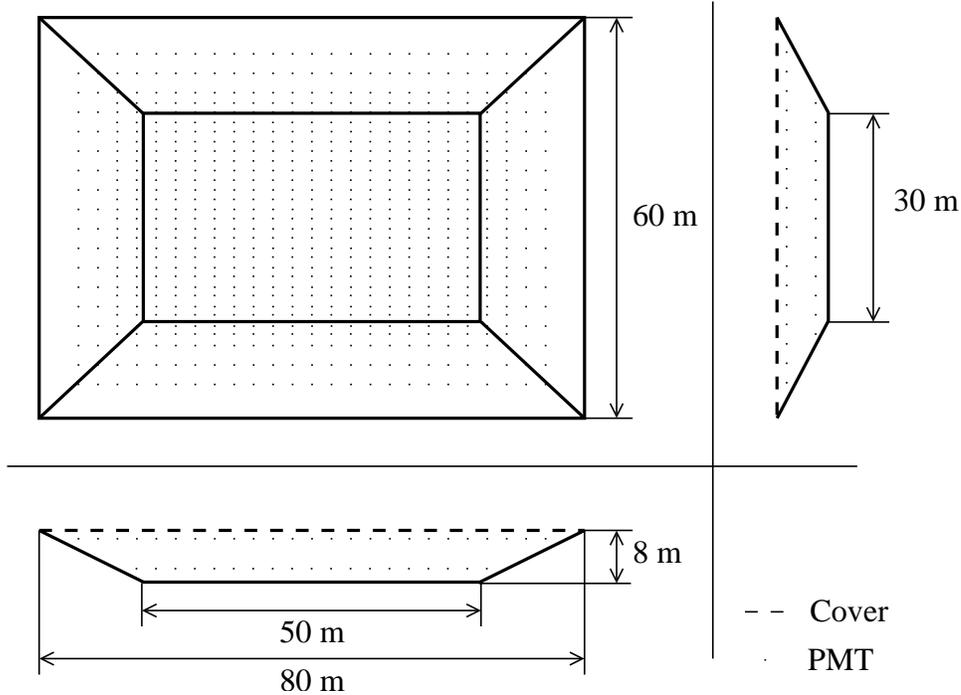}
\caption{Schematic view of the Milagro pond.}
\label{detector:pond}
\end{figure}

The Milagro experiment is a part of what is now known as the Fenton Hill
Observatory located at $35.88^{\circ}$ North latitude and
$106.68^{\circ}$ West longitude in the Jemez Mountains near Los Alamos,
New Mexico. At an altitude of $2650\ (m)$ above the sea level, its
atmospheric overburden is about $750\ (g/cm^2)$. The Milagro detector,
commissioned in June of 1999, records about 1700 extensive air shower
events per second and is sensitive to gamma-showers with energies above
$100\ (GeV)$. The duty-cycle of the detector is about 90\%. The
remaining time the detector is off for scheduled maintenance and/or when
the environmental conditions do not warrant its operation (forest fires,
loss of electrical power, etc). Milagro is built in a pre-existing 21
(metric) kilo-ton trapezoidal water reservoir (see
figure~\ref{detector:pond}) filled with pure water and instrumented with
two horizontal planar layers of photomultiplier tubes (PMT). The top
layer (AS) has 450 PMTs arranged on a $2.8\times 2.8\ (m)$ grid, $1.5\
(m)$ below the water surface. The second layer (MU) has 273 PMTs located
under about $6\ (m)$ of water on an interlaced $2.8\times 2.8\ (m)$
grid. The photo-tube assembly is buoyant with the weight distribution
allowing the photo-cathode to face upward when the assembly is submerged
and anchored to the bottom of the pond with a Kevlar string. A
reflecting conical baffle is installed in each PMT assembly to increase
the light collection area and block horizontal and upward traveling
light. The signals from the PMTs are delivered to the data-acquisition
(DAQ) system for processing and recording. A high-density polypropylene
liner and cover are installed to ensure water-tight bottom and walls of
the pond and light impermeability of the whole detector.

\subsection{Photomultiplier tube.}

As was mentioned above, the Cherenkov radiation produced in the detector
volume is detected by photo-multiplier tubes. Unlike conventional
electro-vacuum tubes where electrons are injected into the tube due to
thermal emission from its cathode, the injection of electrons
(photoelectrons or PE for short) into the photo-tube is caused by light
via the photo-electronic effect. Due to an externally applied voltage,
the electrons travel towards the anode of the tube. However, on the way
they encounter a dynode chain which plays the role of an amplifier. When
electrons hit a dynode, secondary electrons are emitted which bombard
the next dynode on their way to anode. In such a setup, enormous
amplification can be reached with a relatively short dynode chain.

The amplification is not the only important parameter of a PMT, the others
include:
\begin{description}
\item[Spectral Sensitivity:] PMT should be sensitive to the wavelengths 
     produced in the \\ Cherenkov radiation.
\item[Quantum efficiency:] The ratio of the number of photoelectrons
     produced to the number of incident on photocathode photons is called
     quantum efficiency. Ideally it is equal to unity.
\item[Time resolution:] Time resolution of a PMT is thought to be 
     limited by fluctuations in the photoelectron cascade development, 
     especially on its early stages, especially between the photocathode 
     and the first dynode. Lower light intensities generally lead to 
     poorer time resolution.
\item[Pre-pulsing:] Pre-pulses on the PMT output are thought to occur
     when photoelectrons are produced by the first dynode, exposed to the
     incident light. Higher light intensities generally lead to higher
     pre-pulsing probability.
\item[Late pulsing:] Late pulses on the PMT output are thought to occur
     when all photoelectrons are reflected off the first dynode and
     re-enter the dynode chain producing a PMT pulse later than should
     have. Higher light intensities generally lead to lower late pulsing
     probability.
\item[After pulsing:] After pulses on the PMT output are thought to 
     occur when residual gas molecules in the PMT are being ionized by 
     the photoelectrons. The ions hitting the photocathode may cause 
     secondary electron emission which would produce a secondary pulse. 
     Higher light intensities generally lead to higher after pulsing 
     probability.
\item[Saturation:] Saturation is the effect of decreased PMT 
     amplification for higher intensity input. This is caused by 
     inability of dynode chain to accelerate increased numbers of 
     secondary electrons to sufficiently high energy.
\end{description}

After testing several PMT models, the Hamamatsu 8-inch 10-stage R5912SEL
was selected for this application. It has relatively high quantum
efficiency ($0.2-0.25$) at wavelengths of $325-450\ (nm)$, good timing
resolution ($2.7\ (ns)$ at 1PE), relatively low late/pre/after pulsing
rates (about 5\%) and relatively long linear response (up to about 75
PE). For a more detailed description of the PMT characteristics see
references~\cite{milagritoNIM} and~\cite{ileonor}.

\subsection{PMT pulse model, time over threshold.}
\label{detector:tot_model}

Each PMT should provide information about the intensity of light
incident on the PMT photocathode and the time when the light was
registered. Since the total charge in a PMT pulse (number of
photoelectrons) is proportional\footnote{Provided that the PMT
saturation limit is not reached} to the incident light intensity, if the
PMT pulse quickly charges a capacitor which is then slowly discharged
via a load resistor, the total charge in the PMT pulse can be measured
by the capacitor discharge time. This suggests that the time spent by a
pulse over a preset threshold level is associated with the input light
level\footnote{In fact, in this model ToT is proportional to the
logarithm of the number of PEs.} and is the main assumption of the
time-over-threshold (ToT) method employed by Milagro. The PMT signal can
be digitized with logical ``one'' when the PMT pulse exceeds the
discriminator threshold and logical ``zero'' otherwise. A
time-to-digital converter (TDC) attached to such a digital output will
record the ToT. The beginning of the logical ``one'' provides the PMT
pulse arrival time ($T_{start}$). This method of measuring PMT pulse
charge has several advantages over a conventional method when the PMT
pulse is sent to an analog-to-digital converter (ADC). ADCs usually have
narrow dynamic range, and are relatively slow and expensive devices.

Presence of pre- and after- pulses will distort the PMT pulse and it
will not conform to the ToT model described above. Since strong pulses
are more likely to be distorted, two thresholds, high (at the level of
about 7 PE) and low (at about 1/4 PE) are introduced in the Milagro
electronics. Large pulses will therefore cross both thresholds and the
time-over-high-threshold (HiToT) is a much better measure of the pulse
charge. Two close weak pulses will cross only the low threshold leading
to excessively long time-over-low-threshold (LoToT), but absence of
HiToT will flag such signals. To avoid use of two TDCs on a single PMT
channel a logical exclusive OR operation is executed on LoToT and HiToT
digital outputs leading to a train of raising and falling edges
corresponding to each threshold crossing (see
figure~\ref{detector:fig_tot_model}). If a PMT pulse is weak and only
low threshold is crossed, the edge train contains only two edges (2-edge
pulse), if both thresholds are crossed --- four edges are recorded
(4-edge pulse). Each TDC installed in Milagro is capable of recording up
to 16 discriminator level crossings.

The train of edges with their TDC counts constitute the raw PMT signal.

\begin{figure}
\centering
\includegraphics[width=4.5in,height=2.5in]{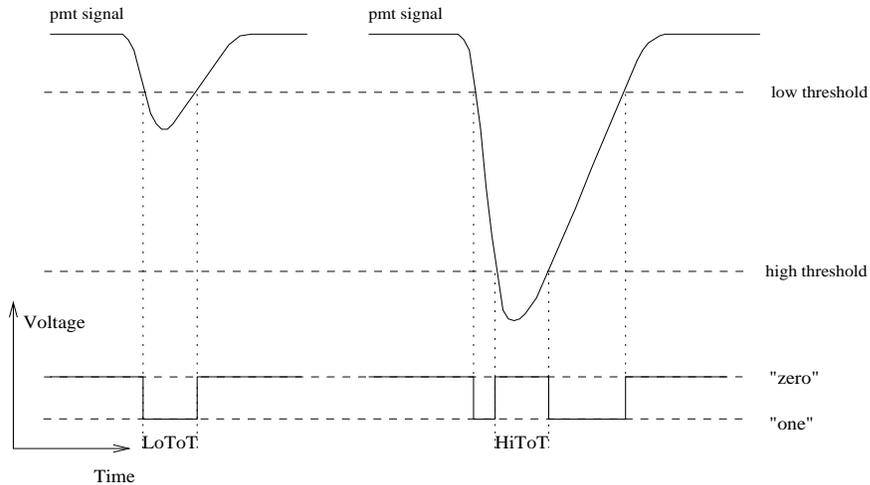}
\caption{Illustration of HiToT, LoToT and edge-train}
\label{detector:fig_tot_model}
\end{figure}

\subsection{The Detector trigger.}

All PMT channels in Milagro were manufactured as uniformly as possible,
facilitating a simple multiplicity triggering logic. Indeed, as an
extensive air shower front hits the detector a majority of PMT signals
arriving at the outputs of the PMT channels will be in close coincidence
with each other. The coincidence window was chosen to be $300\ (ns)$. If
more than 60 PMT signals arrived within the window, a trigger was
generated to the DAQ system. TDC modules are then read out with look
back time of $1.5\ (\mu s)$ and the event is saved. It is desirable to
trigger the detector at a low multiplicity requirement to lower the
detector energy threshold. However, lowering it beyond 60 would increase
the probability of triggering on muon events which is not the goal of
the project. The generated trigger was sent to a Global Positioning
System clock for absolute event time readout.

The TDC readouts from all PMTs channels and the trigger time constitute
the raw event data and are sent for further software conditioning and
processing.

\begin{figure}
\centering
\includegraphics[width=4.5in]{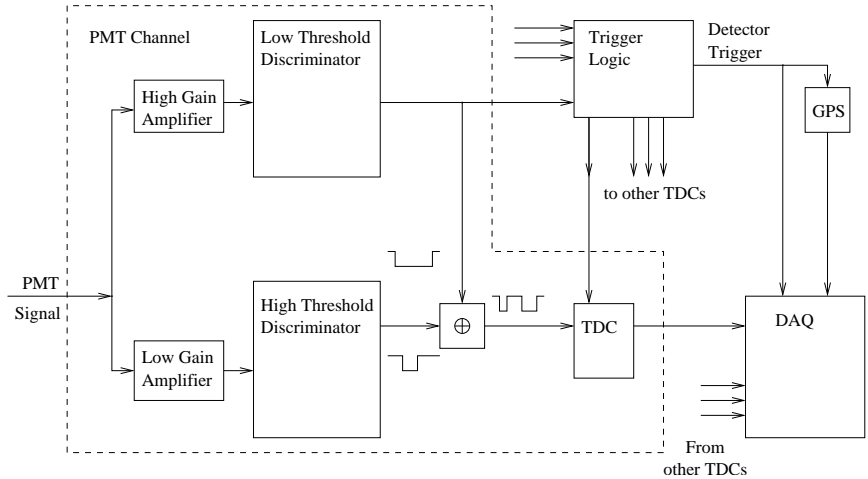}
\caption{Block diagram of detector electronics}
\end{figure}

\section{Event Reconstruction.}

The ultimate goal of any high energy gamma-ray telescope project is to
study the properties of the objects which emitted the particles. This
means that the characteristic parameters of the particles must be
defined. Such parameters are: arrival time and direction, energy of the
particle and its type\footnote{major types are photon and hadron}. (The
shower impact parameter on the detector (core distance) is also an
important parameter, which, however, is not related to the source.) As
was mentioned in chapter~\ref{chapter:air_showers}, the particles of
interest (primary) do not reach the detector level and disintegrate in
the Earth's atmosphere creating extensive air showers of secondary
particles. These secondary particles can be detected and constitute the
observed event. The process of inferring the characteristics of the
primary particle given the observed event is called event
reconstruction. This is a multi-step process which requires deep
understanding of the structure of the extensive air showers, detector
hardware limitations and statistical nature of detection itself.

Currently, the signals from the PMTs in the top layer are used for
shower direction determination and from the bottom --- to distinguish
photon and hadron induced air showers.

\subsection{Pre-processing.}

As was described above, the raw event data contains ``edge-train
information'' registered by each PMT's TDC in the event. These data are
not immediately suitable for primary particle characterization because
the data is tainted by noise and systematic effects in the detector.
Systematic effects include systematic off-sets of TDCs on different PMT
channels (called time pedestals), TDC conversion factors (number of TDC
counts per unit time) and delayed electronics response to lower PMT
signals compared to higher ones (called electronic slewing). These
systematic effects are studied with the help of the calibration system
(see appendix~\ref{chapter:calibration}) and can be taken into account.
Noise effects are random by nature and thus are more difficult to study
and correct. Noise could be due to signals not associated with the main
shower event (thermal/electronic/radioactive noise, non-shower particle
hitting a PMT) or partially recorded edge-trains due to hardware
constraints.

\subsubsection{Noise filtering.}

The main purpose of the edge-finding filter is to check that each PMT
signal in an event conforms to the PMT signal model described in
section~\ref{detector:tot_model}. This should eliminate some
thermal/electronic noise and partially recorded signals. The behavior
of all PMT-electronic channels was studied in great detail and based on
that, a set of criteria was developed which would select viable signals.
(See~\cite{edge_finder}.)

If a PMT signal does not satisfy the criteria, an attempt is made to
convert the signal to the proper form. This is done by checking the number
of edges, their polarity\footnote{polarity means correct sequence of
rising/falling edges} and timing within the PMT signal. This filter is
applied for each PMT in each event registered by the detector. After such
filtering only about eight percent of all PMT signals are considered as
unrecoverable and are discarded.

A completely different problem arises when valid PMT signals from
non-shower particles are recorded with the main shower event. Presence
of these signals will degrade the quality of event reconstruction as
such signals do not carry any useful information about the shower. An
idea of a method for filtration of such PMT signals was first proposed
in~\cite{Todd_causality} and~\cite{Yodh_non-shower} and then used
in~\cite{Bussy_filter} and is based on the fact that PMT signals
produced by a shower must be causally related, i.e. the time interval
between any two PMT pulses multiplied by the speed of light in water
should not be larger than the spatial distance between the PMTs. If a
PMT signal is causally disconnected from the main shower event, it
should be discarded. Unfortunately, at the moment of writing, this idea
is not developed enough to be a part of the standard Milagro event
reconstruction. This filter is applied to calibrated event.

\subsubsection{TDC conversion factors.}

TDC conversion factors were monitored with the help of the calibration
system (see appendix~\ref{chapter:calibration}) and were found to be
stable. Of most importance is the fact that all TDC modules operated at
a common conversion rate of 2 counts per nanosecond with extremely high
precision (see section~\ref{chapter:calibration:tdc_conversion}). This
means that TDC counts can be used as time measure directly and there is
no need to convert TDC counts to time for each PMT channel separately.
This simplified the structure of the reconstruction code.

\subsubsection{Raw to Calibrated event.}

As was mentioned before, time response of PMT-electronic channels is
dependent on the light intensity input. Since each PMT signal remaining
after filtration conforms to the signal model, it is possible to correct
for the effects of electronic delays to signals of varying strength.
Based on that and auxiliary data obtained from the calibration process
(see appendix~\ref{chapter:calibration}) the measured
Time-over-Threshold (LoToT for weak signals and HiToT for strong ones)
can be converted to PMT pulse arrival time ($T_{start}$) and number of
photoelectrons emitted from PMT photocathode (PE). PMT coordinates and
observed $T_{start}$ and PE for each PMT in an event constitute a
calibrated event and contain all information needed for event
reconstruction.

\subsection{Processing: angle, time, energy, type.}
\label{detector:processing}

Even though at this processing stage all information obtainable from the
PMTs is known, to reconstruct the particle characteristics the general
structure of the showers and detector capabilities should be taken into
account. For instance, since the PMT efficiency is only about 20 percent,
there is no guarantee that the observed PMT signal is generated by the
shower-front particles. Particles trailing the front may generate a PMT
pulse too, but if the PMT happened to register the shower-front particles,
the PMT pulse might be discarded as the ToT pulse model does not allow for
more than a single PMT pulse in a shower event. Another example is that it
is almost impossible to differentiate a low energy shower with small
detector impact parameter from a high energy one with large impact
parameter without knowledge of the shower structure. Thus, any method of
event reconstruction must take into account detector and shower features.

\subsubsection{Shower impact parameter.}

As discussed in chapter~\ref{chapter:air_showers} while the primary
particle impact parameter does not provide any information about the
source and the particle it created, it helps to understand the detector
response to the shower produced.

Currently, the PMT PE distribution in the top layer in an event is
analyzed to infer the location of the shower core. If the decision is
made that the core is inside the Milagro pond, a PE-weighted average of
PMT positions is used as the core location, while if it is decided that
the core is outside the pond, it is placed at the distance of $50\
(m)$\footnote{Computer simulations indicate that this is the most
probable core distance for the showers which trigger the detector and
have cores outside of the detector.} from the center of the pond. The
direction to the core, in the later case, is reconstructed by connecting
the center of the pond with the $\sqrt{PE}$-weighted PMT positions. The
decision of whether the core is inside or outside the pound is made
based on the radial profile distribution of the number of PEs observed
in the top layer PMTs.

The information inferred about the shower core is used in the sampling
correction, angular and energy reconstruction. Full details of this
method are described in~\cite{Greg_core}.

\subsubsection{Sampling correction.}

A great care has been take to eliminate systematic and random effect in
the detector on event reconstruction. There is, however, a remaining
one. This has to do with the finite probability of a PMT-electronic
channel to detect light.  Thus, the light, produced by the shower
particles may be lost. The situation is complicated by the fact that
showers have thickness and detection of trailing particles, if
interpreted as the shower front, will degrade the quality of the angular
reconstruction.

Luckily, knowing that the thickness of the shower is a function of impact
parameter and that the number of particles in the shower falls off
quickly with longitudinal distance from the shower front, the amount of
light produced by the trailing particles is generally lower than by the
front of the shower. Using that knowledge, the shower sampling effect
can be observed based on measured light level at a given PMT and its
distance from the shower axis and PMT pulse time can then be corrected
to represent the shower front arrival time.

The Milagro sampling correction has been developed based only on number
of PEs registered by a PMT in~\cite{curve_JoeBussy}
and~\cite{sample_Bussy} and assumes that the shower arrived vertically
when the impact parameter is equal to the core distance.

\subsubsection{Time of event.}

The time of an event is recorded as time of arrival of the PMT
multiplicity trigger and is read from a GPS clock.

\subsubsection{Angular reconstruction.}

After detector sampling effects have been taken into account, the
obtained PMTs' $T_{start}$ times represent the best knowledge of the
shower front. Knowing that the shower front forms a paraboloid, its main
axis can be found and will give the arrival direction of the progenitor
particle.

The algorithm utilized by Milagro first assumes that the shower arrived
vertically and given the shower core position the curvature of the
shower front can be ``taken out'' with what is called ``curvature
correction''.\footnote{The functional form of the curvature correction
was obtained from data and computer simulations
in~\cite{curve_JoeBussy}.} Following that, the shower direction is
sought as the directrix of the plane fitted to the PMTs' $T_{start}$
times (``time-lag'' method) using a weighted $\chi^{2}$-method. (See for
instance~\cite{ZORO_3++}.) The weights for the $\chi^{2}$-fit are
prescribed based on the number of PEs observed, as the quality of PMT
time resolution increases with increase in the input light
level~\cite{curve_JoeBussy}.

\subsubsection{Energy reconstruction.}

Energy estimation is based on the amount of light deposited in the
detector, distance to the core and the angle of the shower arrival and
relies heavily on computer simulations of the shower propagation in
the atmosphere and in the detector. At this time, primary particle
energy is not being inferred in online data processing.

\subsubsection{Primary particle type identification.}

Because of their hadronic cores, air-showers generated by incident
cosmic rays develop differently from purely electromagnetic cascades.
The probability of photons to produce electron-positron pairs is several
orders of magnitude higher than that of any process that might lead to
muon production. In contrast, interactions of high energy hadrons with
atmospheric nuclei lead to the production of charged pions which may
decay into muons. In addition, multi-GeV hadronic particles may also
survive to the ground. Simulations indicate that 80\% of proton and only
6\% of photon induced air showers that trigger Milagro will have at
least one muon or hadron entering the pond.

Hadrons that reach the ground level and produce hadronic cascades in the
detector or muons that penetrate to the bottom layer will illuminate a
relatively small number of neighboring PMTs in that layer. Photon
induced showers, on the other hand, generally will produce rather smooth
light intensity distributions. Based on this simple observation a
technique for identification of photon/hadron initiated showers has been
formulated~\cite{Yodh_gamma_P, crab_paper} and according to computer
simulations can correctly select about 90\% of hadron initiated showers
and about 50\% of photon induced ones.

In a search for sources of high energy photons where hadron initiated
showers represent unwanted background, the proposed identification
scheme will allow increase of signal to noise ratio.

\subsection{Post-Processing: Analysis Techniques.}

After characteristics of the primary particle have been established
further analysis has to be done, based on the concrete task under
investigation. While many different tasks use similar techniques to
answer stated questions, many employ unique methods. For this reason the
discussion of methods and algorithms used in the present work is delayed
until chapter~\ref{chapter:analysis_techniques}.

\section{Detector performance and simulations.}

After a device have been built and tuned, it is desirable to test its
operation and gauge its response. Usually, this is done by comparing the
device's response with the expected one given a known input signal.
Needless to say that such a test is not possible to perform with Milagro
due to unavailability of controllable test sources of high energy
particles above the Earth's atmosphere, and one is forced to resort to
computer simulations to estimate the detector performance. A simulated
extensive air shower is sent to a simulated detector. The output of the
simulated detector is sent for standard analysis and the result is
compared with the input primary particle parameters.

The air shower simulation is done with the CORSIKA
package~\cite{corsika} in the standard US atmosphere down to the
detector level. The simulated shower front is then input into the
GEANT-based detector simulation package. The output of this procedure is
the Milagro ``calibrated'' event which can be sent for the standard
particle characteristics reconstruction described earlier.

The most important parameters of the detector which are obtained based
on computer simulations are angular resolution, energy response, impact
parameter information, particle type identification quality and the
detector's effective area.

Extensive air showers were generated over an energy range of $100\
(GeV)$ to\linebreak $100\ (TeV)$, with zenith angles ranging from $0$ to
$45$ degrees and core locations uniformly distributed over $1000\ (m)$
radius around the detector. Probability of triggering on a shower with
energy outside the selected range or with higher zenith angles is very
small which motivated the choice.

Angular resolution is characterized by the difference between the
reconstructed and the known input particle direction. The overall
accuracy of the angular reconstruction is believed to be $0.75^{\circ}$.
The report~\cite{Yodh_energy} suggests that the energy of an incident
particle can be reconstructed by Milagro with a fractional error of
about 50\% for particles with energies above $1\ (TeV)$.\footnote{The
same report also implies that since the quality of energy reconstruction
relies on the quality of the core reconstruction, it is not possible to
reach the 50\% energy resolution with the current Milagro configuration.
An upgrade with an outrigger array is necessary.} Core location is 
reconstructed with error of about $20\ (m)$ if the shower lands on the 
detector and with error of about $50\ (m)$ otherwise.

\subsection{Effective area.}
\label{detector:effective_area}

As was already mentioned, a shower event can be detected even if its
core lands outside the detector. This leads to the notion of effective
area as the area of imaginary detector which has perfect sensitivity to
events which land on it and zero outside. This parameter describes
sensitivity of the detector to particles of different type, energy and
arrival direction.

If $N_{0}(k,E,\Theta)$ showers induced by particles of type $k$ are
simulated with core locations uniformly distributed over sufficiently
large area $A_{0}$, local arrival directions $(\Theta, \Theta
+\delta\Theta)$ and energies in the range of $(E,E+\delta E)$ then the
effective area $A_{k}(E,\Theta)$ can be computed using the number of
events $N_{t}(k,E,\Theta)$ which satisfy detector trigger condition in
the simulations:

\[ A_{k}(E,\Theta)=\frac{N_{t}(k,E,\Theta)}{N_{0}(k,E,\Theta)}A_{0} \]

Base simulations of proton and photon initiated showers
$A_{k}(E,\theta)$  were obtained where $\theta$ is the local zenith 
angle only.


\chapter{Analysis Techniques}
\label{chapter:analysis_techniques}

\epigraph{George  Orwell ``1984''}{%
\ldots if all others accepted the lie which the Party imposed --- if all
records told the same tale --- then the lie passed into history and 
became truth.}

\section{Coordinates on the Celestial Sphere.}

\subsection{Celestial Sphere.}

Because the stars are distant objects, they {\em appear} to lie on a
sphere concentric with the Earth. This imaginary sphere is known as the
Celestial Sphere. Astronomy uses a number of different coordinate
systems to specify the positions of celestial objects and only  
those relevant to this work ones are discussed here.

The Celestial sphere has North and South Celestial poles as well as the
celestial equator which are projected reference points of the same
positions on the Earth's surface. A coordinate system which is based on
these reference points on the celestial sphere is called the equatorial
celestial coordinate system and is similar to the geographical
coordinate system on the Earth's surface. A point on the celestial
sphere can be described by two coordinates named
``declination''~($\delta$) and ``right ascension''~($\alpha$). The
declination of a star is the analog of the latitude and is the angular
distance from the star to the celestial equator. Right ascension is the
analog of longitude with the zero of right ascension at the point of
vernal equinox.\footnote{Vernal equinox is the point where the Sun
crosses the celestial equator on its south to north path through the sky
and is stationary in space.} Because of the Earth's rotation, right
ascension and declination are not measurable directly by the
ground-based observer and additional local to the observer coordinate
system is introduced. The local coordinates are azimuth~($A$) and zenith
distance~($z$) which can be converted to declination ($\delta$) and hour
angle ($H$). The list below defines the main points and arcs on the
celestial sphere which are illustrated on
figure~\ref{post_processing:coord_definition}.

\begin{figure}
\centering
\psfrag{zz}{$z$}
\psfrag{HH}{$H$}
\psfrag{AA}{$(2\pi-A)$}
\psfrag{aa}{$\alpha$}
\psfrag{dd}{$\pi/2-\delta$}
\psfrag{ve}{$\Upsilon$}
\psfrag{ff}{$\pi/2-\phi$}
\psfrag{f-f}{$\phi$}

\includegraphics[width=3.3in]{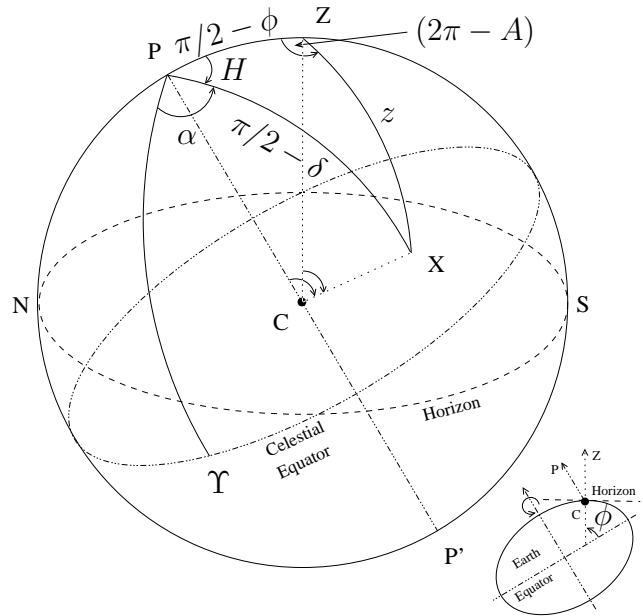}
\caption{Definitions of main points and arcs on the Celestial sphere}
\label{post_processing:coord_definition}
\end{figure}

\begin{description}
\item[$C$]  --- Observer
\item[$CP$] --- Axis parallel to the axis of rotation of the Earth 
            passing through the observer $C$.
\item[$P$, $P'$]  --- North and South Celestial poles.
\item[$Z$]  --- Zenith, CZ is the continuation of the plumb line at
            observer $C$.
\item[Horizon] --- intersection of a plane perpendicular to $CZ$ at point
            $C$ and the celestial sphere.
\item[Local Reference Meridian] --- The arc $PZP'$ of the great 
            circle\footnote{A great circle is a section of a sphere that 
            contains a diameter of the sphere.} containing points $P$, 
            $Z$ and $P'$.
\item[N,S] --- North and South on the horizon as defined by the
            intersection of the great circle $PZP'$ with the horizon.
\item[$\stackrel{\smile}{PZ}$] --- $\angle PCZ =\pi/2 - \phi$, $\phi$ is
            geographical latitude of the observer on the Earth.
\item[Celestial Equator] --- intersection of the plane perpendicular to
            $CP$ at point $C$ with the Celestial sphere.
\item[$X$]  --- A celestial object on the sky (a star).
\item[$\stackrel{\smile}{ZX}$] --- Zenith distance ($z$) of the star X 
            is the angle $\angle ZCX$.
\item[$\angle PZX$] --- Azimuth ($A$)  of the star X is the 
            dihedral\footnote{The dihedral angle is the angle between two 
            planes and is defined as the angle between their normal 
            vectors.} angle between the reference meridian and the $ZCX$ 
            plane measured from North towards East.
\item[$\stackrel{\smile}{PX}$] --- $\angle PCX = \pi/2-\delta$,
            declination ($\delta$) of the star is the angle between
            $\overrightarrow{CX}$ and the Celestial equator.
\item[$\angle ZPX$] --- Hour angle ($H$) of the star is the dihedral 
            angle between the reference meridian $\stackrel{\smile}{PZ}$ 
            and the $PCX$ plane measured from South towards West.
\item[$\Upsilon$] --- Point of vernal equinox
\item[$\stackrel{\smile}{P\Upsilon P'}$] --- Celestial reference 
meridian.
\item[$\Upsilon PX$] --- Right ascension ($\alpha$) of the star is the
            dihedral angle between the $\Upsilon CP$ and $XCP$ planes.
\end{description}

Given the definitions above, the law of cosines for the
trihedron\footnote{Three vectors with common vertex, often called a
trihedral angle since they define three planes.} applied to the
spherical triangle $ZPX$ two times yields the relationship between the
($A,z$) and ($\delta,H$) coordinate systems:

\[ \left\{\begin{array}{l}
   \cos(\pi/2 -\delta)=\cos z\cos(\pi/2 -\phi)+
              \sin z\sin(\pi/2 -\phi)\cos (2\pi-A) \\
   \cos z = \cos(\pi/2 -\delta)\cos(\pi/2 -\phi)+
               \sin(\pi/2 -\delta)\sin(\pi/2 -\phi)\cos H
   \end{array}\right. \]
\[ \left\{\begin{array}{l}
   \sin\delta = \sin\phi\cos z +\cos\phi\sin z\cos A \\
   \tan H = \frac{-\sin z\sin A}{\cos\phi\cos z -sin z \sin\phi\cos A}
   \end{array}\right. \]

Since the local reference meridian is defined relative to the Earth, due
to Earth's rotation the hour angle of a fixed in space point will grow
with time (that is why it is called hour angle) while the local
coordinate declination will remain constant. The hour angle of vernal
equinox $H_{\Upsilon}$ links the local observer's coordinate system
(H,$\delta$) and the celestial equatorial coordinate system
($\alpha$,$\delta$) by providing the position of a fixed point (vernal
equinox) on the celestial sphere in local coordinates: $\alpha =
H_{\Upsilon} - H$. Hour angle of vernal equinox is also called the local
sidereal time since it should be consistent with the observer's
geographical longitude and the time required for one Earth's revolution,
called a sidereal day. In contrast, the solar (or universal) day is
defined as time between two consecutive appearances of the Sun on the
local reference meridian. The solar day is longer than the sidereal one
due to Earth's rotation and orbital motion around the Sun, though both
days are divided into 24 hours.

\subsection{J2000 reference.}

Because the Earth's rotation is not uniform, its axis of rotation is not
fixed in space and even its shape and relative positions on its surface
are not fixed; because the introduced celestial equatorial coordinate
system follows the motion of the Earth's pole and equator, the
coordinate grid ``drifts'' on the surface of the celestial
sphere.\footnote{These drifts include, but not limited to precession,
nutation, celestial pole offset and polar motion.} Therefore, the
introduced coordinate system provides only apparent right ascension and
declination of the stars at the observation moment.

To solve this problem, all coordinates on the celestial sphere are
reported relative to the position of the Earth's pole and equator at
specified moments of time which are called epochs. Each epoch lasts for
50 years and the current one is defined with respect to the Earth's
position at noon on the January 1, 2000. Thus, the apparent celestial
coordinates need to be reduced to the J2000 reference.%
\footnote{%
The major contribution to the ``drift'' of a celestial reference frame
is due to the Earth's pole precession. Newcomb (Newcomb, S.
Astron.J. {\bf 17}, 20 1897) derived the formulae for precession
parameters $\zeta_{A}(t)$, $z_{A}(t)$ and $\theta_{A}(t)$ which specify
the position of mean equinox and the equator of a date with respect to
the mean equinox and equator of the initial epoch. Astronomical Almanac
on page B18 supplies these parameters for the J2000.0 epoch in degrees:

\[ \begin{array}{l}
  \zeta_{A}  = 0.6406161T + 0.0000839T^{2} + 0.0000050T^{3} \\
       z_{A} = 0.6406161T + 0.0003041T^{2} + 0.0000051T^{3} \\
  \theta_{A} = 0.5567530T - 0.0001185T^{2} - 0.0000116T^{3}
  \end{array} \]

where $T$ stands for the time from the basic epoch J2000.0 in Julian
centuries, $T=(\mbox{Julian Day}-2451545.0)/36525$.

If subscript $0$ refers to the coordinates at the epoch J2000.0 and no
subscript to the epoch of the date, the transformation formulae are:

\[ \left\{\begin{array}{rcl}
   \sin(\alpha-z_{A})\cos\delta & = &
                             \sin(\alpha_{0}+\zeta_{A})\cos\delta_{0}  \\
   \cos(\alpha-z_{A})\cos\delta & = &
           \cos(\alpha_{0}+\zeta_{A})\cos\theta_{A}\cos\delta_{0}
                                         -\sin\theta_{A}\sin\delta_{0} \\
   \sin\delta & = & 
           \cos(\alpha_{0}+\zeta_{A})\sin\theta_{A}\cos\delta_{0}   
                                         +\cos\theta_{A}\sin\delta_{0}
   \end{array}\right. \]

\[ \left\{\begin{array}{rcl}
   \sin(\alpha_{0}+\zeta_{A})\cos\delta_{0} & = &
                             \sin(\alpha-z_{A})\cos\delta          \\
   \cos(\alpha_{0}+\zeta_{A})\cos\delta_{0} & = &  
           \cos(\alpha-z_{A})\cos\theta_{A}\cos\delta
                                         +\sin\theta_{A}\sin\delta \\
   \sin\delta_{0} & = &
           -\cos(\alpha-z_{A})\sin\theta_{A}\cos\delta
                                         +\cos\theta_{A}\sin\delta
   \end{array}\right. \]
}

\subsection{Diurnal parallax.}

The Equatorial coordinate system had been defined under the assumption
that the observer is located at its origin --- the center of the Earth.
All observing stations, however, are located on the Earth's surface. Due
to Earth's rotation, the observing station moves and the observation of
a celestial body is being made from different points in space. This will
cause an apparent difference in position of celestial body when made at
different moments of time. The effect is called {\em diurnal parallax}. 
For measurements of distant stars this has a negligible effect, but there
could be a substantial diurnal parallax on objects inside the Solar
system. Diurnal parallax on the Moon, for example, can be as large as 
$0.95^{\circ}$.

\subsection{Milagro event coordinates.}

The local hour angle and declination of an event on the celestial sphere
are calculated from the zenith and azimuth which are provided by the
event reconstruction section~\ref{detector:processing} (see
also~\cite{ZORO_pointing} for a discussion on local coordinates). Local
sidereal time as well as the geographic coordinates of the detector can
be obtained from a Global Positioning System receiver which facilitates
the conversion from local to celestial coordinates. In Milagro, the
coordinates of reconstructed events are reduced from the epoch of date
to the J2000 reference in real time and are saved to disk for further
processing.

As will be clarified in the sections to follow, the signal processing
method employed in this work expects the event coordinates in a local
reference frame. Thus, even though the events are saved in J2000
reference which seems to be convenient, the conversion to J2000 must be
undone during the offline/online signal processing.\footnote{It would be
prudent to save the local hour angle and declination of the registered
events during the online realtime processing. This would force the
coordinates of celestial bodies which are known in J2000 from
catalogues, to be reduced to epoch of date then to be reduced to the
apparent Right Ascension-Declination by application of the parallax
correction (if necessary) then to local hour angle-declination using
local time. This approach would save some computer time during online
and offline data processing because the Milagro event rate is above $1\
(kHz)$ and the detector angular of resolution (several tenths of a
degree) allows for rare (once per several seconds (in 24 seconds the
Earth rotates on $0.1^{\circ}$ of arc)) computation of local coordinates
of the celestial bodies.}


\section{Sky Mapping.}

In counting-type astrophysical experiments, the brightness of a
particular point in the sky is characterized by the number of events
observed from that point during the exposure time. Such experiments
measure the density of events on the surface of the Celestial sphere.
Therefore, the procedure used to generate the sky images (projections of
a sphere onto a plane surface) should conserve the density of events. To
meet this requirement an equal area projection of a sphere onto a plane
must be used. This requirement, however, is not enough to uniquely fix
the projection, and several different projections are available. It is
crucial to understand that any area preserving projection is not
conformal and might distort the distances and/or directions on the map.
That is why different area preserving mappings should be used for
different tasks. For example, the sinusoidal projection
\footnote{The
{\em Sinusoidal Equal Area Projection} is defined as: \[
\left\{\begin{array}{l}
    x=l\cos b \\
    y=b
    \end{array}\right. \]
where $(l,b)$ are galactic coordinates}%
\ has minimal distortions near the equator and that is why is it very
convenient for galactic plane studies. When the same mapping is used for
an object far from the galactic equator, linear distortions become
significant.

In addition to the previously discussed constraints, all events with
identical spatial orientation with respect to the point of interest must
be mapped into a unique location. This is especially important if the
point of interest moves on the Celestial sphere. However, a simple
algebraic difference in Celestial coordinates of any two points does not
define their relative spatial orientation. One way to address this
problem is to introduce auxiliary coordinates on the Celestial sphere:
an analog of latitude ($\chi$) and longitude ($\xi$) which are measured
with respect to the preselected point. (See
figure~\ref{post_processing:chi_xi_concept} and
appendix~\ref{chap_aux_celes_coord} for the definition of the
$(\chi-\xi)$ coordinate system.) Using these coordinates, the sky image
centered on the selected point can be produced with the help of the
Azimuthal Equal Area Projection in the polar case defined as:

\begin{equation}
   \left\{\begin{array}{l}
     x=\sqrt{2(1-\cos\chi)}\sin\xi \\
     y=\sqrt{2(1-\cos\chi)}\cos\xi
   \end{array}\right.
\label{eq:post_process:mapping}
\end{equation}

This mapping not only satisfies the above requirements, it has several
other important features such as conservation of the directions as seen
from its origin, the locus of points equidistant from the center of the
mapping is projected into a circle (see
figure~\ref{post_processing:chi_xi_concept}) and it can be easily 
oriented along the lines of the Earth's magnetic field.

\begin{figure}
\centering
\psfrag{x}{$x$}
\psfrag{y}{$y$}
\psfrag{X}{$X$}
\psfrag{L}{$L$}
\psfrag{M}{$M$}
\psfrag{C}{$C$}
\psfrag{chi}{$\chi$}
\psfrag{xi}{$\xi$}

\includegraphics[width=4.7in]{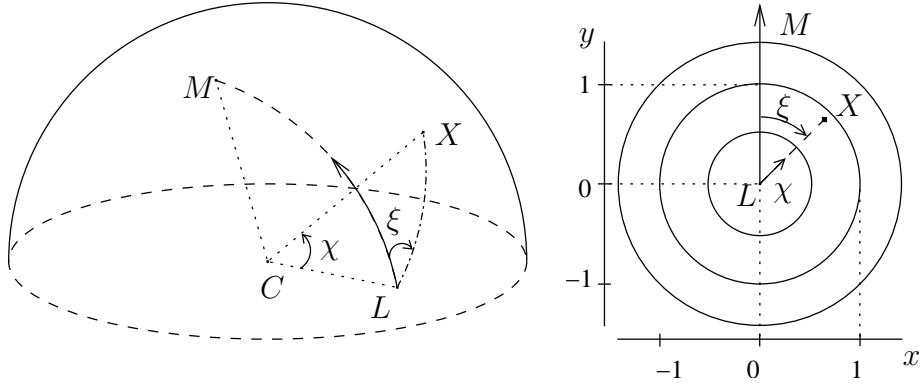}
\caption{Concept of the auxiliary coordinate system on the Celestial
sphere centered on $L$ (left) and corresponding sky projection (right). 
The ``y''-axis of the sky projection always points to a pole $M$ and the
circles are the lines of $\chi=const$ at $\chi=\pi/6 ,\ \pi/3 ,\ \pi/2$.}
\label{post_processing:chi_xi_concept}
\end{figure}


\section{Statistical Nature of Signal Establishment.}

In a typical counting type astrophysical experiment during observation
time $t_{1}$ the number of events observed due to some physical process
is $N_{1}$. Assuming that an event contains no information about any
other one, the number of observed events is a random variable which is
distributed according to the Poisson distribution.\footnote{Some of the
properties of the Poisson distribution are discussed in
appendix~\ref{app_chapter_poisson_distr}} In other words, the
probability to observe exactly $N_{1}$ events during time $t_{1}$ is
given by:

\[ p(N_{1};\lambda)=\frac{(\lambda t_{1})^{N_{1}}}{N_{1}!}e^{-\lambda t_{1}} \]
where $\lambda$ has a meaning of average event rate.

However, an observed event could be due to either a source or
background. Since the average count rate due to background is not known,
based on this one observation it is not possible to decide whether there
were any ``source'' events observed.  Therefore without altering the
conditions of the experiment a second measurement $N_{2}$ during $t_{2}$ 
is
made where it is believed that all observed events are due to background
only. Now, a decision should be made as to whether there is a difference
between these numbers which can be interpreted as a detection of a
source. Since the observed $N_{1}$ and $N_{2}$ are random numbers, this
question should be approached from the statistical point of view. Note
that a statistical test cannot verify that a given
hypothesis\footnote{Any statement concerning the unknown distribution of
a random variable is called a statistical hypothesis.} is true or false,
but can only suggest which of the two or more hypotheses is the more
plausible explanation of the observation.

\subsection{General test construction.}

Suppose\footnote{This subsection is based on the section II of the 
paper~\cite{NeymanPearson}.} that a result of an observation is 
described by the values of $n$ variables:
\[ \{x\} = x_{1},x_{2},\ldots ,x_{n} \]

The $\{x\}$ may represent outcomes of $n$ repeated measurements made
under identical conditions or a sample from a population. All possible
outcomes of a measurement are said to form a sample space $W$. A
hypothesis about the origin of the observed events $H$ defines the
probability of occurrence of every possible observation

\[ p(x_{1},x_{2},\ldots ,x_{n}) \]
and thus, the probability that the observed event will fall into some 
region $w$ of all possible outcomes is
\[ P(w)=\int_{w}p(x_{1},\ldots ,x_{n})\ dx_{1}\ldots dx_{n} \]
Of course, $P(W)=1$. Different hypotheses $H$ with their corresponding
probability distributions $p(\{x\})$ will be endowed by the same 
subscripts, such as $H_{0}$ and $p_{0}$.

A statistical test is formulated so that all prior knowledge strongly
supports $H_{0}$ called the null hypothesis. Hypothesis $H_{0}$ is
rejected if the observed event $\{x\}$ lies within a certain critical
region $w_{c}$ and accepted or doubted otherwise. Such a formulation of
a test implies that it is possible to reject $H_{0}$ when, in fact, it
is true. The danger of falsely rejecting the null hypothesis is
characterized by the error of the first kind or significance $\xi_{c}$
and:

\[ \xi_{c}=P_{0}(w_{c})=
          \int_{w_{c}}p_{0}(x_{1},\ldots ,x_{n})\ dx_{1}\ldots dx_{n} \]

The choice of the value of $\xi_{c}$ depends on the penalty for making the
error, therefore, the risk $\xi_{c}$ must be set in advance, not after the
results of a measurement are available. Even though the error of the first
kind can be chosen to be arbitrary small, the equation
$\xi_{c}=P_{0}(w_{c})$ has, in general, infinitely many solutions on
configuration $w_{c}$ with the same level of significance $\xi_{c}$.

Since $H_{0}$ is being tested, it implies the existence of an
alternative hypothesis $H_{1}$ or there would be no question about
$H_{0}$.\footnote{While it may not be constructive, ``$H_{0}$ is false''
is an admissible alternative hypothesis.} But, as the risk $\xi_{c}$ is
required to be smaller and smaller, the risk $\zeta$ of accepting
$H_{0}$, when $H_{1}$ is true may increase.  This error is called the
error of the second kind and is given by:

\[ \zeta = P_{1}(W\setminus w_{c}) =
 \int_{W\setminus w_{c}}p_{1}(x_{1},\ldots ,x_{n})\ dx_{1}\ldots dx_{n} \]

The two errors $\xi$ and $\zeta$ can rarely be eliminated, and in some
cases it is more important to avoid the first, in others --- the second.
When $H_{0}$ and $H_{1}$ are specified it is the choice of the critical
region which allows control of the errors.

A prescription to resolve the apparent vagueness in the provided
formulation of the test was proposed in~\cite{NeymanPearson}. It is
proposed that given the two hypotheses $H_{0}$ and $H_{1}$ and the 
desired risk level $\xi_{c}$, the corresponding best critical region 
$w_{c}^{best}$ minimizes the error $\zeta$.

If based on the outcome of the experiment, the observed $x$ is inside of
the critical region $w_{c}^{best}$, it is said that the null hypothesis
is rejected in favor of the alternative one with significance $\xi_{c}$
and power $(1-\zeta)$. If, however, $x\notin w_{c}^{best}$, it is said
that the null hypothesis is not rejected in favor of the alternative one
with significance $\xi_{c}$ and power $(1-\zeta)$.

Often, however, rather than use the full data sample $\{x\}$ it is
convenient to define a test statistic\footnote{Statistic is a random
variable which is a function of the observed sample of data.} $U$. Each
hypothesis for the distribution of $\{x\}$ will determine a distribution
for $U$, and a specific range of values of $U$ will be mapped to a
critical region in $W$-space. In constructing $U$ one attempts to reduce
the volume of data without loss of the ability to discriminate between
different hypotheses.

\subsection{Testing a composite hypothesis.}

If the hypothesis being tested does not specify the probability of 
occurrence of every possible observation, it is called a 
composite hypothesis. It will be assumed that the composite hypothesis 
depends on an unspecified parameter $\lambda$ as:

\[ p(x_{1},x_{2},\ldots ,x_{n};\lambda) \]

As before, the null hypothesis should be rejected if the observed event 
lies within a critical region $w_{c}$.

In order to control the error of the first kind $\xi$, the critical
region must satisfy:
\[ \xi_{c}=P_{0}(w_{c})=
 \int_{w_{c}}p_{0}(x_{1},\ldots ,x_{n};\lambda_{0})\ dx_{1}\ldots dx_{n} \]

for every value of the parameter $\lambda_{0}$. In other words, the
error of the first kind should not depend on the unknown value of the
parameter $\lambda_{0}$. If such critical regions exist, it is necessary
to choose the best one which minimizes the error of the second kind. It
should be noted that the error of the second kind may, in general,
depend on the values  $\lambda_{0}$ and $\lambda_{1}$ of the alternative 
hypothesis $p_{1}(x_{1},\ldots ,x_{n};\lambda_{1})$.

This problem has been solved in~\cite{NeymanPearson} for a special class 
of the null hypotheses when $p_{0}(\{x\};\lambda_{0})$ is infinitely 
differentiable function of $\lambda_{0}$ in every point $\{x\}\in W$ and 
the function $p_{0}(\{x\};\lambda_{0})$ satisfies the equation:

\begin{equation}
   \frac{d\phi}{d\lambda_{0}}=A+B\phi, \ \ \ \ 
              \phi=\frac{d\ln p_{0}(\{x\};\lambda_{0})}{d\lambda_{0}}
\label{eq:post_processing:NeymanPearson1}
\end{equation}

and the coefficients $A$ and $B$ are functions of $\lambda_{0}$ only and
do not depend on $\{x\}$. It is shown in~\cite{NeymanPearson} that the
best critical region $w_{c}^{best}$ is constructed of pieces of 
hypersurfaces $\phi=C=const$ such that:

\begin{equation}
   \frac{p_{1}(\{x\};\lambda_{1})}{p_{0}(\{x\};\lambda_{0})} > q,
                                    \ \ \ \forall \{x\}\in w_{c}^{best}
\label{eq:post_processing:NeymanPearson2}
\end{equation}
where $q$ is a constant whose value is governed by $w_{c}^{best}$ chosen 
subject to constraint:

\begin{equation}
   \xi_{c}\int_{\{x\}\in W\cap\phi=C} p_{0}(\{x\};\lambda_{0}) dx=
   \int_{\{x\}\in w_{c}^{best}\cap\phi=C} p_{0}(\{x\};\lambda_{0}) dx
\label{eq:post_processing:NeymanPearson3}
\end{equation}

\subsection{Significance of a measurement.}

In as much as an attempt is being made to identify the presence of a
source, the null-hypothesis $H_{0}$ will be formulated in the following
way:

\begin{quotation}
{\em The source is not present. The results $N_{1}$ and $N_{2}$ of two 
independent observations come from a single Poisson distribution with 
parameter $\lambda$.}
\end{quotation}

with an alternative hypothesis $H_{1}$ that:

\begin{quotation}
{\em The independent counts $N_{1}$ and $N_{2}$ come from Poisson 
distributions with different parameters $\lambda_{1}$ and $\lambda$, 
correspondingly.}
\end{quotation}

Mathematically, if $H_{0}$ is true the probability
$p_{0}(N_{1},N_{2};\lambda)$ to observe $N_{1}$ and $N_{2}$ is:

\[ p_{0}(N_{1},N_{2};\lambda)=
             \frac{(\lambda t_{1})^{N_{1}}}{N_{1}!}e^{-\lambda t_{1}}
             \frac{(\lambda t_{2})^{N_{2}}}{N_{2}!}e^{-\lambda t_{2}} \]
while, if $H_{1}$ is true the probability 
$p_{1}(N_{1},N_{2};\lambda_{1},\lambda)$ is:

\[ p_{1}(N_{1},N_{2};\lambda_{1},\lambda)=
             \frac{(\lambda_{1}t_{1})^{N_{1}}}{N_{1}!}e^{-\lambda_{1}t_{1}}
             \frac{(\lambda t_{2})^{N_{2}}}{N_{2}!}e^{-\lambda t_{2}} \]
where the values of $\lambda$ and $\lambda_{1}$ are unspecified and the 
only requirement is that $\lambda_{1}\neq \lambda$.

The formulated $H_{0}$ satisfies the conditions of a theorem presented
in~\cite{NeymanPearson} which states that there exists the best critical
region $w_{c}^{best}$ corresponding to significance $\xi_{c}$
independent of the value of the parameter $\lambda$.

Following the algorithm for construction of the best critical region
from~\cite{NeymanPearson}, the 
equations~(\ref{eq:post_processing:NeymanPearson1}) and
(\ref{eq:post_processing:NeymanPearson2}) become:

\[ \phi=N_{t}=N_{1}+N_{2}=const \]
\[ \frac{p_{1}}{p_{0}} >q \ \ \Rightarrow\ \ 
   \left(\frac{\lambda_{1}}{\lambda}\right)^{N_{1}}
        e^{-(\lambda_{1}-\lambda)t_{1}} \geq q \ \ \Leftrightarrow\ \
  \left[\begin{array}{ll}
         N_{1}\geq N_{\xi}, & \lambda_{1} > \lambda \\
         N_{1}\leq N_{\xi}, & \lambda_{1} < \lambda
                                                 \end{array}\right.  \]

It is thus clear that the best critical region for testing 
$\lambda_{1}=\lambda$ against $\lambda_{1}\neq\lambda$ does not exist, 
however, it does exist for testing $\lambda_{1}=\lambda$ against 
$\lambda_{1}>\lambda$ or $\lambda_{1}<\lambda$ separately.

The value $N_{\xi}$ corresponding to the error of the first kind $\xi$ 
is found as the solution of the equation
~(\ref{eq:post_processing:NeymanPearson3}):

\[ \left[\begin{array}{ll}
    \xi\sum_{k=0}^{N_{t}}p_{0}(k,N_{t}-k;\lambda)=
\sum_{k=N_{\xi}}^{N_{t}}p_{0}(k,N_{t}-k;\lambda), & \lambda_{1} > \lambda \\ \\
    \xi\sum_{k=0}^{N_{t}}p_{0}(k,N_{t}-k;\lambda)=
    \sum_{k=0}^{N_{\xi}}p_{0}(k,N_{t}-k;\lambda), & \lambda_{1} < \lambda
  \end{array}\right. \]

Immediately, it should be noted that the solution $N_{\xi}$ does not
depend on the values of the parameters $\lambda_{1}$ and $\lambda$ and
the best critical region exists for the $H_{0}$ with regard to all
alternative hypotheses $H_{1}$. After explicitly writing the probability
$p_{0}(N_{1},N_{2})$, one arrives to the following equation on
$N_{\xi}$:

\[ \left[\begin{array}{ll}
       \xi =(1+\alpha)^{-N_{t}}
       \sum_{k=N_{\xi}}^{N_{t}}C_{N_{t}}^{k}\alpha^{k}, 
                                             & \lambda_{1} > \lambda \\ \\
       \xi =(1+\alpha)^{-N_{t}} 
       \sum_{k=0}^{N_{\xi}}C_{N_{t}}^{k}\alpha^{k}, 
                                             & \lambda_{1} < \lambda
    \end{array}\right| 
    \ \ \ \alpha= t_{1}/t_{2} > 0,\ \ \ C_{n}^{m}=\frac{n!}{m!(n-m)!} \]

The error of the second kind $\zeta$ can be computed as:

\[ \left[\begin{array}{cc}
   \zeta=\sum_{N_{t}=0}^{\infty}\sum_{k=0}^{N_{\xi}-1}
                                   p_{1}(k,N_{t}-k;\lambda_{1},\lambda),
                                           & \lambda_{1} > \lambda \\ \\
   \zeta=\sum_{N_{t}=0}^{\infty}\sum_{k=N_{\xi}+1}^{N_{t}}
                                   p_{1}(k,N_{t}-k;\lambda_{1},\lambda),
                                           & \lambda_{1} < \lambda
                                                   \end{array}\right| \]

\[ \left[\begin{array}{ll}
   \zeta= e^{-\lambda_{1}t_{1}-\lambda t_{2}}
   \sum_{N_{t}=0}^{\infty}(\lambda t_{2})^{N_{t}}\sum_{k=0}^{N_{\xi}-1}
      \frac{(\frac{\lambda_{1}t_{1}}{\lambda t_{2}})^{k}}{k!(N_{t}-k)!},
                                           & \lambda_{1} > \lambda \\ \\
   \zeta= e^{-\lambda_{1}t_{1}-\lambda t_{2}}
   \sum_{N_{t}=0}^{\infty}(\lambda t_{2})^{N_{t}}\sum_{k=N_{\xi}+1}^{N_{t}}
      \frac{(\frac{\lambda_{1}t_{1}}{\lambda t_{2}})^{k}}{k!(N_{t}-k)!},
                                           & \lambda_{1} < \lambda 
                                                   \end{array}\right| \]

The explicit solution for the critical region is needed if an ability to
compute the error of the second kind $\zeta$ is desired. As expected,
this error will depend on the values of the parameters $\lambda$ and
$\lambda_{1}$ of the alternative hypothesis $H_{1}$. It is, however,
possible to decide if the null hypothesis should be rejected or not
without the explicit solution. To do this, $N_{\xi}$ must be set to
$N_{1}$ and $\xi$ must be computed from the equations above using
$N_{t}=(N_{1}+N_{2})$. If it is found that the $\xi$ obtained in this
fashion is smaller than the critical value $\xi_{c}$, the null
hypothesis should be rejected and should not be rejected otherwise.

As will be explained below, because the procedure for setting an upper
limit is based on the error of the second kind $\zeta$, expression for
which is not known in a closed form, a ``practical'' statistic which was
proposed in~\cite{LiMa} is considered in this work:

\begin{equation}
   U=\frac{N_{1}-\alpha N_{2}}{\sqrt{\alpha(N_{1}+N_{2})}} \ \ \ \ \ \ \ 
                                                \alpha = t_{1}/t_{2} > 0
\label{eq:post_processing:Udef}
\end{equation}

The denominator in~(\ref{eq:post_processing:Udef}) is the maximum
likelihood estimate on dispersion of $(N_{1}-\alpha N_{2})$ given the
null hypothesis is true. Then, under the null hypothesis the mean value
of the statistic $U$ is zero and the dispersion is equal to unity. If
both $N_{1}$ and $\alpha N_{2}$ have not deviated far from the expected
value of $\lambda t_{1}$, then $N_{1}$ and $\alpha N_{2}$ can be
regarded as coming from Gaussian distributions with the means equal to
$\lambda t_{1}$ and dispersions $\lambda t_{1}$ and $\alpha\lambda
t_{1}$ correspondingly (See discussion on Gaussian limit to Poisson
distribution in appendix~\ref{app_chapter_poisson_distr}.). Hence, the
values of the statistic $U$ are distributed according to a Gaussian
distribution with zero mean and unit variance
$p_{0}(u)=\frac{1}{\sqrt{2\pi}}e^{-\frac{u^{2}}{2}}$. This statement is
valid for all $u\leq u_{0}$\footnote{The value of $u_{0}$ is obtained 
by substituting $k$ with $N_{1,2}$ and $\lambda$ with corresponding 
maximum likelihood estimates $\frac{N_{1}+N_{2}}{t_{1}+t_{2}}t_{1,2}$ 
into the equation~(\ref{eq:poisson:possion2gauss}).}:

\begin{equation}
   |u| \leq |u_{0}| <<
   \min\left(\sqrt[6]{36\alpha(1+\alpha)^{2}(N_{1}+N_{2})},
             \sqrt[6]{36\alpha^{-3}(1+\alpha)^{2}(N_{1}+N_{2})}\right)
\label{eq:post_processing:u0validity}
\end{equation}

If, however, the $H_{1}(\lambda_{1})$ is true, the $U$ will have
approximately Gaussian distribution with unit dispersion and shifted mean:

\[ p_{1}(\lambda_{1})=
     \frac{1}{\sqrt{2\pi}}e^{-\frac{(u-u_{1}(\lambda_{1}))^{2}}{2}} \]
where $u_{1}(\lambda_{1})$ is monotonically increasing function of
$\lambda_{1}$ and is equal to the average value of $U$ computed when 
$H_{1}(\lambda_{1})$ is true.
\[ u_{1}(\lambda_{1})\simeq\frac{(\lambda_{1}-\lambda)t_{1}}
                    {\sqrt{(\alpha\lambda_{1}+\lambda)t_{1}}}\approx
    \frac{\lambda_{1}-\lambda}{\sqrt{\lambda(1+\alpha)}}\sqrt{t_{1}} \]

Let us define the critical range of values of the statistic $U$
corresponding to significance $\xi_{c}$ in the following way (see
figure~\ref{fig:post_processing:critical_region} for an illustration):
\begin{description}
\item[If $\lambda_{1}>\lambda$:] $u>u_{c}$,
                             $\xi_{c}=\int_{u_{c}}^{+\infty}p_{0}(u)du$.
\item[If $\lambda_{1}<\lambda$:] $u<u_{c}$,
                             $\xi_{c}=\int_{-\infty}^{u_{c}}p_{0}(u)du$.
\end{description}

For the reasons of tradition, in astrophysics, it is customary to report
the level of significance not as probability $\xi_{c}$, but as ``number
of sigmas'' $u_{c}$ which motivated the choice of statistic. The
table~\ref{table:post_proc:significance} provides the translation
between the critical value $u_{c}$ and the significance $\xi_{c}$ with 
the approximation error on $\xi$ not exceeding 
$\frac{1}{\sqrt{2\pi}}\int_{u_{0}}^{+\infty}e^{-\frac{u^{2}}{2}}du$.

\begin{table}
\centering
\begin{tabular}{|c|c|} \hline
$|u_{c}|$ & $\xi_{c}$ \\ \hline
 $1.0$  & $1.587\cdot 10^{-1}$ \\
 $2.0$  & $2.275\cdot 10^{-2}$ \\
 $3.0$  & $1.350\cdot 10^{-3}$ \\
 $3.5$  & $2.326\cdot 10^{-4}$ \\
 $4.0$  & $3.167\cdot 10^{-5}$ \\
 $4.5$  & $3.398\cdot 10^{-6}$ \\
 $5.0$  & $2.867\cdot 10^{-7}$ \\ \hline
\end{tabular}
\caption{Significance $\xi_{c}$ and corresponding critical value 
$u_{c}$.}
\label{table:post_proc:significance}
\end{table}

\begin{figure}
\centering
\psfrag{uu}{$u$}
\psfrag{uc}{$u_{c}$}
\psfrag{ul}{$u_{1}(\lambda_{1})$}
\psfrag{pp}{$p(u)$}
\psfrag{p0}{$p_{0}(u)$}
\psfrag{p1}{$p_{1}(u)$}
\psfrag{zeta}{$\zeta$}
\psfrag{xi}{$\xi$}

\includegraphics[width=3.3in]{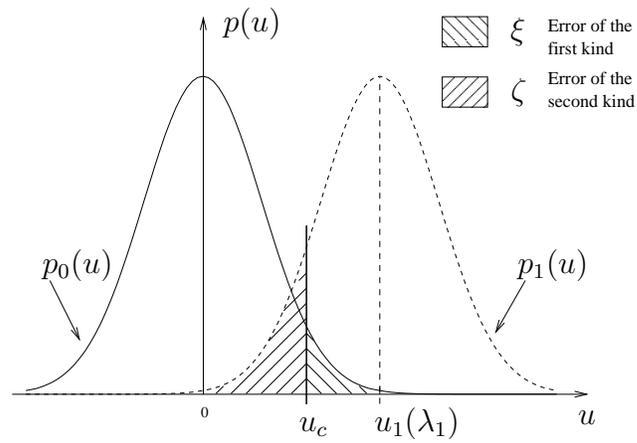}
\caption{Critical region illustration for the statistic $U$ when 
         $\lambda_{1}>\lambda$.}
\label{fig:post_processing:critical_region}
\end{figure}

\subsection{Setting an upper limit.}

Some times, when the null hypothesis can not be rejected based on the
results of a test and there are several alternative hypotheses
available, it is meaningful to ask the question of which of the
alternatives provides error of the second kind larger than $\zeta_{u}$.
For instance, in the case considered here, there are many alternative
hypothesis parametrized by the values of $\lambda_{1}$\footnote{Remember
that $\lambda_{1}$ is not a source strength, but merely the average
event count rate due to possible presence of a source.}. Since each
alternative hypothesis $H_{1}(\lambda_{1})$ defines a probability
distribution $p_{1}(u,\lambda_{1})$ in the sample space, the error
$\zeta(\lambda_{1})$ is (see 
figure~\ref{fig:post_processing:critical_region} for an illustration):

\begin{description}
\item[If $\lambda_{1}>\lambda$:] $\zeta (\lambda_{1})=
             \int_{-\infty}^{u_{c}}p_{1}(u,\lambda_{1})du$.
\item[If $\lambda_{1}<\lambda$:] $\zeta (\lambda_{1})=
             \int_{u_{c}}^{+\infty}p_{1}(u,\lambda_{1})du$.
\end{description}

For the case of $\lambda_{1}>\lambda$, $\zeta (\lambda_{1})$ is
monotonically decreasing function of $\lambda_{1}$. Therefore,
$\lambda_{1}^{u}$ corresponding to the largest allowed error $\zeta_{u}$
is called the upper limit on $\lambda_{1}$ ($\zeta
(\lambda_{1}^{u})=\zeta_{u}$). It means that the probability of making
the error of the second kind by accepting the null hypothesis when in
fact one of the alternative hypotheses with parameter $\lambda_{1}$
$(\lambda_{1} > \lambda_{1}^{u}>\lambda)$ is true is less than
$\zeta_{u}$. For a discussion on the upper limit construction procedure, 
please, see appendix~\ref{chapter:upper_limit}.


\section{Background estimation.}

One method of searching for a source is by counting the number
$N_{on}(\Theta)$ of events from local direction
$(\Theta;\Theta+d\Theta)$ while it was exposed to a source region
$\Omega$ on the sky (the ``on-source'' bin) and comparing it with the
number of background events $N_{on}^{b}(\Theta)$ expected from this
region. The number of background events expected from the ``on-source''
region is given by:

\[ N_{on}^{b}(\Theta)=\int_{\Theta}\left[
          1-\phi(\Theta',t)\right]R_{b}(\Theta',t)d\Theta'dt \]
where $R_{b}(\Theta,t)$ is background event rate from local direction 
$\Theta$ at an observation moment $t$ and
\[ \phi(\Theta,t)=\left\{\begin{array}{ll}
             0, & (\Theta,t)\in\Omega \\
             1, & (\Theta,t)\not\in\Omega  \end{array}\right. \]

Since the function $R_{b}(\Theta,t)$ is not known {\it a priori}, it
should be determined from the observed data. To accomplish that one is
forced to introduce some assumptions about the structure of
$R_{b}(\Theta,t)$. Probably, the most natural simplification comes from
the assumptions that the background events are distributed uniformly on
the sky (their distribution is independent of local coordinates
$\Theta$)\footnote{Charged particles which form the background are
isotropized by galactic and inter-galactic magnetic fields.} and that
the conditions of the experiment (hardware, software, field of view,
everything) remain constant (at least for periods of time long enough to
allow measurement of $R_{b}(\Theta,t)$ to the necessary accuracy). Then,
$R_{b}(\Theta,t)$ can be factorized:

\[ R_{b}(\Theta,t) = G(\Theta)\cdot R_{b}(t) \]

where $R_{b}(t)$ is overall event rate due to background only and is
independent of local coordinates and $G(\Theta)$ is the detection
efficiency of the instrument and does not depend on time. Thus, the
problem of determining $R_{b}(\Theta,t)$ is reduced to the one of
$R_{b}(t)$ and $G(\Theta)$.

Knowing that the number of background events expected from any point on
the sky at some time is:

\[ dN^{b}(\Theta,t)=R_{b}(\Theta,t)d\Theta dt=R_{b}(t)G(\Theta)d\Theta dt \]
the total number of background events $N_{out}^{b}(\Theta)$ which are to
be observed within some large time $T$ from the local direction $\Theta$
outside of the source region\footnote{The outside region should not
contain any known source in the field of view of the detector. The
events from other sources and their source regions should be removed
from the analysis as they would bias the background estimation.}
\footnote{Due to the Earth's rotation, the local direction $\Theta$ will 
fall within the source region $\Omega$ at some times and outside at the 
others. If for a particular source region $\Omega$, $\Theta$ is exposed 
to $\Omega$ only, $N_{on}^{b}$ can not be estimated with the presented
method.} and the total rate $R_{out}^{b}(t)$ from all viewed sky except 
for the source region are:

\[ \left\{\begin{array}{lcl}
   N_{out}^{b}(\Theta)=\int_{T}\phi(\Theta,t)G(\Theta)R_{b}(t)dt &=&
                       G(\Theta)\int_{T}\phi(\Theta,t)R_{b}(t)dt \\
   R_{out}^{b}(t)=\int    \phi(\Theta,t)G(\Theta)R_{b}(t)d\Theta &=&
                          R_{b}(t)\int\phi(\Theta,t)G(\Theta)d\Theta
   \end{array}\right. \]

The functions $N_{out}^{b}(\Theta)$ and $R_{out}^{b}(t)$ are not
distorted by the presence of any source (by assumption) and can be
measured directly. Then, the set of equations can be solved for
$R_{b}(t)$ and $G(\Theta)$ numerically with the initial approximation to
$R_{b}(t)$ taken from the observed total event rate.

Thus, the expected number of background events in the source region can 
be found from:

\[ N_{on}^{b}(\Theta)=\int_{\Theta}\left[1-\phi(\Theta',t)\right]
                                         R_{b}(t)G(\Theta')d\Theta'dt \]


\section{Performing test for a source presence.}
\label{post_process:search_source}

If $N_{on}(\Theta)$ and $N_{on}^{b}(\Theta)$ are the number of events
observed from the local direction $\Theta$ inside of some bin $\Omega$
in the on-source and off-source measurements respectively, the value of 
the statistic $U$ from~(\ref{eq:post_processing:Udef}) is:

\[ U(\Theta)= \frac{N_{on}(\Theta)-N_{on}^{b}(\Theta)}
{\sqrt{N_{on}^{b}(\Theta)+\alpha(\Theta)N_{on}(\Theta)}} \]
where
\[ N_{1}=N_{on}(\Theta) \ \ \ \ \alpha N_{2}=N_{on}^{b}(\Theta) \ \ \ \
             \alpha(\Theta)=N_{on}^{b}(\Theta)/N_{out}^{b}(\Theta) \]

Since the measurements made from different local directions $\Theta$ are
independent, all measurements from $\Theta$'s which fall into the bin
$\Omega$ can be easily combined to obtain compounded statistic of the
measurement in the bin $\Omega$:

\begin{equation}
    U(\Omega)=
      \frac{\sum_{\Theta}N_{on}(\Theta)-\sum_{\Theta}N_{on}^{b}(\Theta)}
                  {\sqrt{\sum_{\Theta}N_{on}^{b}(\Theta)+
\sum_{\Theta}\frac{N_{on}^{b}(\Theta)N_{on}(\Theta)}{N_{out}^{b}(\Theta)}}}
=\frac{N_{on}(\Omega)-N_{on}^{b}(\Omega)}{\sqrt{N_{on}^{b}(\Omega)+
\sum_{\Theta}\frac{N_{on}^{b}(\Theta)N_{on}(\Theta)}{N_{out}^{b}(\Theta)}}}
                                                \ \ \ \Theta\in\Omega
\label{eq:post_process:U_Omega}
 \end{equation}

The critical value $u_{c}$ of the statistic $U(\Omega)$ is set to five. 
If $U(\Omega)$ is greater than five, the null hypothesis will be
rejected and it will be said that the observed counts must have come
from an astrophysical source. A measurement of the source strength can
be performed.

If the observed $U(\Omega)$ is less than five, the null hypothesis will 
not be rejected and an upper limit corresponding to $2.3\%$ error of the 
second kind will be made. A measurement of the source strength can be 
performed only if it is known (from other experiments) that the source 
exists.

It should be remembered for probability interpretation according to the 
table~\ref{table:post_proc:significance} to be valid, the inequality%
~(\ref{eq:post_processing:u0validity}) needs to be satisfied and for
$u_{c}=5$ and typical value of $\alpha=1/15$ the number of events
observed in the observation bin $N_{on}(\Omega)$ should be about $2\cdot
10^{6}$.


\section{Gamma-Ray flux measurement.}
\label{sec:post_processing:flux_measure}

Given the detector response to particles of different types and assumed
source features, it is possible to predict the number of events
$\hat{N}(\tilde{\Omega})$ to be observed in the bin $\tilde{\Omega}$ due
to the source. Then, $\hat{N}(\tilde{\Omega})$ can be compared with the
actually observed number $N(\tilde{\Omega})$ and some statement
regarding the assumed source features can be made. Indeed, let
$P(k,\tilde{k},E,\tilde{E},\Theta,\tilde{\Theta},\vec{r},\vec{R})$ be
the probability that the detector registers a particle of type $k$
coming from local direction $\Theta$ with energy $E$ distance $\vec{r}$
from the apparatus and reconstruction output information about the
particle is $\tilde{k}$, $\tilde{\Theta}$, $\tilde{E}$, $\vec{R}$. Then,
the total number of events due to particles of type $k$ to be observed
from a region $\tilde{\Omega}$ is:

\[ \hat{N}_{k}(\tilde{\Omega})=
      \sum_{\tilde{k}}\int_{\tilde{\Theta}\in\tilde{\Omega}}
      P(k,\tilde{k},E,\tilde{E},\Theta,\tilde{\Theta},\vec{r},\vec{R})
      F(k,E,\Theta)T(\Theta)
 \,dE\,d\tilde{E}\,d\Theta\,d\tilde{\Theta}\,d\vec{r}\,d\vec{R} \]

where $F(k,E,\Theta)$ is the number of particles of type $k$ with energy
$E$ emitted by the source in local direction $\Theta$ per unit area per
unit time, $T(\Theta)$ is the time during which the source is located in
local direction $\Theta$. The integration is performed over all possible
values of energies $E$ and $\tilde{E}$, all possible distances $\vec{r}$
and $\vec{R}$ and all directions in the field of view $\Theta$, but
$\tilde{\Theta}\in\tilde{\Omega}$.

The integration over core distances $\vec{r}$, $\vec{R}$, measured
energy $\tilde{E}$ and the summation over identified particle type
$\tilde{k}$ can be performed:

\[ P(k,E,\Theta,\tilde{\Theta})=\sum_{\tilde{k}}\int
        P(k,\tilde{k},E,\tilde{E},\Theta,\tilde{\Theta},\vec{r},\vec{R})
                                     \,d\tilde{E}\,d\vec{r}\,d\vec{R} \]
\[ \hat{N}_{k}(\tilde{\Omega})=\int F(k,E,\Theta)T(\Theta)\left[
         \int_{\tilde{\Theta}\in\tilde{\Omega}}
                 P(k,E,\Theta,\tilde{\Theta})\,d\tilde{\Theta}\right]
                                                        \,dE\,d\Theta \]

If it is believed that the error in measuring event's direction does not 
depend on particle energy, $P(k,E,\Theta,\tilde{\Theta})$ can be 
factored as:
\[ P(k,E,\Theta,\tilde{\Theta})=
                          A_{k}(E,\Theta)R_{k}(\tilde{\Theta}|\Theta) \]

The function $A_{k}(E,\Theta)$ is known as the ``effective area''
introduced in section~\ref{detector:effective_area};
$R_{k}(\tilde{\Theta}|\Theta)$ is known as the angular resolution (or 
point spread) function. Then, the number of events to be detected from 
the directions in the bin $\tilde{\Omega}$ is:

\begin{equation}
  \hat{N}_{k}(\tilde{\Omega})=\int F(k,E,\Theta)T(\Theta)A_{k}(E,\Theta)
      \left[\int_{\tilde{\Theta}\in\tilde{\Omega}}
      R_{k}(\tilde{\Theta}|\Theta)\,d\tilde{\Theta}\right] \,dE\,d\Theta
\label{eq:post_processing:N_Omega}
\end{equation}

The integration should be performed over the exposure time to the whole
source and given $A_{k}(E,\Theta)R_{k}(\tilde{\Theta}|\Theta)$ can be
done during data processing. Thus, by counting the number of events in
an observation bin $N_{k}(\tilde{\Omega})$ and comparing it with
$\hat{N}_{k}(\tilde{\Omega})$, it is possible to deduce some
properties of the source function $F(k,E,\Theta)$. If the null
hypothesis is rejected, the difference
$N_{on}^{s}(\tilde{\Omega})=(N_{on}(\tilde{\Omega})-N_{on}^{b}(\tilde{\Omega}))$
should be interpreted as $\gamma$-ray count. 

For instance, if a point source is considered with the source function 
$F(\gamma,E,\Theta)=F_{0}\delta(E-E_{0})\delta(\Theta-\Theta_{0}(t))$ 
where $\Theta_{0}(t)$ is the source path in the local coordinates, the
equation~(\ref{eq:post_processing:N_Omega}) gives:

\[ \hat{N}_{on}^{s}(\tilde{\Omega})=
  F_{0}\int\delta(\Theta-\Theta_{0}(t))T(\Theta)A_{\gamma}(E_{0},\Theta)
                         \left[\int_{\tilde{\Theta}\in\tilde{\Omega}}
   R_{\gamma}(\tilde{\Theta}|\Theta)\,d\tilde{\Theta}\right]\,d\Theta = \]
\[ = F_{0}\int A_{\gamma}(E_{0},\Theta_{0}(t))
                            \left[\int_{\tilde{\Theta}\in\tilde{\Omega}}
   R_{\gamma}(\tilde{\Theta}|\Theta_{0}(t))\,d\tilde{\Theta}\right]\,dt \]

\[ F_{0}=\frac{\hat{N}_{on}^{s}(\tilde{\Omega})}
              {\int A_{\gamma}(E_{0},\Theta_{0}(t))
                           \left[\int_{\tilde{\Theta}\in\tilde{\Omega}}
  R_{\gamma}(\tilde{\Theta}|\Theta_{0}(t))\,d\tilde{\Theta}\right]\,dt} \]

If the null hypothesis is not rejected, an upper limit corresponding to 
the error $\zeta_{u}$ can be set as (see 
equation~(\ref{eq:post_process:U_Omega})):

\[ N^{s}_{on}(\tilde{\Omega})<
 u_{1}(\zeta_{u})\sqrt{\sum_{\tilde{\Theta}}N_{on}^{b}(\tilde{\Theta})+
                \sum_{\tilde{\Theta}}
      \frac{N_{on}^{b}(\tilde{\Theta})N_{on}(\tilde{\Theta})}
                                        {N_{out}^{b}(\tilde{\Theta})}} 
                       \ \ \ \ \ \ \tilde{\Theta}\in\tilde{\Omega} \]
leading to:

\[ F_{0}<\frac{u_{1}(\zeta_{u})
                  \sqrt{\sum_{\tilde{\Theta}}N_{on}^{b}(\tilde{\Theta})+
\sum_{\tilde{\Theta}}\frac{N_{on}^{b}(\tilde{\Theta})N_{on}(\tilde{\Theta})}
            {N_{out}^{b}(\tilde{\Theta})}}}
                {\int A_{\gamma}(E_{0},\Theta_{0}(t))
                          \left[\int_{\tilde{\Theta}\in\tilde{\Omega}}
 R_{\gamma}(\tilde{\Theta}|\Theta_{0}(t))\,d\tilde{\Theta}\right]\,dt}  \]


\section{Optimal bin.}

If the Milagro detector had perfect angular resolution, then processing
events from an infinitesimally small region of the sky around a point
source would yield the maximum achievable sensitivity for source search
as described in section~\ref{post_process:search_source}.  However, due
to detector's finite angular resolution, source events should be
expected and accepted from some finite region around it. This, on the
other hand, will increase the number of cosmic ray background events
collected.  Clearly, the optimal source acceptance region (called the
``optimal bin'') should have a configuration which provides the maximum
power for the source search algorithm described in section%
~\ref{post_process:search_source}. That is, for a given detector angular
resolution function and given background event distribution on the sky,
the optimal bin will provide the maximum value of the statistic $U$.

In fact, the procedure for optimal bin construction, like the procedure
for the construction of the best critical region, should be a part of
the statistical test formulation and can not be modified based on
observed data. Since, in the case considered here, the critical region
on the values of the statistic $U$ has been defined, the optimal bin
construction should be considered within the same framework. The optimal
configuration $\tilde{\Omega}_{opt}$ should maximize the value of
statistic $U(\tilde{\Omega})$, thus the equation on the optimal search
region $\tilde{\Omega}_{opt}$ is:

\[ \frac{\delta U(\tilde{\Omega})}{\delta\tilde{\Omega}}=0 \]

Using the definition of $U(\tilde{\Omega})$ from the 
equation~(\ref{eq:post_process:U_Omega}):

\[ U(\tilde{\Omega})=
 \frac{\hat{N}_{on}^{s}(\tilde{\Omega})}
                      {\sqrt{\hat{N}_{on}^{b}(\tilde{\Omega})
                                  \big[1+\big(\sum_{x\in\tilde{\Omega}}
\frac{\hat{N}_{on}^{b}(x)\hat{N}_{on}(x)}{\hat{N}_{out}^{b}(x)}\big)/
                            \hat{N}_{on}^{b}(\tilde{\Omega}) \big]}} \]

and neglecting $\tilde{\Omega}$-dependence of the term in the square 
brackets one arrives at:

\begin{equation}
   \frac{2}{\hat{N}_{on}^{s}(\tilde{\Omega})}\cdot
   \frac{\delta\hat{N}_{on}^{s}(\tilde{\Omega})}{\delta\tilde{\Omega}}
                                          \Big|_{\tilde{\Omega}_{opt}}=
   \frac{1}{\hat{N}_{on}^{b}(\tilde{\Omega})}\cdot
   \frac{\delta\hat{N}_{on}^{b}(\tilde{\Omega})}{\delta\tilde{\Omega}}
                                          \Big|_{\tilde{\Omega}_{opt}}
\label{post_processing:eq:opt_bin}
\end{equation}

The solution of this equation will be performed under the assumptions
that the shape of the observation bin is circular with opening angle
$\tilde{\omega}<<1$; that the bin is centered on a source occupying not
more than a solid angle with opening $\bar{\omega}<<1$ and that the 
number of background events in the bin is proportional to its area (this 
is reasonable in the small angle limit). Then:

\[ \hat{N}_{on}^{b}(\tilde{\omega})\sim \tilde{\omega}^{2}
 \ \ \ \mbox{and} \ \ \ 
   \frac{1}{\hat{N}_{on}^{b}(\tilde{\omega})}\cdot
    \frac{d\hat{N}_{on}^{b}(\tilde{\omega})}{d\tilde{\omega}}
               \Big|_{\tilde{\omega}_{opt}}=\frac{2}{\tilde{\omega}} \]

Also, it will be assumed that the detector's angular resolution is 
described by a 2-D Gaussian with dispersion $\sigma^{2}$. This means 
that the error $\epsilon$ on the reconstructed angle can be parametrized 
by two variables $x$ and $y$ which form a Cartesian coordinate system 
(see figure~\ref{post_processing:angle_res_concept}) and 
$\epsilon=\sqrt{x^{2}+y^{2}}$. The probability of observing event 
$(x,y)$ away from the true direction is:
\[ dR(x,y)=\frac{1}{2\pi\sigma^{2}}e^{-(x^{2}+y^{2})/2\sigma^{2}}dxdy \]
This representation is valid for small errors since the sphere may be 
treated as a plane.

\begin{figure}
\centering
\includegraphics[width=4.3in]{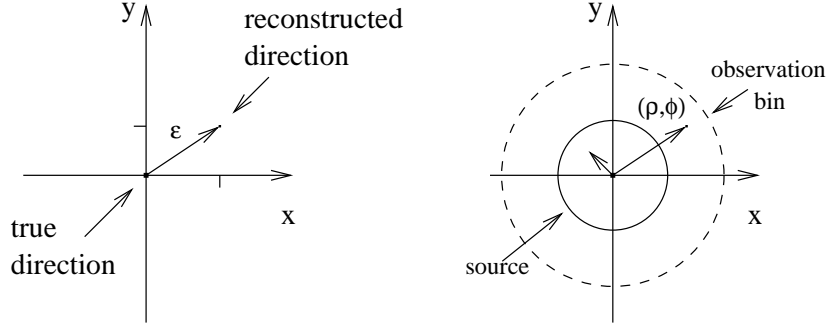}
\caption{Conceptual diagram of a small angular reconstruction error 
         parameterization.}
\label{post_processing:angle_res_concept}
\end{figure}

Under these assumptions, a point source with the source function \\
$F_{\gamma}(E,\Theta)=F(E)\delta(\Theta-\Theta_{0})$
will produce the signal (see 
equation~(\ref{eq:post_processing:N_Omega})):

\[ \hat{N}_{on}^{s}=\left(
  T(\Theta_{0})\int_{E}F(E)A_{\gamma}(E,\Theta_{0})dE\right)\times
   \int_{x^{2}+y^{2}\leq\tilde{\omega}^{2}}\frac{1}{2\pi\sigma^{2}}
                                 e^{-(x^{2}+y^{2})/2\sigma^{2}}dxdy= \]
\[ =\left(T(\Theta_{0})\int_{E}F(E)A_{\gamma}(E,\Theta_{0})dE\right)\times
               \int_{0}^{\tilde{\omega}}\frac{\epsilon}{\sigma^{2}}
                            e^{-\epsilon^{2}/2\sigma^{2}} d\epsilon= \]
\[ =\left(T(\Theta_{0})\int_{E}F(E)A_{\gamma}(E,\Theta_{0})dE\right)\times
                     \Big[1-e^{-\tilde{\omega}^{2}/2\sigma^{2}}\Big] \]

Then, the equation~(\ref{post_processing:eq:opt_bin}) becomes the
equation on the optimal bin opening $\tilde{\omega}$:

\[ \left(\frac{\tilde{\omega}_{opt}}{\sigma}\right)^{2}=2\ln\left[1+
          \left(\frac{\tilde{\omega}_{opt}}{\sigma}\right)^{2}\right]
      \ \ \ \Rightarrow \ \ \ \tilde{\omega}_{opt}\approx 1.585\sigma \]

For a source whose source function $F_{\gamma}(E,\Theta)$ is smooth
within some opening angle $\bar{\omega}<<1$ and zero outside and which 
is located in local direction $\Theta_{0}$, the number of events 
detected from the source in the observation bin is (from equation
(\ref{eq:post_processing:N_Omega})):

\[ \hat{N}_{on}^{s}(\Theta_{0},\tilde{\omega},\bar{\omega})=
                           \frac{1}{2\pi\sigma^{2}}
  \int\int_{x^{2}+y^{2}\leq\bar{\omega}^{2}}T(\Theta_{0}+\vec{\epsilon})
  F_{\gamma}(E,\Theta_{0}+\vec{\epsilon})
                               A(E,\Theta_{0}+\vec{\epsilon})\times \]
\[  \times\left[
           \int_{\tilde{x}^{2}+\tilde{y}^{2}\leq\tilde{\omega}^{2}}
             e^{-((x-\tilde{x})^{2}+(y-\tilde{y})^{2})/2\sigma^{2}}
                                  d\tilde{x}d\tilde{y}\right]dxdydE \]
where $\vec{\epsilon}=(x,y)$ --- describes coordinate of a point inside 
the source bin relative to its center $\Theta_{0}$. Introducing 
the notations $\bar{\omega}^{\sigma}=\bar{\omega}/\sigma$ and
$\tilde{\omega}^{\sigma}=\tilde{\omega}/\sigma$ and substituting the
coordinate system parameterization from Cartesian to polar as
$x=\rho\sigma\cos\phi,\ y=\rho\sigma\sin\phi$ the integration over 
$\tilde{\phi}$ can be performed and one arrives at:

\[ \hat{N}_{on}^{s}(\Theta_{0},\tilde{\omega}^{\sigma},\bar{\omega}^{\sigma})=
         \sigma^{2}\int_{0}^{2\pi}
       \int_{0}^{\bar{\omega}^{\sigma}}\int T(\Theta_{0}+\vec{\epsilon})
         F_{\gamma}(E,\Theta_{0}+\vec{\epsilon})
                             A(E,\Theta_{0}+\vec{\epsilon})dE\times \]
\begin{equation}
   \times\left[\rho e^{-\rho^{2}/2}
              \int_{0}^{\tilde{\omega}^{\sigma}}e^{-\tilde{\rho}^{2}/2}
I_{0}(\rho\tilde{\rho})\tilde{\rho}d\tilde{\rho}\right]\,d\rho\,d\phi
\label{post_process:N_s_example}
\end{equation}
where $I_{0}(\rho)$ is the zeroth order modified Bessel function of the
first kind and $\vec{\epsilon}=(\rho\sigma\cos\phi,\rho\sigma\sin\phi)$.

In the expression~(\ref{post_process:N_s_example}) the integration over
the angle $\phi$ can be performed by expanding the
$T(\Theta_{0}+\vec{\epsilon})F_{\gamma}(E,\Theta_{0}+\vec{\epsilon})
A(E,\Theta_{0}+\vec{\epsilon})$ in to Taylor series:

\[ T(\Theta_{0}+\vec{\epsilon}) F_{\gamma}(E,\Theta_{0}+\vec{\epsilon})
    A(E,\Theta_{0}+\vec{\epsilon})=
                T(\Theta_{0})F_{\gamma}(E,\Theta_{0})A(E,\Theta_{0})+ \]
\[ +\vec{\epsilon}\cdot\nabla_{\Theta}
      \Big(T(\Theta_{0})F_{\gamma}(E,\Theta_{0})A(\Theta_{0})\Big)+
                                               {\cal O}(\epsilon^{2}) \]

All even order corrections in $\vec{\epsilon}$ will give zero
contribution to the integral~(\ref{post_process:N_s_example}) because
the $\phi$ integration is performed over $2\pi$ range. Keeping only the
first order correction in the source size $\bar{\omega}^{\sigma}$ the 
number of events in the observation bin factors as:

\[ \hat{N}_{on}^{s}(\Theta_{0},\tilde{\omega}^{\sigma},\bar{\omega}^{\sigma})=
       \left(\sigma^{2}
\int T(\Theta_{0})F_{\gamma}(E,\Theta_{0})A(E,\Theta_{0})dE\right)\times \]
\[  \times\int_{0}^{\bar{\omega}^{\sigma}}\rho e^{-\rho^{2}/2}
   \left(\int_{0}^{\tilde{\omega}^{\sigma}}e^{-\tilde{\rho}^{2}/2}
            I_{0}(\rho\tilde{\rho})\rho d\tilde{\rho}\right)d\rho \]

Hence, in the equation~(\ref{post_processing:eq:opt_bin}) the
$\Theta_{0}$-dependent factor will cancel and the optimal bin size will
loose its dependence of the source location. The
figure~\ref{post_processing:optimal_bin} shows the solution
$\tilde{\omega}^{\sigma}_{opt}$ of the optimal bin size
problem~(\ref{post_processing:eq:opt_bin}) for the smooth source of size
$\bar{\omega}^{\sigma}$. In the limiting case of the zero source size, 
the solution converges to the previously obtained
$\tilde{\omega}^{\sigma}_{opt}=1.585$. It is also interesting to note
that the fraction $f_{\gamma}^{opt}(\bar{\omega}^{\sigma})$ of the 
signal events retained in the optimal bin is a very weak function of the 
source size:

\[ f_{\gamma}^{opt}(0.0)=  0.715 \ \ \
   f_{\gamma}^{opt}(1.05)= 0.721 \ \ \
   f_{\gamma}^{opt}(1.9)=  0.751 \]

\begin{figure}
\centering
\includegraphics[width=4.0in]{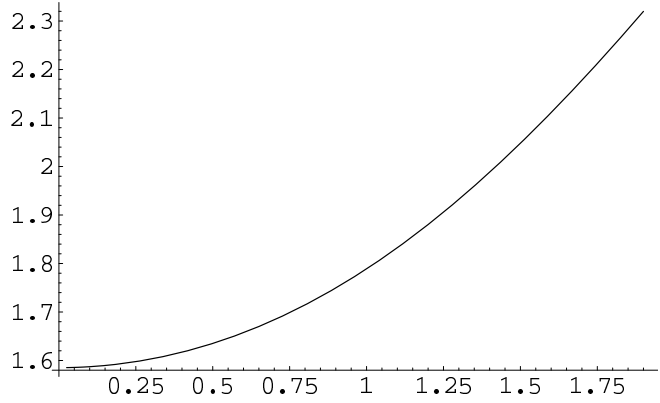}
\caption{Optimal bin size $\tilde{\omega}^{\sigma}_{opt}$ as a function 
         of the size source $\bar{\omega}^{\sigma}$.}
\label{post_processing:optimal_bin}
\end{figure}

\begin{table}
\centering
\begin{tabular}{|c|c|c|c|c|c|c|c|c|c|} \hline
$\bar{\omega}^{\sigma}$ &  $0.00$ & $0.35$ &  $0.40$ &  $0.50$ &  $0.60$ &
                             $0.70$ & $0.80$ &  $0.90$ &  $1.05$ \\ \hline 
$\tilde{\omega}^{\sigma}_{opt}$   & $1.585$ & $1.610$ & $1.617$ &  
       $1.635$ &  $1.657$ &  $1.684$ & $1.715$ & $1.750$ & $1.810$  \\ \hline
\end{tabular}
\caption{Source size $\bar{\omega}^{\sigma}$ and corresponding optimal 
         bin size $\tilde{\omega}^{\sigma}_{opt}$.}
\end{table}

The assumption that the number of background events in a bin is
proportional to the bin's area is a good one for the Milagro data and
the considered examples provide a good guideline for the choice of the
observation bin size. Note that the constructed optimal bin is the
optimal among the source centered circular bins and some other bin shape
could be better. Nevertheless, the circular bin will be used in this
analysis.


\chapter{Photon flux at the Earth due to near solar neutralino 
         annihilations}
\label{chapter:simulations}

\epigraph{George  Orwell, ``1984''}{%
   `Yes,' I says to 'er, `That's all very well,' I says. `But if you'd
been in my place you'd of done the same as what I done. It's easy to
criticize,' I says, `but you ain't got the same problems as what I got.'

   `Ah,' said the other, `that's jest it. That's jest where it is.'
}

While some introduction to the main goal of the current work has been
done, the proper formulation of the problem is long overdue.

There is overwhelming evidence that the Universe, and the galaxies in
particular, are full of the ``dark matter''. There is no reason to
assume that the Milky Way is any different. In this work, it is supposed
that the weakly interacting particle (neutralino), predicted by
super-symmetric theories, is the solution of the ``dark matter''
problem. If this is indeed the case, the neutralinos will form a halo
around the Milky Way Galaxy and at the location of the Solar System the 
density of the halo neutralinos is often assumed to be $\rho_{0}=0.3\ 
(GeV/cm^{3})$.

The neutralinos entering the Solar system may loose energy via elastic
scattering with ordinary matter scatterers and become trapped in the
Solar system. For simplicity, the Solar system is assumed to consist of
the uniform density Sun only. This means that only the particles whose
orbits cross the Sun can be captured on near-solar bound orbits and
their orbits will always cross the Sun. Due to the capture and repeated
scatterings in the Sun, there will be a near-solar enhancement in the
density of the neutralinos. This process is responsible for dark matter
accretion in the Solar system. The dark matter diminution is due to
neutralino-neutralino annihilations. The annihilation can not happen
faster than accretion, otherwise, all dark matter would have annihilated
by now. On the other hand, if accretion happens faster than
annihilation, the Sun would constantly increase its mass. Thus, it is
reasonable to assume that the Solar system has reached dynamic
equilibrium and that the capture rate is equal to the annihilation one.
Also, it is reasonable to assume that all possible elliptical orbits,
not crossing the Sun have annihilated by now, as there is no mechanism
to populate them by the incoming particles. Given that one of the
annihilation channels is $\gamma$-ray production, one might expect an
enhanced $\gamma$-ray signal from the neighborhood of the Sun due to
neutralino annihilations.  The purpose of this chapter is to estimate
the $\gamma$-ray flux due to this process.

\section{General formulation of the problem.}

The problem of finding the density distribution of the particles in the
Solar system can be addressed by kinetic theory.\footnote{The
hydro-dynamic approach is not justifiable since the neutralinos do not
interact with each other.} If $g(p,x)$ is the density of particles in
phase space, then it should satisfy the Boltzmann
equation\footnote{Summation over repeated indices is assumed.}:

\[ \frac{d}{dt}g(p,x)=\frac{\partial g(p,x)}{\partial p^{i}}\dot{p_{i}}+
         \frac{\partial g(p,x)}{\partial x^{i}}\dot{x_{i}} =C[g(p,x)] \]

where $C[g(p,x)]$ is the collision integral and the explicit dependence
of $g(p,x)$ on time has been dropped since in the considered model the
process is assumed to be stationary: $\partial g(p,x)/\partial t=0$. The 
spatial density of particles $n(x)$ is:

\[ n(x)=\int g(p,x) dp \]

The collision integral consists of two terms: one is due to neutralino
annihilations $C_{a}[g(p,x)]$ and the other is due to scattering in the
Sun $C_{s}[g(p,x)]$.

\[ C_{a}[g(p,x)]=-\int \sigma(p,p')g(p,x)g(p',x)dp'=
                 -\langle\sigma v\rangle_{a}g(p,x)n(x) \]

\[ C_{s}[g(p,x)]=\Theta(x,R_{\odot})\int\left\{ W(p+q,q)g(p+q,x) -
                  W(p,q)g(p,x)\right\} dq \]
\[  \Theta(x,R_{\odot})=\left\{\begin{array}{ll}
                             1, & |x|\leq R_{\odot} \\
                             0, & |x|>    R_{\odot}
                        \end{array}\right. \]

where $q$ is the particle momentum change in a collision,  $W(p,q)$ is 
the probability that a particle will change its momentum from $p$ to 
$(p-q)$ in a collision.

A simplification can be made by noting that neutralino mass is much 
greater than that of any scatterer in the Sun and relative energy loss 
and momentum change of neutralino in a scattering are small. Thus, 
$W(p,q)$ is a quickly decreasing function of $|q|$ and diffusion 
approximation can be made:

\[ W(p+q)g(p+q,x)\approx  \]
\[ \approx W(p,q)g(p,x) +
         q_{i}\frac{\partial}{\partial p^{i}}\Big( W(p,q)g(p,x)\Big)+
  \frac{1}{2}q_{i}q_{j}\frac{\partial^{2}}{\partial p^{i}\partial p^{j}}
         \Big( W(p,q)g(p,x)\Big)+\cdots  \]

\[ C_{s}[g(p,x)]\approx
      \frac{\partial}{\partial p^{i}}\Big\{\tilde{A}_{i}(p)g(p,x)+
      \frac{\partial}{\partial p^{j}}\Big(B_{ij}(p)g(p,x)\Big)\Big\} \]
where
\[ \tilde{A}_{i}(p)=\int q_{i}W(p,q)dq \ \ \
   B_{ij}(p)=\frac{1}{2}\int q_{i}q_{j}W(p,q)dq \]
The function $W(p,q)$ can be constructed by considering a structure-less 
elastic scattering process where the angle of deflection of the 
incident particle in the center of mass reference frame is uniformly 
distributed between $0$ and $\pi$.

The boundary condition for the problem can be formulated by assuming a
Maxwellian distribution of velocities of galactic neutralinos. Then, in 
the Sun's reference frame the distribution will be shifted by the 
velocity of the Sun $V_{0}$ in the Galactic disk:

\[ \lim_{x\rightarrow\infty}g(p,x)=g_{\infty}(p)=
        \rho_{0}\left(\frac{1}{2\pi v_{0}^{2}m_{\chi}^{2}}\right)^{3/2}
     e^{-\frac{(p-m_{\chi}V_{0})^{2}}{2v_{0}^{2}m_{\chi}^{2}}}d^{3}p \]
where $m_{\chi}$ is the neutralino mass. It will also be assumed that 
$\sqrt{2v_{0}^{2}}=\sqrt{V_{0}^{2}}=220\, (km/s)$.

The sought for annihilation rate density at a point $x$ is simply:
\[ I_{a}(x)=-\int C_{a}[g(p,x)]dp= \langle\sigma v\rangle_{a}n^{2}(x) \]

Needless to say, the task of solving the stated problem analytically or
numerically is daunting even when $W(p,q)$ has a simple structure. 
Direct computer simulations of the system will require enormous amounts
of computer time. However, the distribution function $g(p,x)$ is not the
immediate goal of the project and only the distribution of the
annihilation points is of interest. Therefore, it is proposed to perform
computer simulations of annihilating particles only. This poses two
problems: how to know that the particle will annihilate and what the
boundary condition on the annihilated particles is.

\section{General idea of the solution.}

The stated problem is solved with ``backward in time'' simulation. The
particles generated at the annihilation points (so, it is known that the
particles annihilated) are then propagated backward in time gaining
energy in each scattering in the Sun until they exit the Solar system. 
If the distribution of the annihilation points is correct, the correct
distribution of the annihilating particles at the boundary will be
restored automatically. Thus, an algorithm should be devised to adjust
an {\it a priori} distribution of the annihilation points in such a way
as to reconstruct the correct distribution of annihilating particles at
the Solar system boundary. This is possible if the particles carry
information about their origin right until the annihilation point. The
last assumption is reasonable since the neutralinos interact in the Sun
only and their momentum does not change much in each scattering.

Since two particles are required in the act of annihilation, the
particle pairs will be considered. Let $X$ describe the state of a pair
of particles, then $X_{0}$ will denote the pair state at the
annihilation point and $X_{\infty}$ --- the state at the boundary.  The
trajectory pairs can be divided into two classes: the first class
containing the trajectories with both particles trapped in the Solar
system via scattering in the Sun and the second --- the trajectories
when at least one of the trajectories was not trapped and annihilated in
its first flight through the Solar system. Supposing that the time of
flight through the Solar system is much smaller than the mean life time
of the particle, the contribution from the second class can be neglected
and only captured trajectory pairs can be considered. Let the joint
probability that a pair of captured particles annihilated at $X_{0}$ and
entered the Solar system at $X_{\infty}$ be $P(X_{\infty},cap,X_{0})$.
The task of the project is to find the distribution of the annihilation
points or $P(X_{0},cap)$ and according to Bayes' theorem, the joint
probability $P(X_{\infty},cap,X_{0})$ is:
\begin{equation}
  P(X_{\infty},cap,X_{0})=P(X_{0}|X_{\infty},cap)P(X_{\infty},cap)=
                                P(X_{\infty}|cap,X_{0})P(X_{0},cap) \\
\label{eq:simulations:bayes}
\end{equation}
\begin{equation}
 \int P(X_{\infty},cap,X_{0})\, dX_{0}=P(X_{\infty},cap)=
             \int P(X_{\infty}|cap,X_{0})P(X_{0},cap)\, dX_{0}
\label{eq:simulations:bayes2}
\end{equation}
and
\[ P(X_{\infty},cap)=P(cap|X_{\infty})P(X_{\infty}) \]

where $P(X_{\infty})$ is the known distribution of all trajectories
entering the Solar system at infinity and $P(cap|X_{\infty})$ is the
conditional probability that the trajectories $X_{\infty}$ will be
captured.

The probabilities $P(X_{\infty})$ and $P(cap|X_{\infty})$ can be found 
analytically (see\linebreak sections~\ref{sec:simulations:PX_infty}
and~\ref{sec:simulations:PcapX_infty} correspondingly) and
$P(X_{\infty}|cap,X_{0})$ can be constructed using a backward in time
computer simulation (see section~\ref{sec:simulations:PX_infty_cap_X0}). 
Also a method of solving the equation~(\ref{eq:simulations:bayes2}) for
$P(X_{0},cap)$ is described in section~\ref{sec:simulations:weighting}.


\subsection{Solution of the Fredholm equation.}
\label{sec:simulations:weighting}

The equation~(\ref{eq:simulations:bayes2}) is the Fredholm equation of 
the first kind with respect to the sought for function $P(X_{0},cap)$:

\begin{equation}
 P(X_{\infty},cap)=\int P(X_{\infty}|cap,X_{0})P(X_{0},cap)\, d(X_{0},cap)
\label{eq:simulations:fredholm}
\end{equation}

The kernel $P(X_{\infty}|cap,X_{0})$ of this equation is not known
analytically and is constructed in simulations (see
section~\ref{sec:simulations:PX_infty_cap_X0}). It should be noted that
due to the nature of the kernel construction method, the kernel
$P(X_{\infty}|cap,X_{0})$ represents a ``list'' of ``$(X_{0},cap)$''s
and ``$(X_{\infty},cap)$''s which are ``connected'' by the simulation
process.  Because the initial state $(X_{0},cap)$ is random and the
propagation process is stochastic, it is statistically improbable to
have repeated pairs in the list. This implies that all knowledge
regarding the kernel can be expressed as:

\[ P(X_{\infty}|cap,X_{0})=\left\{\begin{array}{ll}
          1, & \{(X_{\infty},cap),(X_{0},cap)\}\in\mbox{list} \\
          0, & \{(X_{\infty},cap),(X_{0},cap)\}\not\in\mbox{list}
                                                   \end{array}\right. \]

Thus, the equation~(\ref{eq:simulations:fredholm}) becomes:
\[ P(X_{0},cap) = P(X_{\infty},cap) \]

In other words, when annihilation state $(X_{0},cap)$ of a captured
particle is associated with a state at the boundary $(X_{\infty},cap)$
by the propagation process from
section~\ref{sec:simulations:PX_infty_cap_X0} the relative contribution
$P(X_{0},cap)$ from the state $(X_{0},cap)$ is $P(X_{\infty},cap)$.

Thus, generating the initial states $(X_{0},cap)$ such that all final 
states $(X_{\infty},cap)$ are sampled will lead to the solution of the 
equation~(\ref{eq:simulations:fredholm}).

To construct the histogram of the radial distribution of the 
annihilation points it should be noted that the state 
$X_{0}=(\vec{r}_{0},\vec{v}_{1},\vec{v}_{2})$ is the position and the 
velocities of the two particles at the annihilation point, thus:

\[ P(\vec{r}_{0},cap)= \int_{\omega(\vec{r}_{0})} P(X_{0},cap)
                                     \, d\vec{v}_{1}\, d\vec{v}_{2}=
  \int_{\omega(\vec{r}_{0})}P(\vec{r}_{0},\vec{v}_{1},\vec{v}_{2})
                                     \, d\vec{v}_{1}\, d\vec{v}_{2} \]
where $\omega(\vec{r}_{0})$ is the velocity volume over which the 
integration is being performed. This volume should include all particles 
which are captured and is finite. If the above integration is performed 
by the means of the Monte Carlo method with uniform sampling in the 
velocity volume $P(\vec{r}_{0},cap)$ becomes:

\[ P(\vec{r}_{0},cap)=\frac{\omega^{2}(\vec{r}_{0})}{N_{v}(\vec{r}_{0})} 
   \sum_{\vec{v}_{1},\vec{v}_{2}}P(\vec{r}_{0},\vec{v}_{1},\vec{v}_{2}) \]
where $N_{v}(\vec{r}_{0})$ is the number of sampled points.

Because the histogram of the annihilation points $H(\vec{r}_{0})$ is 
defined is the average of $P(\vec{r}_{0},cap)$ in the $\vec{r}_{0}$ bin, 
$H(\vec{r}_{0})$ is:

\[ H(\vec{r}_{0})=\frac{1}{N_{r_{0}}}
                               \sum_{\vec{r}_{0}}P(\vec{r}_{0},cap)=
                  \frac{1}{N_{r_{0}}}\sum_{\vec{r}_{0}}
                    \frac{\omega^{2}(\vec{r}_{0})}{N_{v}(\vec{r}_{0})}
   \sum_{\vec{v}_{1},\vec{v}_{2}}P(\vec{r}_{0},\vec{v}_{1},\vec{v}_{2}) \]
where $N_{r_{0}}$ is the number of entries in the $\vec{r}_{0}$ bin.

The obtained expression for the histogram $H(\vec{r}_{0})$ of the 
annihilation points may be simplified by choosing a fixed large volume 
of the velocity space $\omega(\vec{r}_{0})=\Omega$ and by noting that 
in any finite random sample the probability to observe two different 
pairs of orbits passing through the same point $\vec{r}_{0}$ is zero 
leading to $N_{v}(\vec{r}_{0})=1$.

\[ H(\vec{r}_{0})=\frac{\Omega^{2}}{N_{r_{0}}}
                  \sum_{\vec{r}_{0},\vec{v}_{1},\vec{v}_{2}}
                               P(\vec{r}_{0},\vec{v}_{1},\vec{v}_{2}) \]

\subsection{Transition tables $\{(X_{\infty},cap),(X_{0},cap)\}$.}
\label{sec:simulations:PX_infty_cap_X0}

The transitions from $X_{\infty}$ to $X_{0}$ for captured particles are
found with backward in time simulations and was mentioned before. 
Because the particles evolve independently, each particle is propagated
from its initial state (annihilation point) backward in time until it
encounters an interaction in the Sun. At this point, momentum is changed
according the rules of elastic scattering, the energy is gained, and the
new angular momentum is computed. Afterwards, the particle is propagated
until it encounters the next scattering. This process repeats until
the accumulated energy is greater than zero, which means that the
particle is no longer bound to the Solar system and its state at
infinity is found and recorded.

Because the particle may spend long periods of time between interactions
in the Sun, one must solve the equations of particle motion analytically
and use the results. The motion in central potential fields is
integrable and each trajectory is defined by integrals of motion: the
total energy of neutralino $E$ and its angular momentum $\vec{J}$.
However, the kinematics depends only on mass densities of these 
quantities: ${\cal E}=E/m$ and $\vec{\cal J}= \vec{J}/m$ and mostly 
those will be used in the calculations.

The major simplification in the trajectory calculation comes from the
fact that the energy and angular momentum are conserved between the
scatterings. This means that between the scatterings the motion is
executed in one plane and only rotation of the whole orbit is possible.
The Runge-Lenz vector $\vec{\cal K}$, which fixes the orientation of
the particle orbit and is conserved outside the Sun may change its
direction only when particle passes through the Sun. Since between the
scatterings the trajectory inside the Sun does not change its
properties, the particle passage through the Sun can be tracked by
rotation of the Runge-Lenz vector. Any point on the particle trajectory
can be specified by its angle relative to the current direction of the
Runge-Lenz vector.

The trajectory length inside the Sun plays the main role in the
propagation process because it is the quantity which defines when the
next scattering should occur. Given an initial point and the path-length
inside the Sun until the next scattering, the point of next scattering
can be found by rotating the Runge-Lenz vector on the appropriate angle.
Thus, the task is to convert the trajectory length inside the Sun into
the angle of rotation.  This problem can be solved analytically for the
specified solar model.

The act of scattering can also be described as rescaling and the
rotation of the velocity. Thus the whole process of particle propagation
from its annihilation point back to infinity can be described as a
sequence of rotations applied to the Runge-Lenz vector and the velocity
rotation with rescaling.

The details of the simulation process are described in the 
appendix~\ref{chap:simulations_app:simulations_app}.


\subsection{Distribution of neutralinos at infinity $P(X_{\infty})$.}
\label{sec:simulations:PX_infty}

Because the state $X_{\infty}$ describes the state of two identical
particles at the boundary, the $P(X_{\infty})$ function is constructed
as the product of two identical distributions describing a single
particle:

\[ P(X_{\infty})=g_{\infty}(x_{1})g_{\infty}(x_{2}) \]

The expression for $g_{\infty}(x)$ is a simple generalization of the
results obtained earlier in~\cite{Gould1987} and~\cite{PressSpergel}.
However, a self-consistent derivation is provided here for completeness.

Let us choose a sphere of a large radius $R$ around the Sun so that the
effects of Sun's gravity are negligible and the velocity distribution of
the particles is a known function $f(\vec{v})d^{3}\vec{v}$ and does not
depend on the point on the sphere. Let $n_{\chi}$ be the concentration
of particles at the sphere. The number of particles entering the Solar
system per unit time with velocity $\vec{v}$ from the surface element
$d\vec{A}$ is then:
\[ dN =n_{\chi}f(\vec{v})\,(\vec{v}\cdot d\vec{A})\,d^{3}\vec{v}\,dt \] 
where we are interested in the particles with $(\vec{v}\cdot
d\vec{A})<0$ since the particles should enter the sphere.

Since we are considering the sphere $dA=R^{2}\sin\theta\,d\theta\,d\phi_{J}$ 
and $d\vec{A}\uparrow\uparrow\vec{R}$ we can choose the coordinate system
$(\theta,\,\phi_{J})$ so that $\theta$ is counted from the direction of
the velocity $\vec{v}$, then:

\[ (\vec{v}\cdot d\vec{A})= (\vec{v}\cdot\vec{R})R\sin\theta d\theta d\phi_{J}
  =\frac{1}{2v}d\left(v^{2}R^{2}\sin^{2}\theta\right)d\phi_{J}=
   \frac{d{\cal J}^{2}d\phi_{J}}{2v} \]

Then, the number of particles entering the Solar system is:
\[ dN=n_{\chi}f(\vec{v})
                  \frac{dt\,d^{3}\vec{v}\,d{\cal J}^{2}d\phi_{J}}{2v},
                  \ \ \ (\vec{v}\cdot\vec{R})<0  \]

If we are interested in the total distribution, we must note that since
the velocity distribution does not depend on the spatial point, for every
particle with $(\vec{v}\cdot\vec{R})<0$ there will be exactly one
particle with $(\vec{v}\cdot\vec{R})>0$. Hence, the number of particles
entering the Solar system with the velocity $\vec{v}$ and magnitude of
angular momentum ${\cal J}$ per unit time is:

\begin{equation}
   dN=
   n_{\chi}f(\vec{v})\frac{dt\,d^{3}\vec{v}\,d{\cal J}^{2}d\phi_{J}}{4v},
                           \ \ \ \ 
   P(X_{\infty})=\frac{dN}{dt\,d^{3}\vec{v}\,d{\cal J}^{2}d\phi_{J}}=
                 \frac{n_{\chi}f(\vec{v})}{4v}
\label{PXinfty}
\end{equation}

Consider the case when the velocity distribution at infinity is
spherically symmetric as in~\cite{PressSpergel},
\[ f(v)v^{2}dv=4\pi(2\pi\sigma^{2})^{-3/2}e^{-v^{2}/2\sigma^{2}}v^{2}dv \]
then after integration over the spherical coordinates of the velocity
and $d\phi_{J}$ one arrives at:
\[ \frac{dN}{dt}=2\pi^{2}n_{\chi}f(\vec{v})v\,dv\,d{\cal J}^{2}
                =2\pi^{2}n_{\chi}f(\vec{v})\,d{\cal E}\,d{\cal J}^{2} \]
This is the formula (2.7) from ~\cite{PressSpergel}.

If the motion of the Sun with speed $V_{0}$ in the locally isotropic 
Galactic halo is taken into account, then, as in~\cite{Gould1987}:

\[ \tilde{f}(v)v^{2}dv =
      \int_{\theta,\phi}f(\vec{v})\frac{d^{3}\vec{v}}{v}=
      \frac{\sinh\frac{vV_{0}}{\sigma^{2}}}{\frac{vV_{0}}{\sigma^{2}}}
      e^{-V_{0}^{2}/2\sigma^{2}}
\left[4\pi(2\pi\sigma^{2})^{-3/2}e^{-v^{2}/2\sigma^{2}}v^{2}dv\right] \]
and the rate at which the particles enter the Solar system per 
angular momentum per speed is:
\[ dN=\pi n_{\chi}dt\,d{\cal J}^{2}
                  \int_{\theta,\phi}f(\vec{v})\frac{d^{3}\vec{v}}{2v}=
            \frac{\pi}{2}n_{\chi}dt\,d{\cal J}^{2}\tilde{f}(v)v^{2}dv \]
This is the expression which will be used for calculation of the 
distribution of the particles at infinity.


\subsection{Capture probability $P(cap|X_{\infty})$.}
\label{sec:simulations:PcapX_infty}

Because the state $X_{\infty}$ describes the state of two identical
particles at the boundary and because each particle is captured
independently, the $P(cap|X_{\infty})$ function is constructed as the
product of two identical capture functions describing a single
particle:

\[ P(cap|X_{\infty})=g_{cap}(x_{1})g_{cap}(x_{2}) \]

Also it will be assumed that the capture happens in one collision. In
other words, after the first collision (forward time) the particle will
have negative energy\footnote{The particle may scatter twice or more in
the first pass through the Sun, but it is assumed to become trapped
after the first collision.}. This will greatly simplify calculations,
while higher order corrections can be considered. The argument for this
is that the mean-free-path of neutralinos in the Sun $\lambda
=1/n_{p}\sigma_{p\chi}$ is of the order of $10^{4}-10^{9}\ (R_{\odot})$
for expected values of $\sigma_{p\chi}$. The energy loss in a collision
is (see section~\ref{sec:simulations_app:scatter}):

\[ \Delta{\cal E}=
     \frac{2\eta(1-\cos\theta)}{(\eta +1)^{2}}\frac{v^{2}}{2}=
     \frac{2\eta(1-\cos\theta)}{(\eta +1)^{2}}\Big({\cal E}-U(r)\Big)=
         \nu\Big({\cal E}-U(r)\Big) \ \ \ \eta=\frac{m_{\chi}}{m_{p}} \]
where ${\cal E}$ is the energy before the collision and $U(r)$ is the
potential energy at the collision. Since $\theta$ is the scattering angle
in the center-of-mass system and no cross-section structure is assumed,
$\cos\theta$ is distributed uniformly between $-1$ and $1$, this leads to
the fact that $\nu$ is uniformly distributed between zero and
$\frac{4\eta}{(\eta +1)^{2}}$.

\[ {\cal E}_{after}={\cal E}-\Delta{\cal E} <0 \ \ \Rightarrow\ \
             {\cal E}-\nu ({\cal E}-U(r)) <0   \ \ \Rightarrow\ \
            \left\{\begin{array}{ll}
            \nu > \frac{\cal E}{{\cal E}-U(r)} & \mbox{need for capture}\\
            0< \nu < \frac{4\eta}{(\eta +1)^{2}} & \mbox{allowed range}
            \end{array}\right. \]
Thus, the probability that a particle will be captured in one collision 
at the distance $r$ from the center of the Sun is:

\[ p_{cap}({\cal E},r)=\frac{(\eta +1)^{2}}{4\eta}
  \left[\frac{4\eta}{(\eta +1)^{2}} - \frac{\cal E}{{\cal E}-U(r)}\right]
\Theta\left[\frac{4\eta}{(\eta +1)^{2}} - \frac{\cal E}{{\cal E}-U(r)}\right] \]
\[ \Theta(z)=\left\{\begin{array}{ll}
               1,& z\geq 0 \\
               0,& z < 0
             \end{array}\right. \]

The probability that a particle will travel distance $y$ without 
scattering and scatter in $y,y+dy$ is:
\[ dp=\frac{1}{\lambda}e^{-y/\lambda}dy \approx \frac{dy}{\lambda} \]
Again, this approximation is valid since the mean free path of 
neutralinos inside the Sun $\lambda$ is much greater than the particle 
trajectory length inside the Sun.

The probability that the particle with energy ${\cal E}$ and angular
momentum ${\cal J}$ will loose energy in one collision to become
captured on the bound orbit is:

\[  g_{cap}(x)=
      \int_{0}^{L({\cal E,J})}\frac{1}{\lambda}p_{cap}({\cal E},r(y))dy \]
where $L({\cal E,J})$ is the path-length inside the Sun.

From the energy conservation law:
\[ {\cal E}= \frac{v^{2}}{2} +U(r) =
           \frac{\dot{r}^{2}}{2}+ \frac{{\cal J}^{2}}{2r^{2}}+U(r) \]
\[ \frac{dy}{dt}=\sqrt{2[{\cal E}-U(r)]}, \ \ \ 
             \frac{dr}{dt}=\sqrt{2[{\cal E}-U(r)]-{\cal J}^{2}/r^{2}} \]

\[ g_{cap}(x)=\frac{2}{\lambda}\int_{r_{min}}^{r_{max}}
\sqrt{\frac{2[{\cal E}-U(r)]}{2[{\cal E}-U(r)]-{\cal J}^{2}/r^{2}}}\left\{
      \frac{(\eta +1)^{2}}{4\eta}\left[\frac{4\eta}{(\eta +1)^{2}} -
                       \frac{\cal E}{{\cal E}-U(r)}\right]\right\}\, dr \]
The $r_{min}$ is the minimal distance from the Sun's center to the orbit 
and can be found from the equation of motion:
\[ {\cal E}={\cal J}^{2}/2r_{min}^{2}+U(r_{min})=
            {\cal J}^{2}/2r_{min}^{2}-\frac{\alpha}{2R_{\odot}}
        \left(3-r_{min}^{2}/R^{2}_{\odot}\right)  \ \Rightarrow\ \ \]
\[ r_{min}^{2}=R_{\odot}^{2}\frac{(3\alpha R_{\odot}+2{\cal E}R_{\odot}^{2})-
\sqrt{(3\alpha R_{\odot}+2{\cal E}R_{\odot}^{2})^{2}-4\alpha R_{\odot}{\cal J}^{2}}}%
{2\alpha R_{\odot}} \]

The $r_{max}$ is computed from the restriction on maximum capturable 
energy and should not be greater than $R_{\odot}$ or less than 
$r_{min}$.

\[ \frac{4\eta}{(\eta +1)^{2}} - \frac{\cal E}{{\cal E}-U(r)} >0 \
\Rightarrow\ {\cal E} < -\frac{4\eta}{(\eta -1)^{2}}U(r)
\]

\[ r_{max}^{2}=R_{\odot}^{2}\left(
      3-\frac{R_{\odot}}{\alpha}\frac{(\eta-1)^{2}}{2\eta}{\cal E}\right),
      \ \         r_{min}\leq r_{max}\leq R_{\odot},\ \ \
{\cal E}<\frac{4\eta}{(\eta -1)^{2}}\cdot\frac{3\alpha}{2R_{\odot}} \]

The integrals which need to be executed to find $g_{cap}(x)$ are
elliptical integrals. $g_{cap}(x)$ can be written in the
form:

\[ g_{cap}(x)=\frac{1}{\lambda}\int_{r_{min}^{2}}^{r_{max}^{2}}
         \sqrt{\frac{a-by}{ay-by^{2}-c}}dy \ - \ \ \ \ \ \ \ \]
\[ \ \ \ \ \ \ \ \ \   -\ \frac{(\eta +1)^{2}}{2\eta}\cdot
         \frac{\cal E}{\lambda}\int_{r_{min}^{2}}^{r_{max}^{2}}
         \frac{dy}{\sqrt{(ay-by^{2}-c)(a-by)}}  \]
where
\[  a=2{\cal E}+\frac{3\alpha}{R_{\odot}}, \ \ \ 
    b=\frac{\alpha}{R_{\odot}^{3}},       \ \ \
    c={\cal J}^{2} \]
From Gradshtein and Rizhik(Russian 3.141-2 page 245):
\[  \int_{r_{min}^{2}}^{r_{max}^{2}}\sqrt{\frac{a-by}{ay-by^{2}-c}}dy=
                \int_{C}^{r_{max}^{2}}\sqrt{\frac{y-A}{(y-B)(y-C)}}dy=
                                            2\sqrt{(A-C)}\,EE(\gamma,q) \]

From Gradshtein and Rizhik(Russian 3.131-3 page 233):
\[ \int_{r_{min}^{2}}^{r_{max}^{2}} \sqrt{\frac{1}{(ay-by^{2}-c)(a-by)}}dy=
   \frac{1}{b}\int_{C}^{r_{max}^{2}}\sqrt{\frac{1}{(y-A)(y-B)(y-C)}}dy= \]
\[ =\frac{2}{b\sqrt{(A-C)}}EF(\gamma,q) \]

\[ A>B\geq r_{max}^{2}>C, \ \ \ 
   A=\frac{a}{b}, \ \ \
   B=\frac{a+\sqrt{a^{2}-4bc}}{2b}, \ \ \ \
   C=r_{min}^{2}=\frac{a-\sqrt{a^{2}-4bc}}{2b} \]
\[ \gamma = \arcsin\sqrt{\frac{r_{max}^{2}-C}{B-C}},\ \ \
   q=\sqrt{\frac{B-C}{A-C}} \]
Further simplification comes from the fact that $A-C=B$

Where the elliptical integrals are:
\[ EF(\phi,k)=\int_{0}^{\phi} 1/\sqrt{1-k^{2}\sin^{2}(t)}\,dt=
           \int_{0}^{\sin\phi}\frac{dt}{\sqrt{(1-t^{2})(1-k^{2}t^{2})}},
                                                      \ \ \ \ |k|<1 \]
\[ EE(\phi,k)=\int_{0}^{\phi}   \sqrt{1-k^{2}\sin^{2}(t)}\,dt=
           \int_{0}^{\sin\phi}\sqrt{\frac{1-k^{2}t^{2}}{1-t^{2}}}\,dt,
                                                      \ \ \ \ |k|<1 \]


\section{Predicted photon flux.}

\begin{table}
\centering
\begin{tabular}{||c||c||c||}                     \hline
$m_{\chi},\ (TeV)$  & $f_{out}$ &
     $I(m_{\chi})\times\frac{\sigma_{p\chi}}{10^{-43}cm^{2}}
                           \frac{\rho_{0}}{0.3GeV/cm^{3}},\ (s^{-1})$
                                                        \\  \hline\hline
 $0.1$  & $0.195$ & $1.65\ 10^{18}$   \\ \hline
 $0.2$  & $0.195$ & $4.17\ 10^{17}$   \\ \hline
 $0.5$  & $0.196$ & $6.72\ 10^{16}$   \\ \hline
 $1.0$  & $0.199$ & $1.68\ 10^{16}$  \\ \hline
 $2.0$  & $0.201$ & $4.22\ 10^{15}$  \\ \hline
 $5.0$  & $0.2$ & $6.72\ 10^{14}$  \\ \hline
$10.0$  & $0.2$ & $1.69\ 10^{14}$  \\ \hline
\end{tabular}
\caption{Summary of the simulation/computation results.
Capture integral $I$ and the fraction $f_{out}$ of annihilations between 
$1$ and $2$ solar radii as a function of neutralino mass $m_{\chi}$.}
\label{table:simulations:results}
\end{table}

The results of the computer calculation are summarized in the
table~\ref{table:simulations:results} for several selected neutralino
masses. About $40-50\%$ of particles annihilate outside the Sun, but
their distribution is a sharply falling function of distance from the
Sun (see figures~\ref{fig:simulations:annihilation_radial1} and
\ref{fig:simulations:annihilation_radial2}), so only the annihilations
happening between one and two solar radii will be considered to produce
detectable signal.  The fraction of neutralinos annihilating between one
and two radii of the Sun is denoted as $f_{out}$ in the table.


Only a small fraction of the annihilated particles will produce photon
signal. If photon yield for producing a photon with energy $E_{\gamma}$
per neutralino in neutralino-neutralino annihilation is
$b_{\gamma}(E_{\gamma},m_{\chi})$, the total number of photons produced
per second is:

\[ d\tilde{\Phi}_{0}= I(m_{\chi})\cdot f_{out}\cdot 
                        b_{\gamma}(E_{\gamma},m_{\chi})\ dE_{\gamma} \]

Some of the produced photons will be absorbed by the Sun. The fraction
of the photons escaping the Sun is $f_{escape}$ and is of the order of
$1/2$. The distance between the Earth and the Sun is $L_{\oplus}$ which
leads to the flux of number of photons per unit area per time at Earth 
from neutralino annihilations as:

\[ d F_{\chi}(E_{\gamma})= I(m_{\chi})\cdot f_{out}\cdot 
               b_{\gamma}(E_{\gamma},m_{\chi})\cdot 
                         f_{escape}/4\pi L_{\oplus}^{2}\ dE_{\gamma} \]

\[ d F_{\chi}(E_{\gamma})=\]
\begin{equation}
           =\frac{\rho_{0}}{0.3\ (GeV/cm^{3})}\cdot
            \frac{\sigma_{p\chi}}{10^{-43}\ cm^{2}}\cdot 
                 b_{\gamma}(E_{\gamma},m_{\chi})\cdot
            \frac{f_{out}\cdot f_{escape}}{0.2\cdot 1/2}\cdot
  \frac{I(m_{\chi})}{2.8\cdot 10^{28}}\ dE_{\gamma} \ \ \ \ (cm^{-2}s^{-1})
\label{eq:simulations:predicted_flux}
\end{equation}

The photon yield may have the following structure:

\[ b_{\gamma}(E_{\gamma},m_{\chi}) =
    b_{\gamma}^{\delta}(m_{\chi})\delta\Big(E_{\gamma}-m_{\chi}\Big) +
     b_{\gamma}^{c}(m_{\chi}) P\Big(\frac{E_{\gamma}}{m_{\chi}}\Big) \]

where $P\Big(\frac{E_{\gamma}}{m_{\chi}}\Big)$ is the probability to
produce photon with energy $E_{\gamma}$, $E_{\gamma}<m_{\chi}$ in the 
continuous spectrum neutralino to photon annihilation process.

There are indications (see~\cite{Bergstrom3}) that the continuum 
spectrum probability has the form:

\[ P\Big(\frac{E_{\gamma}}{m_{\chi}}\Big)\sim\frac{1}{m_{\chi}}\cdot
\Big(\frac{E_{\gamma}}{m_{\chi}}\Big)^{-1.5}e^{-7.8E_{\gamma}/m_{\chi}} \]

\begin{figure}
\centering
\includegraphics[width=3.3in]{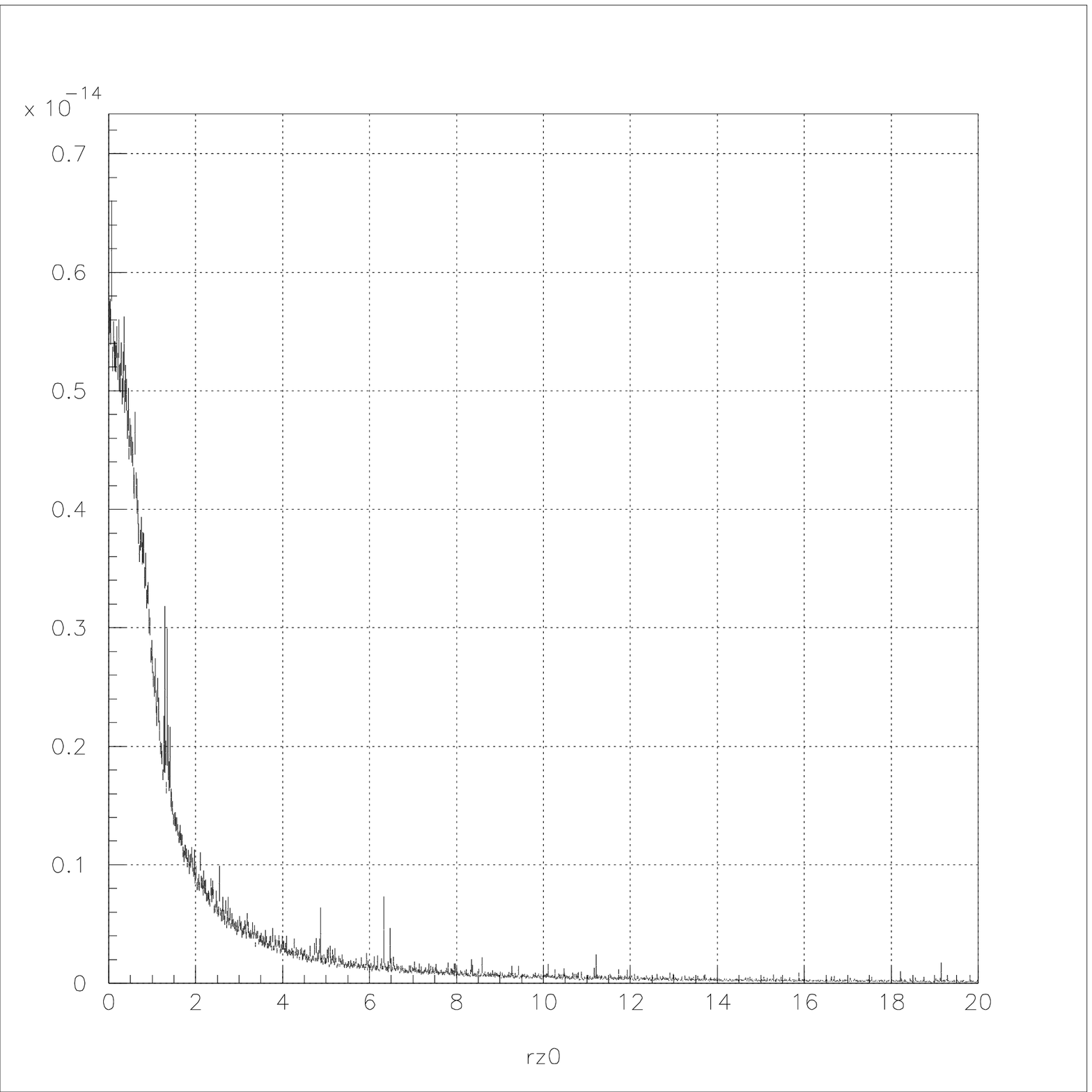}
\caption{Radial distribution of the annihilation points for 
$m_{\chi}=200\ (GeV)$ and $\sigma_{p\chi}=10^{-43}\ (cm^{2})$. Vertical 
scale is in arbitrary units, horizontal scale is in $R_{\odot}$.
above $25\cdot 10^{6}$ neutralino annihilations}
\label{fig:simulations:annihilation_radial1}
\end{figure}

\begin{figure}
\centering
\includegraphics[width=3.3in]{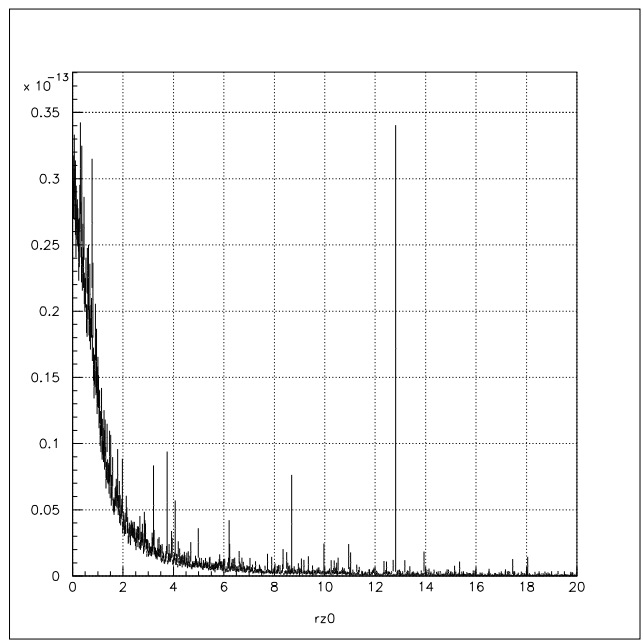}
\caption{Radial distribution of the annihilation points for 
$m_{\chi}=1000\ (GeV)$ and $\sigma_{p\chi}=10^{-43}\ (cm^{2})$. Vertical 
scale is in arbitrary units, horizontal scale is in $R_{\odot}$.
above $22\cdot 10^{6}$ neutralino annihilations}
\label{fig:simulations:annihilation_radial2}
\end{figure}

%
%

\chapter{Outcome of the test for presence of the photon flux from the 
         Sun and its implications}
\label{chapter:results}

\epigraph{George  Orwell, ``1984''}{%
His courage seemed suddenly to stiffen of its own accord.
}

The gamma ray signal from neutralino annihilations near the Sun should
appear as an excess number of events from the direction of geometrical
center of the Sun over the expected background. Observation of the Solar
region can be performed by tracking the Sun on the Celestial sphere
using the one-arc-minute precision formulae for the Sun's Celestial
coordinates from~\cite{FlandernPulkkinen}. The interpretation of the
observed signal, however, is not an easy problem. Largely, this is due
to the fact that the cosmic ray background is not expected to be
uniform; the Sun absorbs the cosmic rays impinging on it and forms a
cosmic-ray shadow. The situation is complicated by the magnetic fields
of the Earth and the Sun. Due to bending of charged particles
trajectories in magnetic fields, the Sun's shadow will be smeared and
shifted from the geometrical position of the Sun in the TeV range of
particle energies. On the other hand, in the presence of strong Solar
magnetic fields, lower energy particles can not reach the surface of the
Sun and are reflected from it. Such particles are not being removed from
the interplanetary medium and may not even form a cosmic-ray shadow of
the Sun.  Therefore, it is difficult to ascertain the exact shape of the
cosmic-ray shadow at the Sun's position and deduce excess above it.

The effect of the Earth's magnetic field and the Solar wind can be
studied by observing the shadow of the Moon during solar day. If the
solar magnetic field is weak, the shadows of the Sun and the Moon should
be very similar because of the geometry of the problem. The Sun and the
Moon cover similar size regions on the celestial sphere and traverse
similar paths on the local sky in one year of observation. In addition,
the Moon is far enough from the Earth to be considered outside of the
effect of the Earth's magnetic field, so is the Sun.

\section{The data set.}

\begin{table}
\centering
\begin{tabular}{|c|c|c|c|}              \hline \hline
        & $N_{on}$ & $N_{on}^{b}$ & U      \\ \hline \hline
Sun     & 137211  & 137728  & -1.35  \\ \hline
Moon    & 49762   & 50064   & -1.31  \\ \hline
\end{tabular}
\caption{Number of events in the optimal bin centered on the Sun and the
         Moon (see section~\ref{post_process:search_source}).}
\label{table:results:Nevents}
\end{table}

\begin{figure}
\centering
\begin{tabular}{cc}
\includegraphics[width=2.7in]{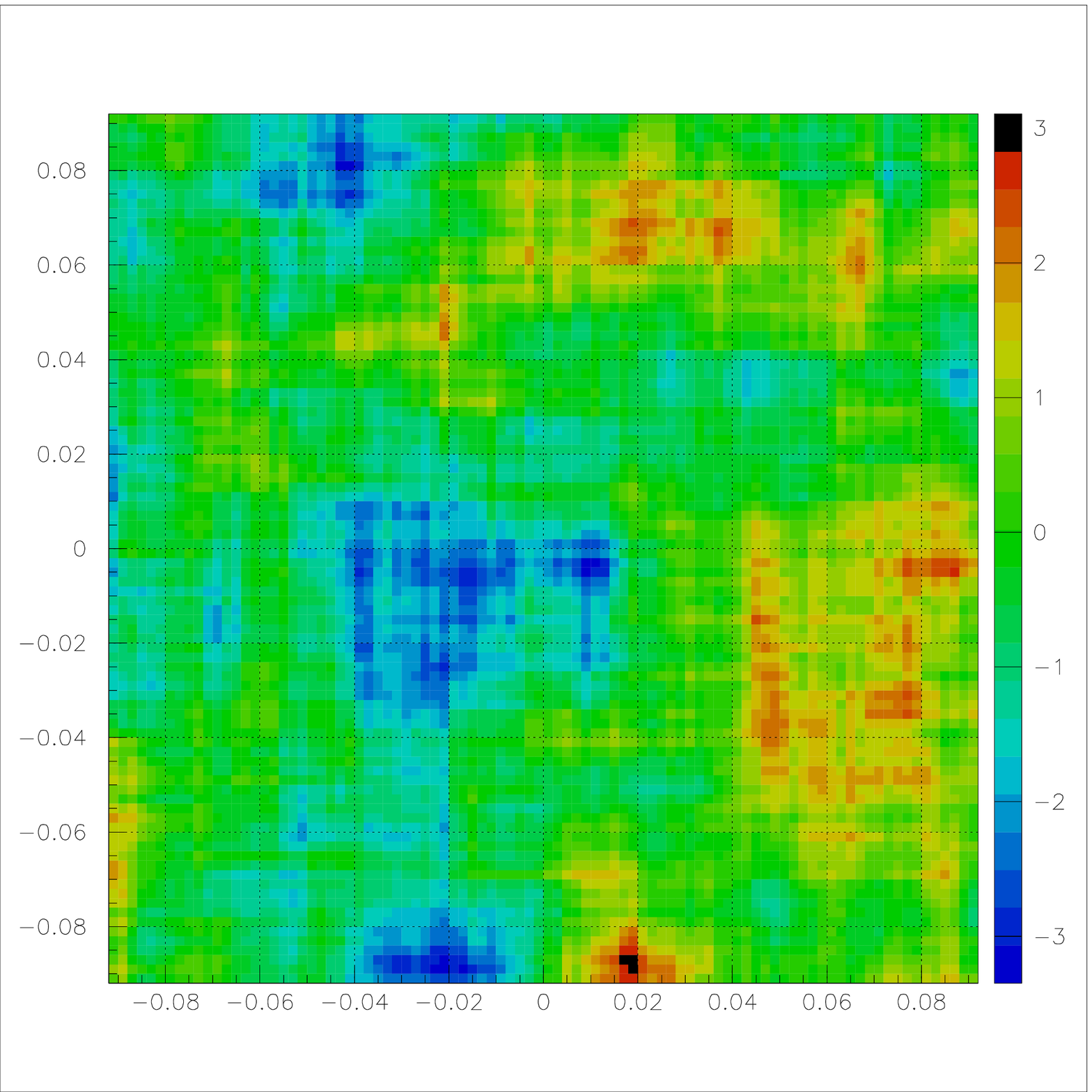} &
\includegraphics[width=2.7in]{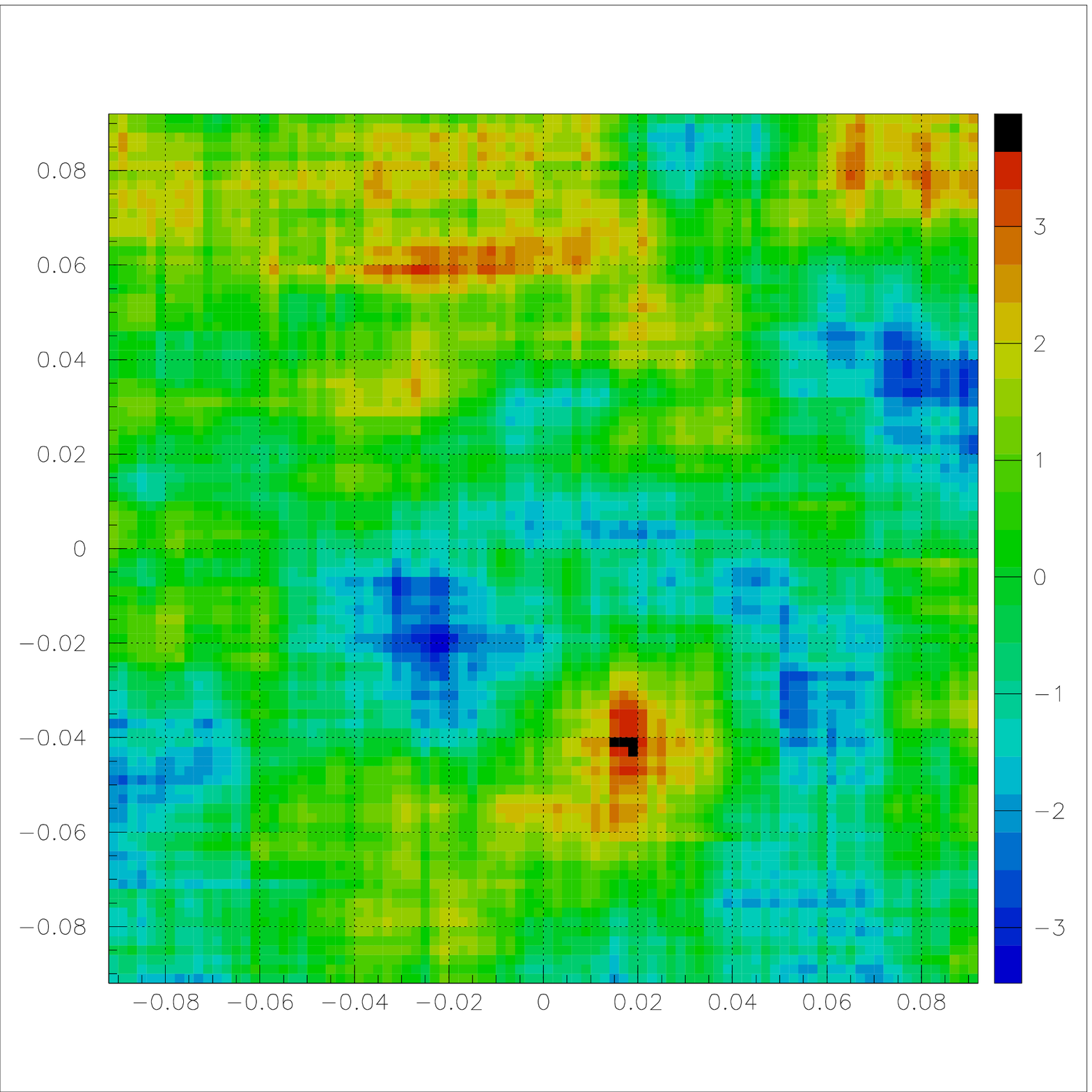} \\
\includegraphics[width=2.7in]{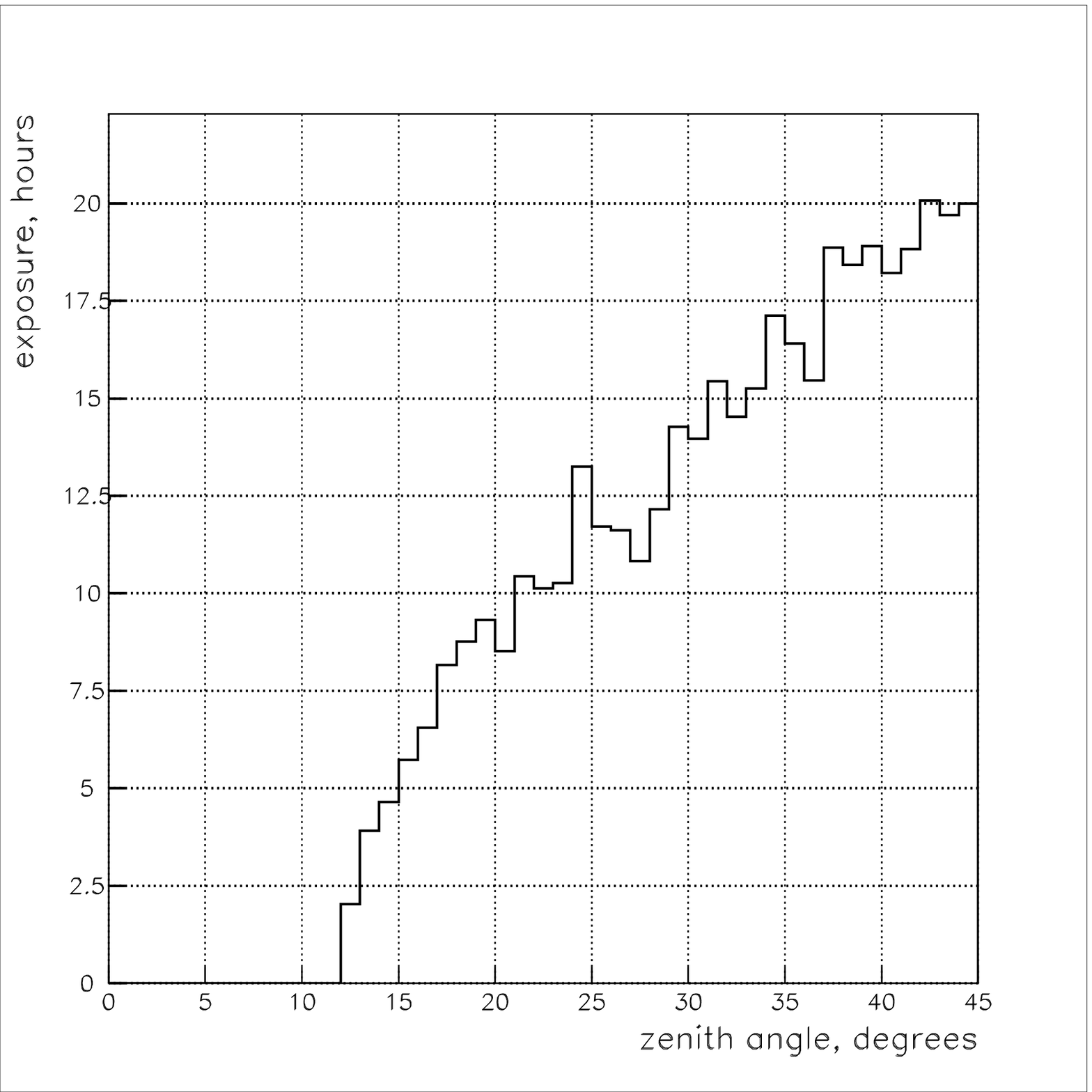} &
\includegraphics[width=2.7in]{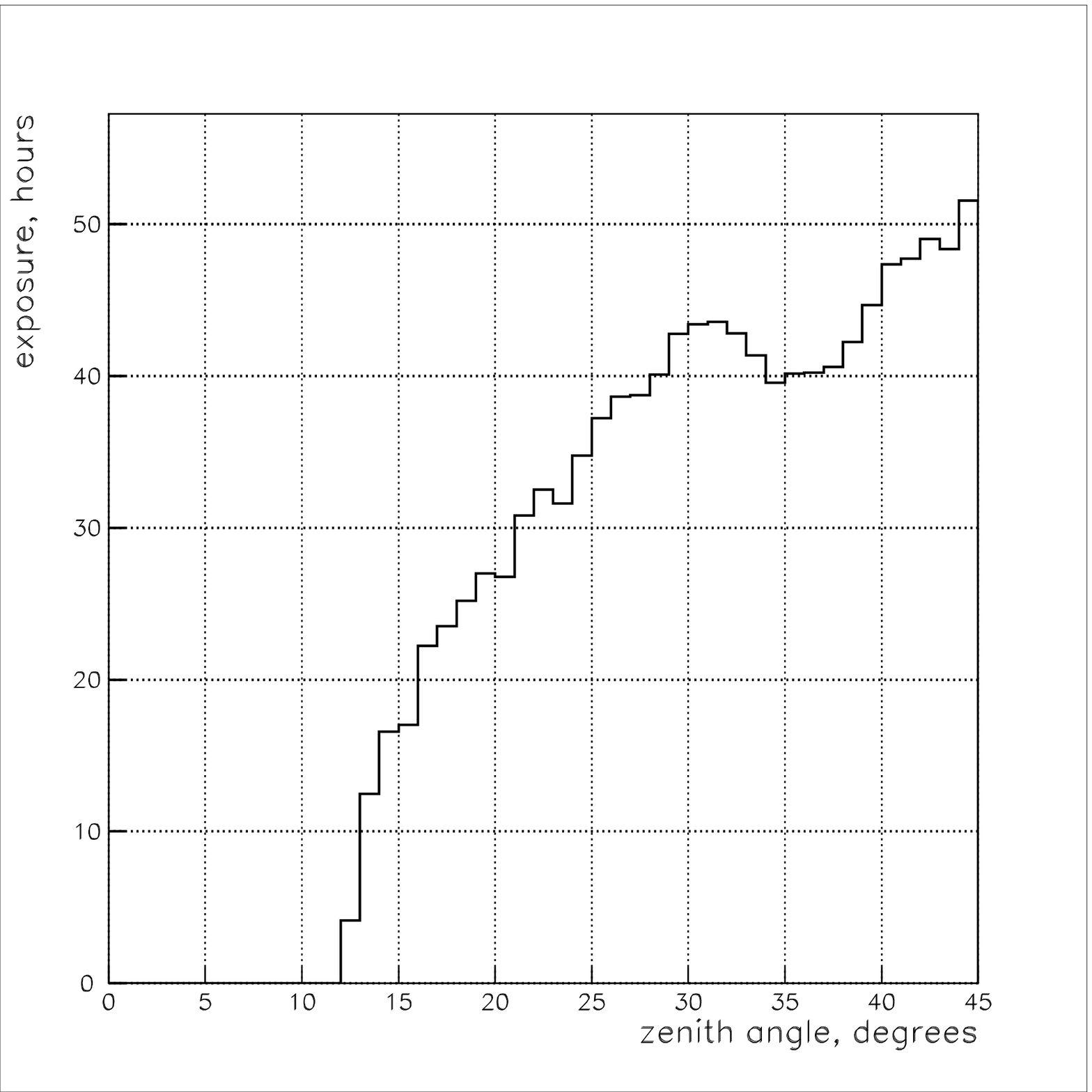} \\
\end{tabular}
\caption{Significance maps of the regions of the sky around the daytime 
Moon(left) and the Sun(right) and the corresponding source exposure as 
function of zenith angle in hours per degree. The color code is the 
value of $U$ (see equation~(\ref{eq:post_process:U_Omega})).}
\label{fig:results:maps_and_exposure}
\end{figure}

The data to be used in this work was chosen to satisfy the following
conditions: online reconstruction, the number of photo-tubes required to
trigger the detector greater than 60, the number of photo-tubes used in
the angular reconstruction (Nfit) greater than $20$, zenith angles
smaller than $45$ degrees and all events should pass the gamma/hadron
separation cut. The data used were collected between July 19 2000 and
September 10 2001. The dates are motivated by introduction of the hadron
separation parameter into the online reconstruction code on July 19,
2000 and detector turn-off for major repairs on the 11th of September
2001. Several data runs were disregarded from the dataset which included
calibration runs and the data when the online DAQ was in an unstable
regime.

For the Sun analysis a $\pm 5^{\circ}$ regions around the Moon and the
Crab nebula were vetoed from the data set. For the Moon analysis, same
size regions around the Sun and the Crab were vetoed. Overall, 1164.7
hours of exposure on the Sun and 423.5 hours of exposure on the Moon
during the day time is obtained in this data set.  The number of events
in the optimal circular bin of $1.26^{\circ}$ radius centered on the
Moon and the Sun is given in table~\ref{table:results:Nevents} and the
sky maps with corresponding exposure graphs are presented in
figure~\ref{fig:results:maps_and_exposure}. The sky maps are generated
according to the equation~(\ref{eq:post_process:mapping}) with the 
vertical axis pointing to the Geocentric Geomagnetic North dipole pole.

\section{A limit on possible gamma-ray flux due to near-Solar 
neutralino annihilations.}

\begin{table}
\centering
\begin{tabular}{|c|c|c|} \hline\hline
$m_{\chi}\ (TeV)$ & $\Delta\ (cm^{2}s)$ & $\Sigma\ (cm^{2}s)$ \\ \hline\hline
0.1  & $1.055\cdot 10^{11}$  &  $0.000$             \\ \hline
0.2  & $8.772\cdot 10^{11}$  &  $4.969\cdot 10^{7}$ \\ \hline
0.5  & $6.070\cdot 10^{12}$  &  $2.634\cdot 10^{9}$ \\ \hline
1.0  & $3.389\cdot 10^{13}$  &  $2.127\cdot 10^{10}$ \\ \hline
2.0  & $1.600\cdot 10^{14}$  &  $1.280\cdot 10^{11}$ \\ \hline
5.0  & $5.942\cdot 10^{14}$  &  $1.208\cdot 10^{12}$ \\ \hline
10.0 & $1.608\cdot 10^{15}$  &  $5.575\cdot 10^{12}$ \\ \hline
20.0 & $3.684\cdot 10^{15}$  &  $2.136\cdot 10^{13}$ \\ \hline
50.0 & $8.030\cdot 10^{15}$  &  $1.035\cdot 10^{14}$ \\ \hline
\end{tabular}
\caption{Coefficients of the flux limit calculation (see 
         equation~(\ref{eq:results:flux_constraints})).}
\label{table:results:flux_limit_coef}
\end{table}


\begin{figure}
\centering
\begin{tabular}{cc}
\includegraphics[width=2.7in]{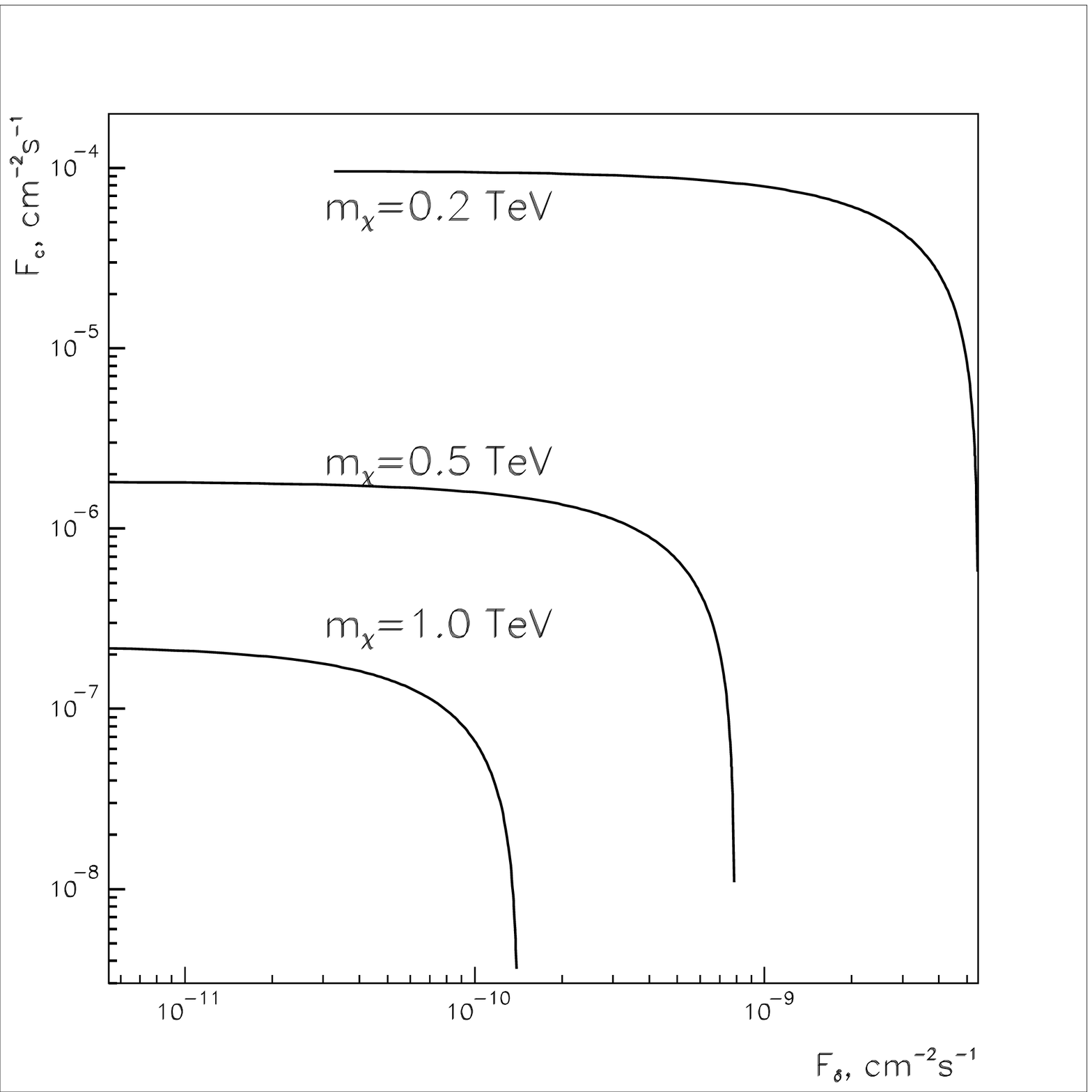} &
\includegraphics[width=2.7in]{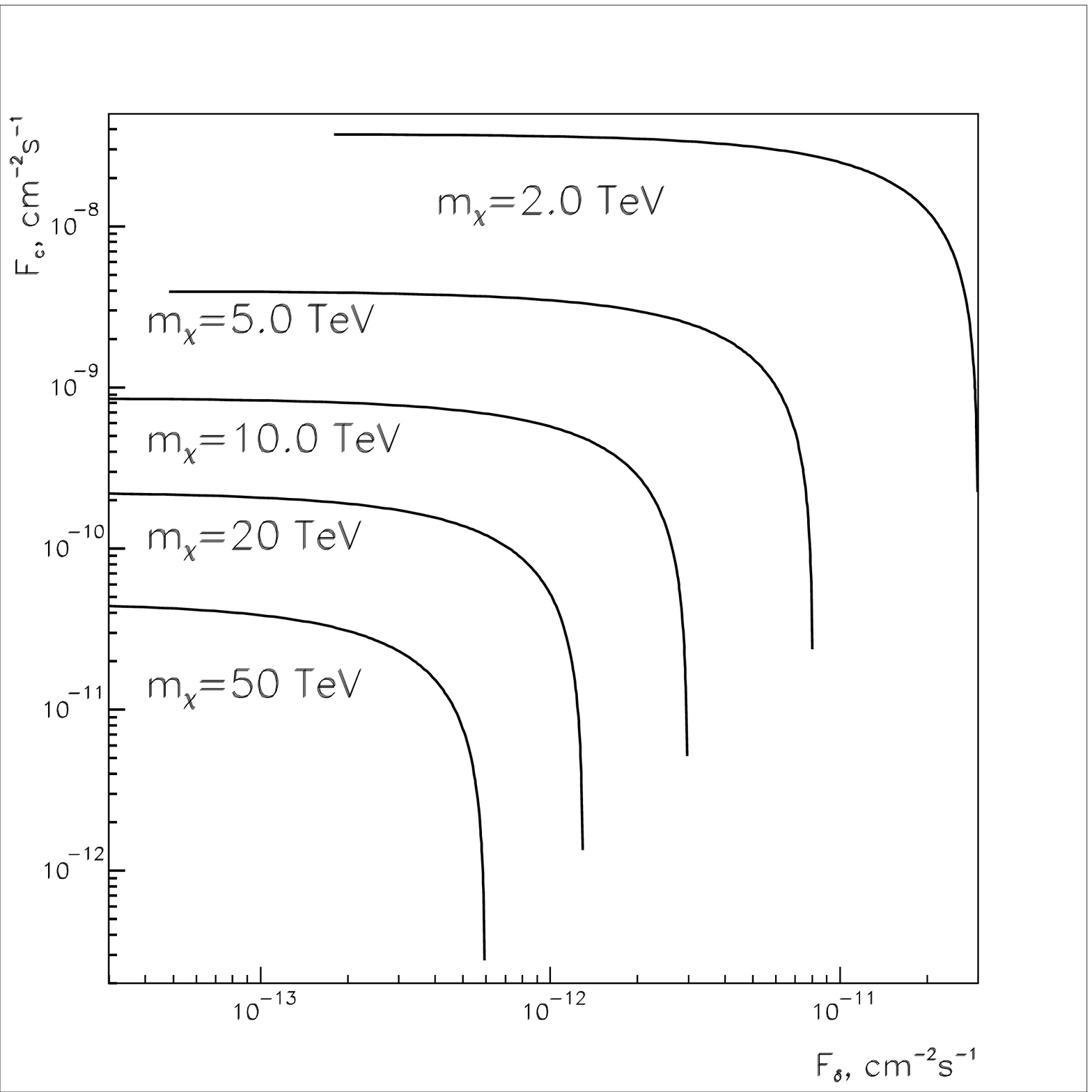} 
\end{tabular}
\caption{The values of ($F_{\delta}$,$F_{c}$) below the lines are 
         allowed based the constructed upper limit for corresponding 
         neutralino masses.}
\label{fig:results:flux_limit_1}
\end{figure}



\begin{table}
\centering
\begin{tabular}{|c|c|c|}              \hline \hline
$m_{\chi}\ (TeV)$ & $F_{\delta}<\ (cm^{-2}s^{-1})$ & 
        $\frac{\sigma_{p\chi}}{10^{-43}cm^{2}}
         \frac{\rho_{0}}{0.3GeV/cm^{3}}b_{\gamma}^{\delta}<$ \\ \hline \hline
0.1    & $4.54\cdot 10^{-8}$  & $770$    \\ \hline
0.2    & $5.46\cdot 10^{-9}$  & $351$    \\ \hline
0.5    & $7.89\cdot 10^{-10}$ & $328$    \\ \hline
1.0    & $1.41\cdot 10^{-10}$ & $234$    \\ \hline
2.0    & $2.99\cdot 10^{-11}$ & $204$    \\ \hline
5.0    & $8.06\cdot 10^{-12}$ & $339$    \\ \hline
10.0   & $2.98\cdot 10^{-12}$ & $512$    \\ \hline

\end{tabular}
\caption{The upper limit on the monochromatic photon flux due to 
near-solar neutralino annihilations and corresponding upper limit on the 
$\sigma_{p\chi}\rho_{0}b_{\gamma}^{\delta}$.}
\label{table:results:limits}
\end{table}


\begin{table}
\centering
\begin{tabular}{|c|c|c|}              \hline \hline
$m_{\chi}\ (TeV)$ & $F_{c}<\ (cm^{-2}s^{-1})$ & 
            $\frac{\sigma_{p\chi}}{10^{-43}cm^{2}}
             \frac{\rho_{0}}{0.3GeV/cm^{3}}b_{\gamma}^{c}<$ \\ \hline 
\hline
0.1    &      ---             &      ---            \\ \hline
0.2    & $9.64\cdot 10^{-5}$  & $6.47\cdot 10^{6}$   \\ \hline
0.5    & $1.82\cdot 10^{-6}$  & $7.58\cdot 10^{5}$   \\ \hline
1.0    & $2.25\cdot 10^{-7}$  & $3.75\cdot 10^{5}$   \\ \hline
2.0    & $3.74\cdot 10^{-8}$  & $2.48\cdot 10^{5}$   \\ \hline
5.0    & $3.97\cdot 10^{-9}$  & $1.65\cdot 10^{5}$   \\ \hline
10.0   & $8.59\cdot 10^{-10}$ & $1.42\cdot 10^{5}$   \\ \hline

\end{tabular}
\caption{The upper limit on the continuum photon flux due to 
near-solar neutralino annihilations and corresponding upper limit on the 
$\sigma_{p\chi}\rho_{0}b_{\gamma}^{c}$.}
\label{table:results:limits2}
\end{table}

Because the shape of the solar shadow is not known, not to claim a false
signal the null hypothesis is formulated as cosmic-ray background is
uniform and there is no $\gamma$-ray emission from the solar region. The
mean value of the statistic $U$ (see
equation~(\ref{eq:post_processing:Udef})) is equal to zero under this
hypothesis. Based on the results of the measurement (see
table~\ref{table:results:Nevents}) the formulated null hypothesis can
not be rejected with significance $2.867\cdot 10^{-7}$ (see
table~\ref{table:post_proc:significance}) and an upper limit on the
possible $\gamma$-ray flux from the solar region should be obtained.

The deficit of events from the direction of the Sun can not be greater
than that produced by the Moon because of Sun/Moon similarities. To be
conservative in setting the upper limit, the strongest event deficit
produced by the Moon in $5^{\circ}$ radius from its position should be
used as a correction for possible presence of the shadow of the Sun. The
smallest value of the statistic $U$ observed in the sky map centered on
the geometrical position of the Moon is $-3.3$ (see
figure~\ref{fig:results:maps_and_exposure}). The exposure on the Sun is
about $2.75$ greater than that on the Moon, leading to the estimated
maximal deficit in the Sun's direction computed in terms of $U$ as:
$U_{sun}^{max}=-3.3\sqrt{2.75}=-5.5$.

Thus, the upper limit on the photon flux from the region of the Sun
corresponding to the significance $2.867\cdot 10^{-7}$ with error of the
second kind $2.275\cdot 10^{-2}$ (see
table~\ref{table:post_proc:significance}) is computed based on the value
of the statistic $U$:
\[ u_{1}=5.0+5.5+2.0=12.5 \]
leading to the upper limit on the mean number of the gamma counts:
\[ N < N_{u}=u_{1}\sqrt{N_{b}+\alpha N_{s}}=4791 \]

The differential photon flux due to neutralino annihilations taken 
from~\cite{Bergstrom3} has the form of:

\[ \frac{dF(E)}{dE}=F_{\delta}\delta(E-m_{\chi})+
                 \frac{F_{c}(\frac{E}{m_{\chi}}>0.01)}{m_{\chi}}\cdot
      \frac{\Big(\frac{E}{m_{\chi}}\Big)^{-3/2}e^{-7.8E/m_{\chi}}}%
                                {\int_{0.01}^{1}x^{-3/2}e^{-7.8x}dx} \]
where $F_{\delta}$ is the integral flux due to a $\delta$-function-like
photon annihilation channel and $F_{c}(\frac{E}{m_{\chi}}>0.01)$ is the 
integral flux for $\frac{E}{m_{\chi}}>0.01$ due to continuum photon 
spectrum annihilation channel of neutralinos with mass 
$m_{\chi}$.\footnote{$\int x^{-3/2}e^{-ax}dx=-\frac{2e^{-ax}}{\sqrt{x}}-
2\sqrt{a\pi}Erf(\sqrt{ax})$}

Computing the number of events to be observed by the detector using the
formula~(\ref{eq:post_processing:N_Omega}) for the given spectrum and
following the procedure for setting an upper limit (from
section~\ref{sec:post_processing:flux_measure}), it is possible to set a
constraint on the integral fluxes in the form:

\begin{equation}
   F_{\delta}\cdot\Delta + F_{c}\cdot\Sigma < 4791
\label{eq:results:flux_constraints}
\end{equation}

$F_{\delta}$ and $F_{c}$ are in the units of $cm^{-2}s^{-1}$ and the
coefficients $\Delta$ and $\Sigma$ are given in the
table~\ref{table:results:flux_limit_coef} for different neutralino
masses. The figure~\ref{fig:results:flux_limit_1} shows the region of 
parameter space $(F_{c},F_{\delta})$ restricted by the upper limit.

The upper limit on the monochromatic photon flux due to neutralino
annihilations and corresponding limit on
$\sigma_{p\chi}\rho_{0}b_{\gamma}^{\delta}$ (see
equation~(\ref{eq:simulations:predicted_flux})) are presented in
table~\ref{table:results:limits}. The upper limit on the continuous
photon flux with energies above $0.01m_{\chi}$ due to neutralino
annihilations and corresponding limit on
$\sigma_{p\chi}\rho_{0}b_{\gamma}^{c}$ (see
equation~(\ref{eq:simulations:predicted_flux})) are presented in
table~\ref{table:results:limits2}. The 
figure~\ref{fig:results:cross_limit} depicts the combined limit on
$\sigma_{p\chi}\rho_{0}b_{\gamma}^{c}$ and 
$\sigma_{p\chi}\rho_{0}b_{\gamma}^{\delta}$.

\begin{figure}
\centering
\includegraphics[width=4.2in]{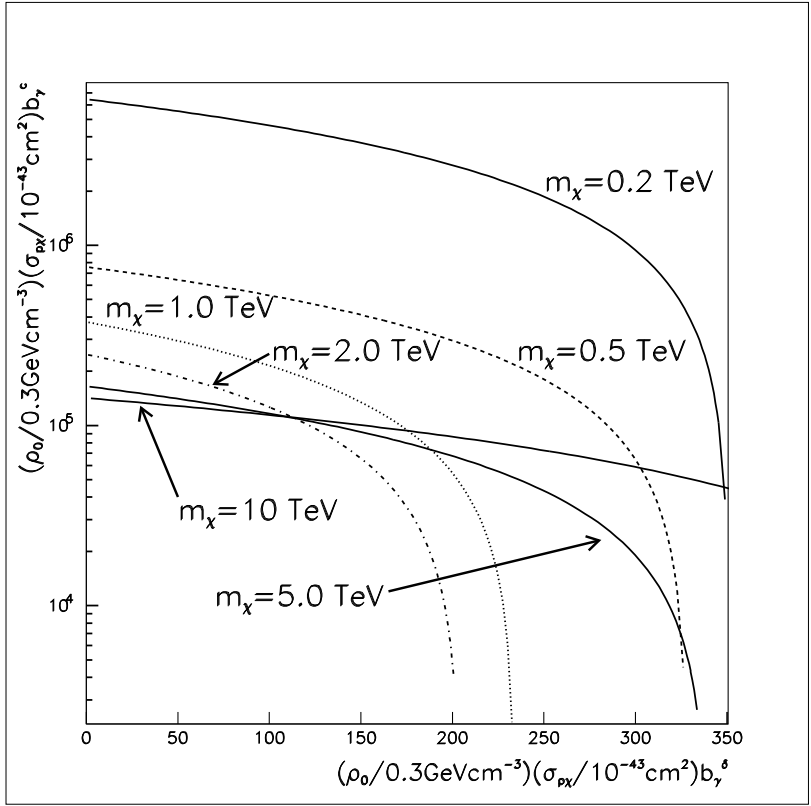}
\caption{The values of 
        $\Big(\rho_{0}\sigma_{p\chi}b_{\gamma}^{\delta}$,
         $\rho_{0}\sigma_{p\chi}b_{\gamma}^{c}\Big)$ below the lines are
         allowed based the constructed upper limit for corresponding
         neutralino masses.}
\label{fig:results:cross_limit}
\end{figure}


\chapter{Conclusion}
\label{chapter:conclusion}

\epigraph{George  Orwell ``1984''}{%
The landscape that he was looking at recurred so often in his dreams that
he was never fully certain  whether or not he had seen it in the real
world.}

The Milagro data set collected during 2000-2001 has been analyzed and
searched for the evidence of a steady near-1TeV $\gamma$-ray flux from
near-solar neutralino annihilations. As a result of the analysis, it was
argued that no evidence for the gamma-ray signal due to such a process has
been found. The upper limit on the gamma-ray flux due to such a process
with significance $2.867\cdot 10^{-7}$ and the power $(1-2.275\cdot
10^{-3})$ has been set. Even in the absence of a clear signal the
constructed upper limit may constrain the values of free parameters of
supersymmetric models.

The interpretation of the constructed limit on the gamma-ray flux is
highly model dependent. It is based, for instance, on assumptions
regarding the shape of the velocity distribution of the dark matter in the
halo and the assumed structure of the Solar System. The current work
presents a calculation of the neutralino annihilation rate density as a
function of distance from the Sun and the neutralino capture rate onto
near-solar bound orbits. It is shown that in the considered model only
about 50\% to 60\% of annihilations happen inside the Sun. The calculation
allowed translating the established limit on the gamma-ray flux from
neutralino annihilations to the limit on the product of the
neutralino-proton scattering crossection $\sigma_{p\chi}$, the integrated
photon yield per neutralino in neutralino-neutralino annihilation
$b_{\gamma}$ and the local galactic halo dark matter density $\rho_{0}$.

To the knowledge of the author the current work presents a first attempt
to set a constraint on the parameters of supersymmetric models by
observing high energy gamma rays from the region of the Sun. Continuous
improvements in reconstruction algorithms, detector modifications and
longer observation times will led to a better upper limit.  One of the
factors which lead to a deterioration of the constructed upper limit is
the inability to compensate for presence of the Solar cosmic-ray shadow
due to the intricate structure of the Solar magnetic fields. Once the
cosmic-ray shadow of the Sun is understood quantitatively, it may be
possible to improve upon the limit.

\appendix                           
\chapter{Poisson distribution}
\label{app_chapter_poisson_distr}

\section{Definition}

The Poisson distribution arises in many problems as the distribution of
the number of occurrences of some event over an interval of time or
region of space. The distribution assumes that an event can occur at
random at any time or point in space and that the probability of event
occurrence does not depend on any other event. The Poisson distribution
is defined as:

\begin{equation}
        p(k;\lambda)=\frac{\lambda^{k}}{k!}e^{-\lambda}
\label{eq:poisson:poisson}
\end{equation}

\[ \sum_{k=0}^{\infty}p(k;\lambda) =
               e^{-\lambda}\sum_{k=0}^{\infty}\frac{\lambda^{k}}{k!}=
               e^{-\lambda}e^{\lambda} =1 \]
\[ <k>=\sum_{k=0}^{\infty}kp(k;\lambda)=
               e^{-\lambda}\sum_{k=0}^{\infty}\frac{k\lambda^{k}}{k!}=
     e^{-\lambda}\lambda\sum_{k=1}^{\infty}\frac{\lambda^{k-1}}{(k-1)!}=
               \lambda \]
\[ D(k)=<k^{2}>-<k>^{2}= e^{-\lambda}\sum_{k=0}^{\infty}
                           \frac{k^{2}\lambda^{k}}{k!} -\lambda^{2}= \]
\[ =e^{-\lambda}\sum_{k=1}^{\infty}\left[
        \frac{(k-1)\lambda^{k}}{(k-1)!}+\frac{\lambda^{k}}{(k-1)!}
        \right]-\lambda^{2}=e^{-\lambda}\left[
     \lambda^{2}e^{\lambda}+\lambda e^{\lambda}\right]-\lambda^{2}=\lambda \]

\section{Gaussian Limit of Poisson Distribution}

Substituting the $k!$ in the Poisson distribution
(\ref{eq:poisson:poisson}) by the approximate expression using the
Stirling formula:
\[ n! = \sqrt{2\pi n}\left(\frac{n}{e}\right)^{n}e^{\theta(n)} \]
the Poisson distribution becomes:
\[ p(k;\lambda) \approx \sqrt{\frac{1}{2\pi\,k}}
               \left(\frac{e\lambda}{k}\right)^{k}e^{-\lambda} \]
or 
\[ \ln\left[p(k;\lambda)\sqrt{2\pi\,k}\right] \approx
                   k\ln\left[\frac{e\lambda}{k}\right] - \lambda =
                   k\left[1-\ln\frac{k}{\lambda}\right] - \lambda \]

Let $k=\lambda+\delta$ than 
\[ \ln\left[p(k;\lambda)\sqrt{2\pi\,k}\right] \approx
(\lambda +\delta)\left[1-\ln\frac{\lambda +\delta}{\lambda}\right] - \lambda = \]
\[ =(\lambda +\delta)\left[1-
\sum_{n=1}^{\infty}(-1)^{n+1}\frac{1}{n}\left(\frac{\delta}{\lambda}\right)^{n}
\right] -\lambda = -\frac{1}{2}\frac{\delta^{2}}{\lambda} +
      \frac{1}{6}\frac{\delta^{3}}{\lambda^{2}} +
                         {\cal O}(\frac{\delta^{3}}{\lambda^{2}}) \]
Thus:
\[ p(k;\lambda) \approx \sqrt{\frac{1}{2\pi\,(\lambda+\delta)}}\,
                e^{-\frac{\delta^{2}}{2\lambda}}\cdot
                e^{ \frac{\delta^{3}}{6\lambda^{2}} + 
                              {\cal O}(\frac{\delta^{3}}{\lambda^{2}})} \]

For sufficiently small $\delta$ such that
$\frac{\delta}{\lambda}<<1$ and
$\left|e^{\frac{\delta^{3}}{6\lambda^{2}}}-1\right|<<1$
the Poisson distribution approaches the Gaussian distribution with mean 
and dispersion equal to $\lambda$:

\[ \left|e^{\frac{\delta^{3}}{6\lambda^{2}}}-1\right|<<1\ \ \  \Rightarrow \ \ \
 \left|\frac{\delta^{3}}{6\lambda^{2}}\right| << 1 \ \ \ \Rightarrow\ \ \ 
   \frac{\delta}{\lambda} <<\sqrt[3]{\frac{6}{\lambda}} \leq 1,\
   \forall\ \lambda\geq 6 \]

\begin{equation}
   p(k;\lambda)= \frac{\lambda^{k}}{k!}e^{-\lambda}\rightarrow
    \sqrt{\frac{1}{2\pi\lambda}}\,e^{-\frac{(k-\lambda)^{2}}{2\lambda}}
   \ \ :\ \  \left|\frac{(k-\lambda)^{3}}{6\lambda^{2}}\right| << 1,\
                                              \forall\ \lambda\geq 6
\label{eq:poisson:possion2gauss}
\end{equation}


\chapter{Calibration}
\label{chapter:calibration}

\epigraph{George  Orwell, ``1984''}{%
For, after all, how do we know that two and two make four?
}

The desire to reconstruct the position of events on the Celestial Sphere
with systematic errors much less than $1^{\circ}$ dictates that the
times registered by PMTs have to have resolution about $1\ (ns)$ and the
locations of the photo-tubes be known to about $10\ (cm)$ accuracy in
horizontal and about $3\ (cm)$ in vertical directions. To meet the
latter requirement photographic and theodolite surveys of the pond were
performed. At the end of the construction period, when the pond was
filled with water, an ``as-built'' measurement of the PMT elevation was
done.

Even though great care has taken to construct all PMT channels as
uniformly as possible, remaining systematic differences in PMT channels
should be studied and removed. This includes synchronization of all TDC
modules (find TDC time offsets and conversion factors) and compensation
for the PMT-pulse amplitude dependence of TDC measurements (known as the
slewing correction).

To correctly reconstruct the shower front, shower size and, ultimately,
to estimate the energy of the primary particle, the relative
``pulse-height'' to photo-electron~(PE) conversion must be determined to
interpret all PMT amplitude measurements in a common unit for each
event. This is then translated to an absolute scale of the energy
deposited in the water.

Full description of the calibration system with its components, 
operation, data analysis and other comments is available 
in~\cite{ZORO_manual} and references cited therein.

\section{Calibration system setup.}

\begin{figure}
\centering
\includegraphics[width=4.0in]{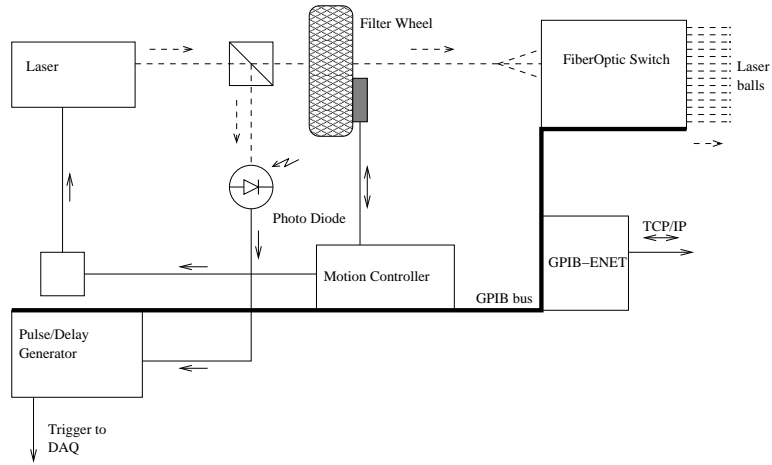}
\caption{Calibration system setup}
\label{fig:int_setup}
\end{figure}

The Milagro calibration system has been designed to reflect all the above
goals and is based on the laser -- fiber-optic -- diffusing ball concept
used in other water-Cherenkov detectors (See, for
instance,~\cite{Becker-Szendy}).  A computer operated motion controller
(Newport MM3000) drives a neutral density filter wheel to attenuate a
pulsed nitrogen dye laser (Laser Photonics LN120C) beam. The beam is
directed to one of the thirty diffusing laser balls through the
fiber-optic switch (DiCon MC606) as shown on Fig~\ref{fig:int_setup}.  
Part of the laser beam is sent to a photo-diode. When triggered by the
photo-diode, the pulse-delay generator (Stanford Research DG535) sends a
trigger pulse to the data acquisition system. A laser fire command is
issued by the motion controller, providing full automation of the
calibration process. The balls are floating in the pond so that each PMT
can register signals from more than one light source. Such a redundant
setup allows calibration of the PMTs and the electronics.

Calibration data was collected by stepping the filter wheel through a full
circle in $10$ degrees increments for each laser ball. On each laser ball
- filter wheel setting about $2000$ laser triggers at $20\ (Hz)$ and about
$1600$ ``random'' triggers (with no light input) at the rate of $400\
(Hz)$ were sent to the data acquisition system. Raw data from all PMT
channels was recored and analyzed. Only 2- and 4-edge events with correct
polarity were selected to ensure proper ToT determination. For purposes of
occupancy measurements all data was used without any edge selection.

\section{Timing calibration.}

The importance of accurate time readings from the PMT channels can not 
be over stressed as the quality of event reconstruction depends on it. 
The issues which need to be addressed are described in this section.

\subsection{TDC Conversion Factor.}
\label{chapter:calibration:tdc_conversion}

The time of PMT pulse threshold crossings is read by LeCroy 1887 FASTBUS
TDC modules. These digital devices measure time in the units of
``counts'' and, according to specifications, each count corresponds to
0.5 nanoseconds. Introduction of known variable delays in the
calibration-DAQ trigger logic\footnote{Special TDC calibration data
should be taken for this procedure.} allowed observation of common time
shifts in all PMT channels to verify the TDC conversion factors at
$2.0000\pm 0.0003$ counts per nanosecond. This assured that all TDC
modules operate on a common scale and allowed for a simple
interpretation of TDC measurements.

\subsection{Electronic slewing correction.}

\begin{figure}
\centering
\includegraphics[width=0.6\textwidth]{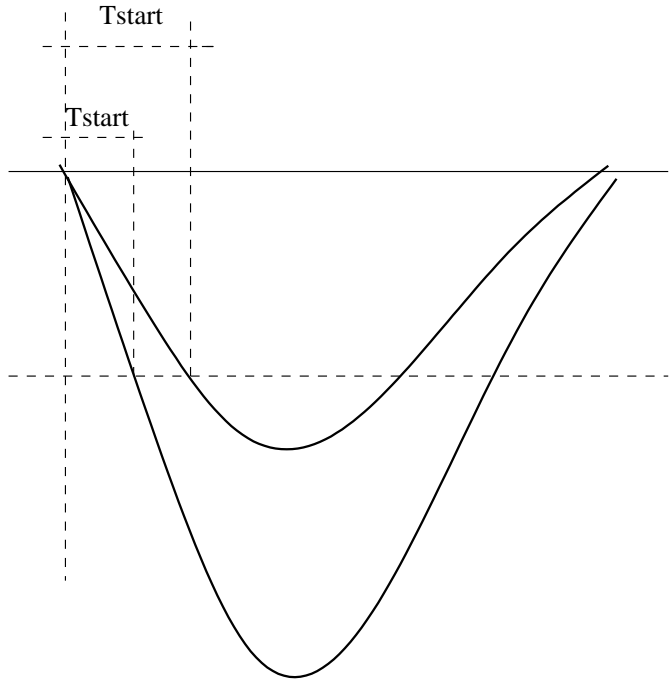}
\caption{Illustration of the electronic slewing. Stronger pulses cross 
the discriminator threshold earlier than the weaker ones.}
\label{fig:calibration:tstart}
\end{figure}

\begin{figure}
\centering
\includegraphics[width=0.4\textwidth]{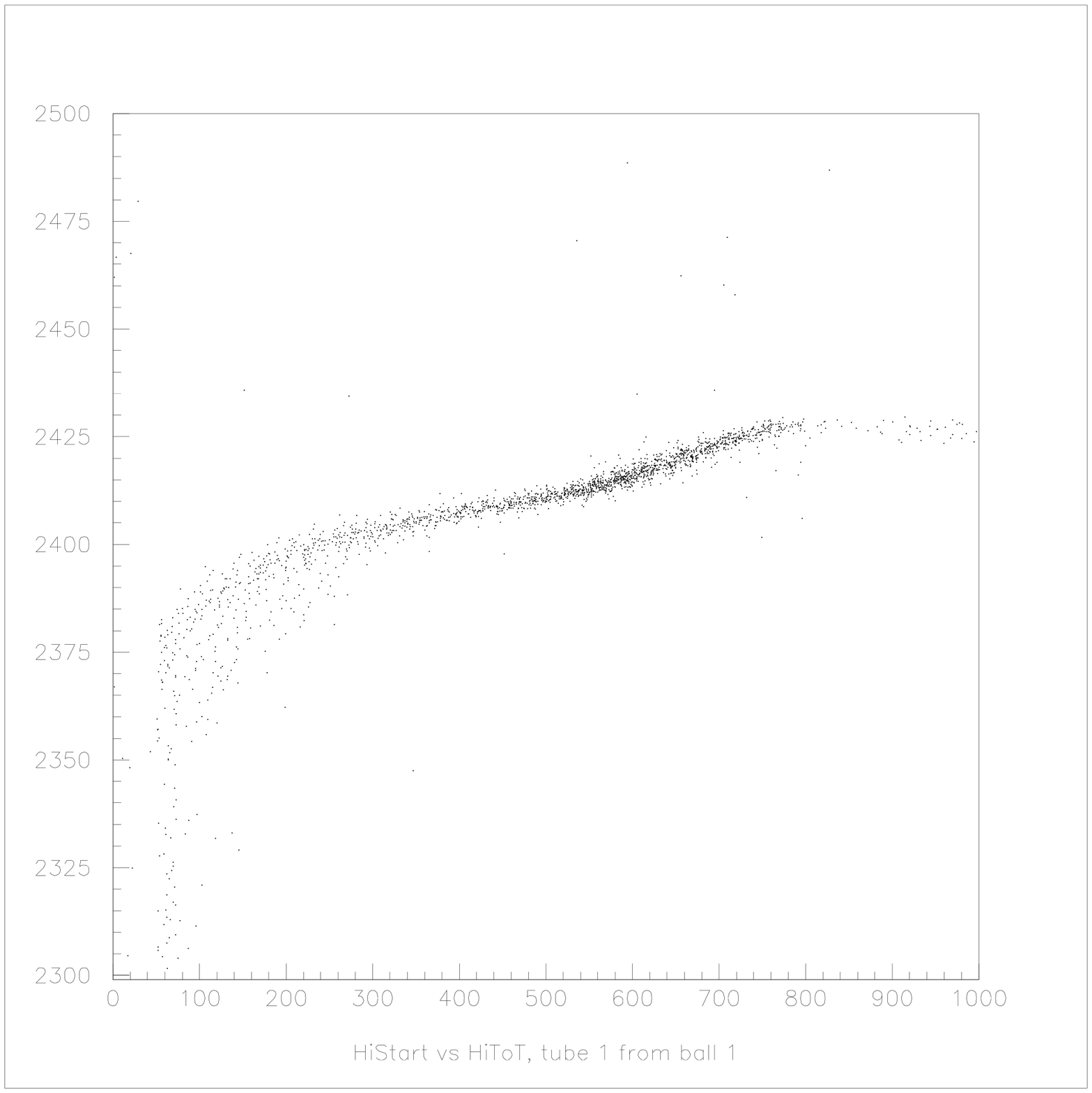} 
\hspace{0.02\textwidth}
\includegraphics[width=0.4\textwidth]{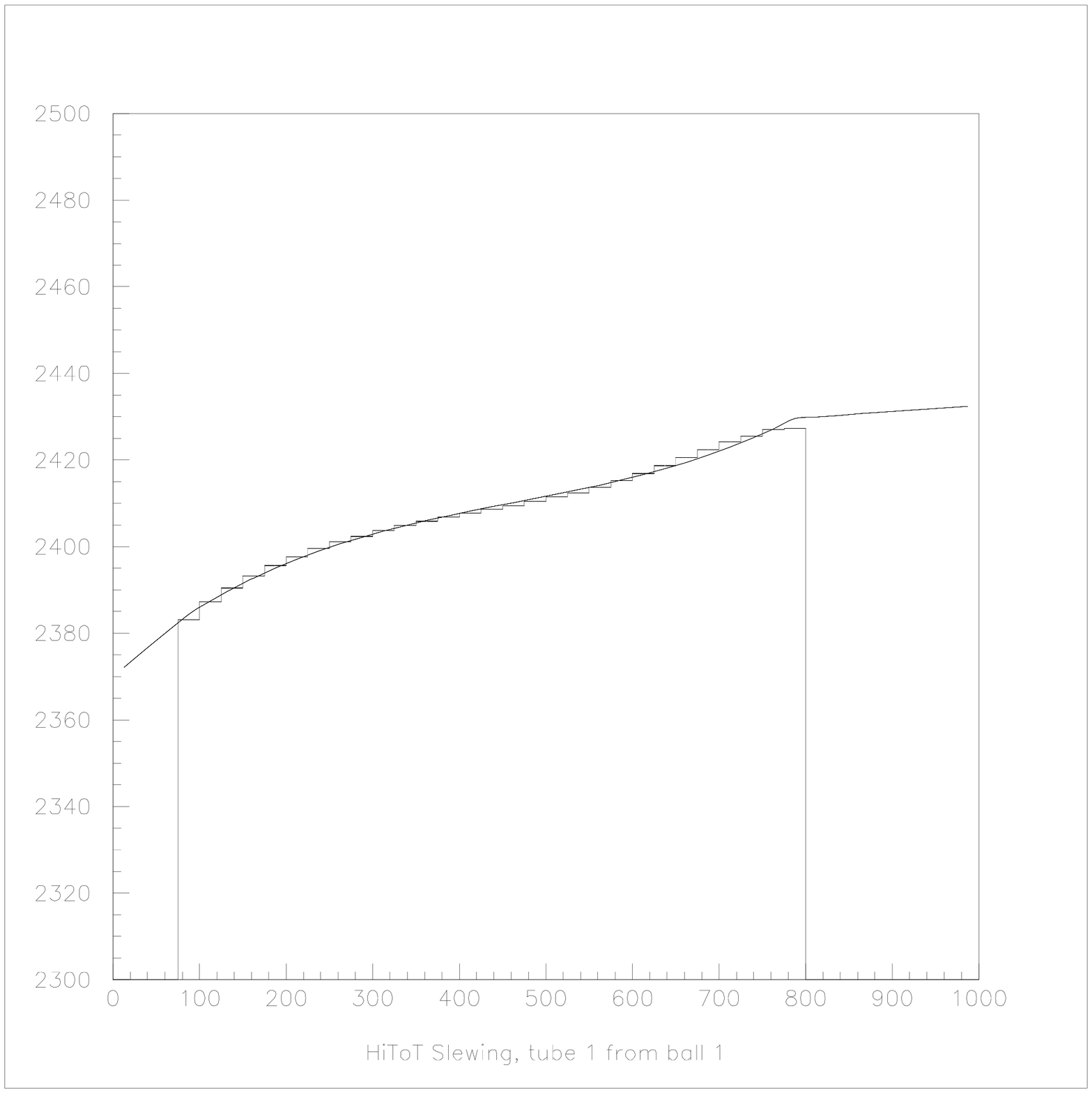}
\caption{Plots show $T_{start}$ vs $ToT$ data obtained for calibration 
(left) and the polynomial fit to the data (right). The units of both 
axes are TDC counts.}
\label{fig:slewing}
\end{figure}

Time response of a PMT channel depends on the input light intensity and
is called electronic slewing. Indeed, a weaker pulse will cross the
discriminator threshold later than a strong one arriving at the same
time (see figure~\ref{fig:calibration:tstart}). Based on the ToT PMT
pulse model described in~\ref{detector:tot_model}, when the size of the
pulse is described by the time over threshold, a slewing correction can
be devised by studying the PMT pulse arrival time ($T_{start}$) as a
function of $ToT$ for different filter wheel transparency settings. The
time of light pulse emission is supplied by the photo-diode and is
believed to be free of slewing effects since the light level incident on
the diode is constant. The slewing correcting curve is found by fitting
the obtained calibration data with a polynomial (see
figure~\ref{fig:slewing}).

Note, however, that obtained $T_{start}$ includes time propagation of 
the laser pulse in the detector medium and optical fibers. These may 
vary from PMT to PMT and should be taken into account to produce true 
$T_{start}$ versus $ToT$ dependence.

The procedure described above should be performed independently for both 
threshold levels HiToT and LoToT .

\subsection{Speed of light in water, fiber delay.}

In order to correct for propagation time of light in the detector,
coordinates of PMTs and laser balls as well as the speed of light in water
must be known. PMT and laser ball coordinates are known from the survey.
Only a typical index of refraction of water is found in reference tables
and fiber optic delays may vary from laser ball to laser ball, thus, all
these parameters need to be measured with the calibration system.

Interestingly enough, this problem can be easily
solved~\cite{ZORO_peds2} if several PMTs can register light from two
laser balls (cross-calibration). The times measured from two different
laser balls $T_{start}^{1}$ and $T_{start}^{2}$ on the same PMT after
slewing correction should be identical. The non-zero difference
between the times $T_{start}^{1}$ and $T_{start}^{2}$ can be attributed
to an error in the water propagation time $\Delta_{propagation}$ and/or
difference in the laser balls' fiber optic delay $\Delta_{fiber}$. If 
$\tau$ is defined as:

\[ \tau =T_{start}^{1}-T_{start}^{2}-\Delta_{fiber}-\Delta_{propagation} \]
it will be zero in absence of errors in water propagation time and fiber 
delays.

The distribution over all PMTS of observed $\tau$ from a given laser
ball pair can be constructed and studied. Since the $\Delta_{fiber}$ is
constant for the given laser ball pair and $\Delta_{propagation}$
depends on the relative PMT ball positions, the use of correct speed of
light will yield the minimal width of the $\tau$ distribution. After
that, the mean of the distribution can be interpreted as the fiber optic
difference $\Delta_{fiber}$. Note that a PMT in close proximity to any
one of the laser balls will have enhanced sensitivity to speed of light
variation, while PMTs located half way between the laser balls will have
enhanced sensitivity to the fiber optic difference.

Needless to say that the procedure described in this subsection can be
used to test the self consistency of the timing calibration. The use of
wrong coordinates of laser balls or PMTs will reveal itself as mismatch
in the slewing curves ($\tau$). In fact, using this procedure, it was
discovered that coordinates of several PMTs were interchanged. An
extension of the procedure described here is discussed in
~\cite{ZORO_peds1} where coordinates of the laser balls themselves are
allowed to vary and can be restored.

Correcting the slewing curves by corresponding fiber optic delays and 
water propagation times yields the final calibration curves of 
$T_{start}$ as a function of $ToT$.

\section{Photo-Electron calibration.}

The main purpose of the photo-electron calibration is to find a
relationship between the observed ToT and the number of photo-electrons
emitted inside the photo-tube. The calibration procedure is based on a
well known occupancy method described in the literature (see for
instance~\cite{Becker-Szendy} or~\cite{ISABEL_occ}) and proceeds in two
general steps. First, the ToT-PE conversion is established for low input
light levels using the occupancy method and then a different procedure
is applied to calibrate PMTs at high light levels given the
characteristics of the calibration filter wheel. The whole procedure
relies on the assumptions that the number of photo-electrons produced in
a PMT is proportional to the light intensity at the PMT's photocathode,
that the input light level into the calibration system is constant, and
that all light level modulation is due to a controlled change in the
transmittance of the filter wheel only.

The calibration data required for photo-electron calibration is the same
as for the timing calibration which is obtained with laser light passing
through a filter at different transparency settings. While it is
difficult to establish the light level stability of the laser output, it
was found that if probability of the laser to produce no light when it
is triggered is less than 2.5\%, the PE calibration results are self
consistent.

This section presents the main ideas of the PE calibration followed by a
description of innovations in the method implementation. Full
description of the occupancy method applied for Milagro calibration is
presented in~\cite{ZORO_occ}, \cite{ISABEL_occ} and~\cite{ZORO_manual}.

\subsection{Low light level calibration and the Occupancy method.}

The occupancy method is based on the assumption that the number of
photo-electrons produced at a PMT's photocathode obeys a Poisson
distribution: $P(n;\lambda)=\frac{\lambda^{n}}{n!}e^{-\lambda}$. Here
$\lambda$ is the mean number of PEs produced at the photocathode.  This
is justified by the assumption that the emission of a photo-electron is
not related to emission of a different one which is true if the
photo-tube did not reach its saturation.

The probability $\eta$ that a photo-tube registered the light pulse 
(which means at least one photo-electron was emitted from the 
photo-cathode) is called the occupancy and is given by:

\[ \eta =P(n>0;\lambda)=1-P(n=0;\lambda)=1-e^{-\lambda} \ \ \ \
   \Rightarrow\ \ \ \ \lambda = -\ln(1-\eta) \]

Based on its definition, occupancy $\eta$ can be easily measured if a
PMT is illuminated many times by the light pulses of identical
intensity:

\[ \eta=\frac{\mbox{number of observed pulses}}%
             {\mbox{number of sent pulses}}  \]

As the intensity of input light is varied, it is possible to find a
relationship between the number of PEs and the observed ToT directly.
However, for high light levels when $\lambda >2$, it is not possible to
measure $\lambda$ reliably based on $\eta$, since the error of the
measurement on $\lambda$ increases exponentially with error on $\eta$:

\[ \Delta\lambda =\frac{1}{1-\eta}\Delta\eta =e^{\lambda}\Delta\eta \]

\subsection{High light level calibration.}

The high light level calibration is based on the assumption that there
is no saturation of the PMT channel and the mean number of
photo-electrons produced is proportional to the input light level
intensity. If the transmittance of the filter wheel $T$ is known, then:

\[ \lambda =a\cdot T \]
where $a$ is some parameter which is constant, but different for
different laser ball-PMT pairs. It can be found from this equation at
low light levels because $T$ is known and $\lambda$ can be measured with
the occupancy method. Thus, given the transmittance properties of the
filter wheel, the ToT to PE conversion can be found at high light levels
with a linear error on $\lambda$.

For some PMT laser ball pairs, even the lowest light level possible was
relatively high for the occupancy method to be used. For these PMTs, a
farther away laser ball was used to establish the ToT-to-PE conversion
for lower light levels. The obtained conversion curve was then extended 
by the data from the nearest laser ball to the highest possible light 
level.

If, contrary to the assumption, saturation of the PMT channel is
present, the number of PEs can not be established using this method. 
However, since the goal of calibration is to study the PMT response to
different light levels, this is not a problem and the method described
here allows one to infer the light intensity at the PMT cathode from the
observed ToT as the effective number of PEs which should have been
emitted from the photocathode provided the PMT response were linear.

\subsection{Filter calibration.}

As was mentioned earlier, transmittance properties of the filter wheel
are important for high light level PE calibration. The transmittance
properties can be obtained from the manufacturer of the filter or can be
measured in laboratory. A method to calibrate the filter wheel using the
same calibration data as for slewing and PE calibration was proposed and
used. This method has the advantage that the filter is calibrated as it
is being used in detector calibration. The method employs the occupancy
method with additional supposition that for any two sufficiently close
transmittances of the filter $T_{1}$ and $T_{2}$ there exist a PMT for
which the occupancy method can be used at both light intensities. Then,
given two corresponding measurements of mean number of photo-electrons:

\[ \frac{T_{2}}{T_{1}}=\frac{\lambda_{2}}{\lambda_{1}} \]

The transmittance $T_{3}$ can be related to $T_{2}$ in the analogues 
manner and so forth, leading to the restoration of the levels of 
transmittance for all filter settings. Because the absolute calibration 
of the filter is not required, it is always possible to set $T_{0}=1$.

\subsection{Dynamic Noise Suppression.}

\begin{figure}
\centering
\includegraphics[width=0.7\textwidth]{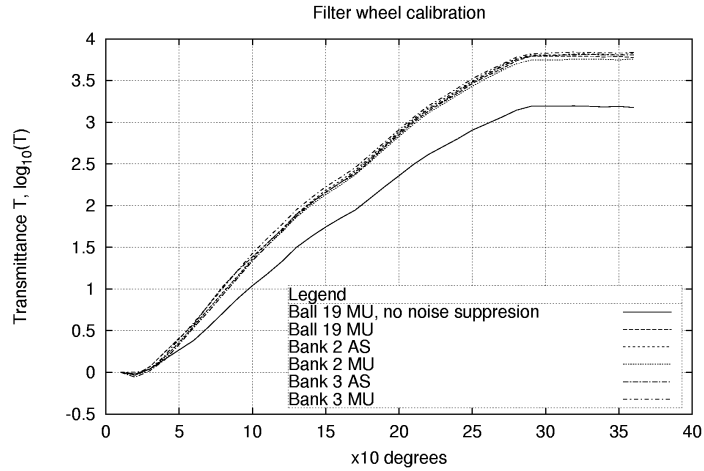}
\caption{Filter wheel calibration with and without noise suppression.
(Bank 2,3 represent laser balls 11-20 and 21-30 respectively, while AS 
and MU represent PMTs from ``top'' and ``bottom'' layers used for filter 
calibration.)}
\label{fig:noise_suppression}
\end{figure}

The PE and filter calibration procedures rely on correct knowledge of PMT
occupancy. PMT thermo-electron emission, Cherenkov light from the shower
particles and other sources can cause a signal on the PMT output not
related to the input calibration light. This noise will increase the
measured PMT occupancy and damage the calibration accuracy.

Dynamic noise suppression is a technique which allows correction of the 
apparent occupancy on a tube by tube basis and is based on the 
assumption that the arrival of the laser light is not correlated with 
the noise pulses. Then, the probability to observe anything (apparent 
occupancy $P(any)$) is:

\[ P(any) = P(laser) + P(noise) - P(laser)\cdot P(noise) \]

\[ \eta = P(laser) = \frac{P(any) - P(noise)}{1 - P(noise)} \]
where $P(laser)$ is the probability to observe laser light (true
occupancy $\eta$) and $P(noise)$ is probability to observe noise pulse.

$P(any)$ can be measured by sending laser pulses to a PMT, as before, 
while $P(noise)$ can be estimated by sending uncorrelated triggers to 
data acquisition system without any light input which can be interlaced 
with the laser data taking.

The dynamic noise suppression is an important step in calibration
process and was used for the ToT-to-PE and filter calibrations. The
effect of the noise suppression is shown on
figure~\ref{fig:noise_suppression} where the filter wheel calibration 
curves are presented with and without the noise suppression.

\subsection{Statistical error of the occupancy method.}

\begin{table}
\centering
\begin{tabular}{||c||c||c||}                  \hline
$\eta$  & $\delta(\eta)$  &   pass/fail       \\  \hline 
 0.1    &   0.009         &     fail          \\  \hline
 0.2    &   0.009         &     pass          \\  \hline
 0.3    &   0.008         &     pass          \\  \hline
 ...    &    ...          &     ...           \\  \hline
 0.8    &   0.004         &     pass          \\  \hline
 0.9    &   0.002         &     fail          \\  \hline
\end{tabular}
\caption{Occupancy accuracy test results to satisfy the error 
$q=\frac{\Delta\lambda}{\lambda}=0.01$ on measured $\lambda$ with 
$n=2000$ laser shots.}
\label{table:allowed_occ}
\end{table}

In order to address the question of the accuracy of the occupancy
method, suppose that $n$ shots of the laser beam were sent, out of which
$m$ were detected by the PMT. Occupancy $\eta$ is estimated as:

\[ \eta=\frac{m}{n} \]

The question of the accuracy of this estimation is that of the
confidence interval. Since the probability of a PMT registering a signal
in a shot is equal to $\eta$, then the number of detected pulses $m$ is
distributed according to binomial distribution:

\[ P_{\eta}(m,n)=C^{m}_{n}\eta^{m}(1-\eta)^{n-m},  \ \ \ \ 
                                        C^{m}_{n}=\frac{n!}{m!(n-m)!} \]

The upper bound of the confidence interval $\eta_{upper}$ corresponding
to the significance $\alpha$ is defined so that probability of detecting
$k \leq m$ pulses is less than $(1-\alpha)$:

\begin{equation}\label{calib:eq_upper_bound}
   P_{\eta_{upper}}(k \leq m,n)=
     \sum_{k=0}^{m} C^{k}_{n}\eta_{upper}^{k}(1-\eta_{upper})^{n-k}
                                                      \leq 1-\alpha
\end{equation}

Correspondingly, lower bound $\eta_{lower}$ is such that

\[ P_{\eta_{lower}}(k \geq m,n)=
     \sum_{k=m}^{n} C^{k}_{n}\eta_{lower}^{k}(1-\eta_{lower})^{n-k}
                                                      \leq 1-\alpha \]
or
\begin{equation}\label{calib:eq_lower_bound}
   \sum_{k=0}^{m-1} C^{k}_{n}\eta_{lower}^{k}(1-\eta_{lower})^{n-k} 
                                                        \geq \alpha
\end{equation}

If the required relative error on the value of occupancy is
$\delta(\eta)=\frac{\Delta \eta}{\eta}$ then $\eta_{upper}$ should be no
greater than $(1+\delta(\eta))\eta$ and $\eta_{lower}$ should be no less
than $(1-\delta(\eta))\eta$.  Thus, for given number of laser shots and
number of registered pulses it is possible to check if the required
accuracy is met. The requirements on the accuracy $\delta(\eta)$ are
governed by the requirement on the relative error
$q=\frac{\Delta\lambda}{\lambda}$ on PE ($\lambda$) determination:

\[ \lambda=-\ln(1-\eta)  \ \ \ \Rightarrow \ \ \ 
                        \Delta\lambda = \frac{1}{1-\eta}\Delta \eta \]
\[  q=\frac{\Delta\lambda}{\lambda}=
                           -\frac{\Delta \eta}{(1-\eta)\ln(1-\eta)}
 \ \ \ \Rightarrow \ \ \ \delta(\eta)=-q \frac{1-\eta}{\eta}\ln(1-\eta)  \]

Now, the task is to estimate the allowed range of occupancies $\eta$ 
with error not exceeding the specified value $q$ provided that $n$ 
laser pulses were sent.

For $\eta<0.5$ the equation~\ref{calib:eq_lower_bound} will be 
automatically satisfied if:

\[ \sum_{k=0}^{\eta*n} C^{k}_{n}\eta_{upper}^{k}(1-\eta_{upper})^{n-k}
               \leq 1-\alpha, \;\;\; \eta_{upper}=(1+\delta(\eta))\eta \]

and for $\eta>0.5$ the~\ref{calib:eq_upper_bound} is satisfied if:
\[ \sum_{k=0}^{\eta*n-1} C^{k}_{n}\eta_{lower}^{k}(1-\eta_{lower})^{n-k} 
               \geq \alpha,   \;\;\; \eta_{lower}=(1-\delta(\eta))\eta \]

Therefore, for relative error $q=0.01$, confidence $\alpha=95\%$
and $n=2000$ the allowed range of occupancies to be used is
$0.2<\eta<0.8$ (see table~\ref{table:allowed_occ}). For the phototube
noise measurements ($\eta \sim 0.03$) the requirements are eased:
$q=0.1$. We also verify that using 60000 random ``shots'' is just enough
to reach the goal. (The requirement of $q=0.01$ is met at 66\%
confidence level only.)


\subsection{Threshold effect on the occupancy measurement.}

The problem addressed here is that of a finite threshold 
which a PMT pulse should cross in order to be recorded by the
electronics. The presence of the finite threshold leads to an under
estimation of the occupancy and if the final effect on the number of PEs 
is bigger than $q$ a correction should be made.

The assumption of PMT operation is that electron multiplication in each
stage is a statistical process with some average gain $g$. Then, the
number of electrons $k$ on the output of the first amplification stage
obeys Poisson distribution with average $gw$ where $w$ is the number of
photo-electrons emitted from the PMT photocathode (This assumption is
similar to that of occupancy method and is based on proposition that
probability of emitting an ``amplified'' electron does not depend on
other electrons being emitted.) and is distributed according to:

\[ P(k;gw )=\frac{(gw)^{k}}{k!}e^{-gw}  \]

It is the number of electrons on the output of the first stage that
dominates the fluctuations on the output of the entire cascade and is
responsible for ability of the PMT pulse to cross the discriminator
threshold. The PMT electronic channel was constructed in such a way that
only $\rho g$ or more electrons after the first stage will produce
signal strong enough to be detected. Thus, the probability that PMT 
signal of strength $w$ PEs will not be detected is:

\[ \beta (w) =\sum_{k=0}^{k<\rho g}P(k;gw)=
         \sum_{k=0}^{k<\rho g}\frac{(gw)^{k}}{k!}e^{-gw} \]

and the occupancy given average number of PE $\lambda$ is decreased:

\[ \eta(\lambda)=
            1-P(0;\lambda)-\sum_{w=1}^{\infty}P(w;\lambda)\beta (w)=
                 1-e^{-\lambda}-\lambda e^{-\lambda}\beta(1)-
                   \frac{\lambda^{2}}{2}e^{-\lambda}\beta(2)-\cdots \]

For estimation purposes Milagro PMT is assumed to have uniform stage to
stage amplification with typical gain of the entire PMT's cascade of
$2\cdot10^7$ (see~\cite{milagritoNIM} for the measured PMT's gain at the
operated voltage). Since the PMT consists of 10 stages, the gain of a
single stage is $g=(2\cdot10^{7})^{\frac{1}{10}}=5.3$ and the threshold
level is set at $\rho =0.25$ (signals with more then 0.25 equivalent PEs
will cross the discriminator threshold and will be detected.) the
function $\beta(w)$ falls off rapidly and for given PMT parameters a one
PE input will be lost with probability $\beta(1)=3.1\cdot 10^{-2}$ while
2PE --- with $\beta(2)=2.9 \cdot 10^{-4}$ and can be neglected. Hence,
taking into account only the loss of 1 PE signals, the measured
occupancy is equal to:

\[  \eta\approx 1-e^{-\lambda}- \beta(1)\lambda e^{-\lambda} \]

The magnitude of the correction $ \beta(1)\lambda e^{-\lambda}$ has to
be compared with the required accuracy on $\eta$ which leads to the
direct comparison of $\beta(1)$ and $q$. 

Therefore, it is concluded that the systematic error on occupancy is
comparable to the statistical one and for the desired accuracy of PE
determination (few per cent) the threshold effect can be neglected. If
the probabilities $\beta(1)$ are known for each PMT the correction,
could be done with the following approximate formula:

\[ \lambda\approx -\frac{1}{1-\beta(1)}\ln(1-\eta) \]

Also note, the filter wheel calibration is quasi-immune from this
problem because it is based on the ratio of the $\lambda$'s for a given
PMT and the $\frac{1}{1-\beta(1)}$ factor cancels.

\section{Calibration Extrapolation.}

The maximum light level at which the calibration data was available
often was lower than could be observed in the shower data rendering
strong PMT pulses unusable. To cope with this problem, extrapolation was
used to infer the values of calibration parameters beyond the
calibrated range based on the known values and trends. Indeed, typically
it was required to extend HiToT calibrated range by about $100\ (ns)$ to
interpret shower data.\footnote{LoToT was never extrapolated as it is 
prudent to switch to the use of HiToT as soon as it becomes meaningful.}

\subsection{Slewing extrapolation.}

The shape of the slewing correction function depends on the discriminator
threshold level, amplification coefficients, gains of PMTs, wiring and so
on. Instead of trying to take all the unknown parameters into account and
putting together a physical model of the slewing, a statistical one was
built taking the following approach.

It is believed that all PMT channels (PMTs themselves and electronic
boards) were designed and manufactured to meet common characteristics. 
Therefore, the study of the channels' responses (calibration) can be
viewed as a multiple (about 700 times) measurement of a singe function:
$T_{start}$ vs $ToT$.  The fact that the curves obtained for different
channels are slightly different can be attributed to the ``manufacturing
imperfections'' and the channels differ only due to unavoidable
uncontrollable reasons such as spread of characteristics of electronic
components and/or actual slewing measurement errors. Thus, a slewing
curve for a PMT can be viewed as a particular realization of some random
function. All slewing curves together form one slewing function family
characterized by its mean dependence\footnote{Here, $t$ and $t'$ denote
the time over threshold.} $m(t)$ and correlation function $K(t,t')$, by 
analogy with the mean and dispersion for a random variable.

The characteristics $m(t)$ and $K(t,t')$ of the random function were
deduced from the observed high range slewing calibration data and, based
on the random function framework carried out to the first order of
canonical expansion, the value of the slewing correction $x(t)$ given
the last known calibrated value $x(t_{1})$ at time $t_{1}$ is:

\[ x(t)=m(t)+\frac{x(t_{1})-m(t_{1})}{K(t_{1},t_{1})}K(t,t_{1}) \]

with the root mean square error of:

\[ rms(t)=\sqrt{K(t,t)-\frac{(K(t,t_{1}))^{2}}{K(t_{1},t_{1})}} \]

Using this method slewing curves were extrapolated only to the point
where real data for at least 50 PMTs existed with estimated error on
extrapolation of the order of $0.7\ (ns)$.\footnote{The comparison of
extrapolated data from a calibration run with calibrated data obtained
independently~\cite{ZORO_SlewCompare} yielded the measured extrapolation
error of $0.55\ (ns)$ in good agreement with the expectations.} Beyond
that, linear extrapolation was used with the slope of
$0.0381\frac{T_{start}}{HiToT}$.

Details describing the extrapolation method used to extend slewing curves
for the HiToT calibration can be found in report~\cite{ZORO_SlewExtra}. A
brief review of the notion of random functions can be found in
memo~\cite{ZORO_Random}.

\subsection{PE extrapolation.}

Contrary to slewing calibration where the quality of the data increases
with the input light level, the quality of ToT-to-PE conversion degrades
due to exponential relation between ToT and PE and possible PMT
saturation. This lead to the conclusion that sophisticated random
function extrapolation is not justified and the extrapolation was
developed based on a simple physical argument that $\log PE\sim ToT$.
Thus, the PE vs ToT data was fit to a third order polynomial of the
form:

\[ \ln PE = a_{0}+a_{1}ToT+a_{2}ToT^{2}+a_{3}ToT^{3} \]
and the values of the polynomial were used as the extrapolation. 
Needless to say that beyond approximately $PE=100$ the error grows very 
fast and any algorithm relying on PE should treat the values beyond 
$PE=100$ as logical ``big'', ``Big'' and ``BIG''.

\section{Energy calibration.}

It is planned that absolute energy calibration measurements will be done
using through-going muons. The imaging capabilities of the detector will
be exploited in order to find, fit and select well-defined through-going
muon tracks. Once the geometry of the track is known, the Cherenkov energy
deposit will be estimated and compared against the photo-electron
distribution in the event.  This was the primary absolute energy
calibration method used in the IMB detector~\cite{Becker-Szendy}.

\chapter{Auxiliary Celestial Coordinate system}
\label{chap_aux_celes_coord}


\begin{figure}
\centering
\begin{minipage}[b]{0.5\textwidth}
\centering
\psfrag{M}{$M$}
\includegraphics[width=2in]{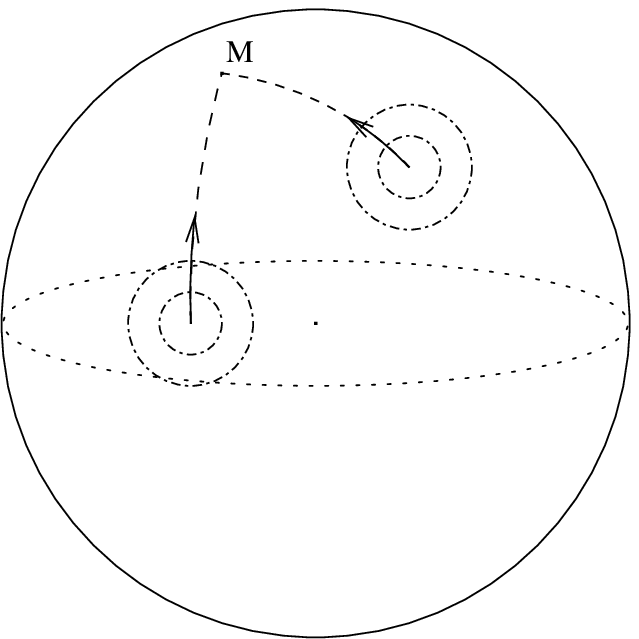}
\end{minipage}%
\begin{minipage}[b]{0.5\textwidth}
\centering
\psfrag{chi}{$\chi$}
\psfrag{xi}{$\xi$}
\psfrag{HH}{$H$}
\psfrag{L}{$L$}
\psfrag{X}{$X$}
\psfrag{C}{$C$}
\psfrag{P}{$P$}

\includegraphics[width=2.0in]{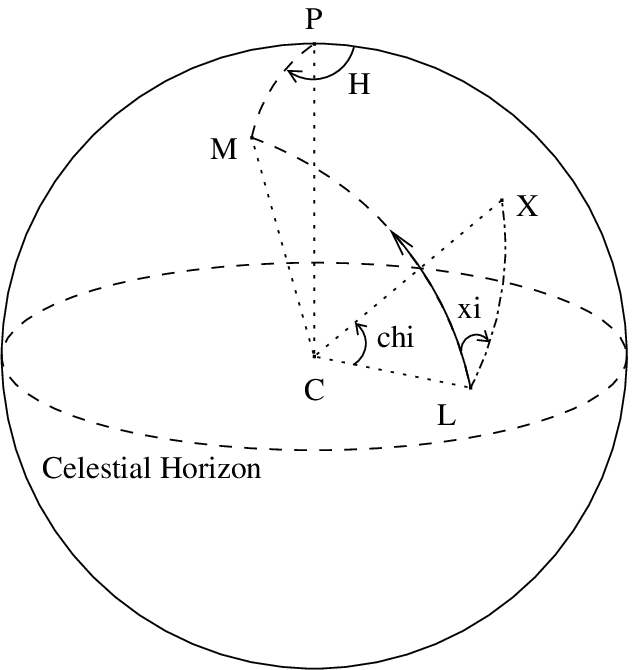}
\end{minipage}\\[-10pt]%
%
%
\begin{minipage}[t]{0.5\textwidth}
\caption{Schematic of the map}
\end{minipage}%
\begin{minipage}[t]{0.5\textwidth}
\caption{Auxiliary coordinates on the Celestial sphere}
\label{sky_map_appendix:aux_coor_def}
\end{minipage}%
\end{figure}

An additional coordinate system on the Celestial sphere which is used in 
this work is defined relative to a preselected celestial object (point 
on the celestial sphere). The object is the origin of the coordinate 
grid. The ``zero'' axis of this coordinate system can be chosen to point 
to any point $M$ on the celestial sphere.

Let $L=(\delta_{L},\alpha_{L})$ be the point on the Celestial sphere
which should be the center of the map directed to point
$M=(\delta_{M},\alpha_{M})$, then the point $X=(\delta_{X},\alpha_{X})$
on the Celestial sphere will have coordinates $(\chi,\xi)$ relative to
the point $L$ defined as (see
figure~\ref{sky_map_appendix:aux_coor_def}):

\[ \chi =\stackrel{\smile}{LX}=\angle LCX, \ \ \ \xi=\angle MLX\]

$\xi$ is measured clockwise from the $LM$ line and $\xi\in(-\pi;\pi)$,
$\chi\in(0;\pi)$

From spherical triangle $\triangle LPX$:
\[ \cos\chi =\cos(\stackrel{\smile}{LX})=\sin\delta_{L}\sin\delta_{X}+
          \cos\delta_{L}\cos\delta_{X}\cos (\alpha_{L}-\alpha_{X}) \]

From spherical triangle $\triangle PMX$:
\[ \cos(\stackrel{\smile}{MX})=\\sin\delta_{M}\sin\delta_{X}+
          \cos\delta_{M}\cos\delta_{X}\cos (\alpha_{M}-\alpha_{X}) \]

From spherical triangle $\triangle PML$:
\[ \cos(\stackrel{\smile}{ML})=\sin\delta_{M}\sin\delta_{L}+
          \cos\delta_{M}\cos\delta_{L}\cos (\alpha_{M}-\alpha_{L}) \]

From spherical triangle $\triangle MLX$:
\[ \cos(\stackrel{\smile}{MX})=
        \cos(\stackrel{\smile}{ML})\cos(\stackrel{\smile}{LX})+
      \sin(\stackrel{\smile}{ML})\sin(\stackrel{\smile}{LX})\cos(MLX) \]
\[ \cos\xi = \cos(\angle MLX)= \frac{\cos(\stackrel{\smile}{MX})-
              \cos(\stackrel{\smile}{ML})\cos(\stackrel{\smile}{LX})}%
             {\sin(\stackrel{\smile}{ML})\sin(\stackrel{\smile}{LX})} \]

There is no problem in defining the $\chi$, however, points symmetric
with respect to $MCL$-plane are indistinguishable. A way to solve this
problem is to introduce an auxiliary vector
$\vec{n}=\overrightarrow{CM}\times\overrightarrow{CL}$.\footnote{Since
Right Ascensions of the objects come only in differences, the derived
formulae will work in the local coordinates of $(\delta, H)$ by 
substituting $\alpha_{*}\rightarrow H_{*}$. However,
the $(\delta,H)$ coordinate system is left-handed and the cross-product
is defined in the right-handed system. Beware!} Then if
$(\vec{n}\cdot\overrightarrow{CX})>0$ then $\xi\in(0;\pi)$ else
$\xi\in(-\pi;0)$.


\chapter{Kinematics of the particles in the Solar system
         (Simulations appendix)}
\label{chap:simulations_app:simulations_app}


\section{ {\tt propagate\_infinity()}.}

The task of this function is to find a particle state at infinity given 
its state at the annihilation point by ``backward-in-time'' propagation. 
The function, however, should return the state in ``forward-time'' frame 
since only these quantities have physical meaning. The sequence of 
actions to be performed is summarized below:

\begin{enumerate}

\item Find orientation of the orbit in space 
      (section~\ref{sec:simulations_app:conserved_quant}).
\item Check that a particle crosses the Sun 
      (section~\ref{sec:simulations_app:cross_Sun}). If "no", go 
      to~\ref{finalize}.
\item Generate column density which should be accumulated before next
      scattering (section~\ref{sec:simulations_app:generate_path_in_sun}). 
      Compute parameters of the inside and outside orbits (if orbit at 
      least partially exits the Sun) (see 
      section~\ref{sec:simulations_app:traject_param}). Propagate the 
      particle to its first scattering point or go to \ref{finalize} if 
      the particle does not scatter. (The last situation is not possible 
      within the considered model, but is implemented for future 
      development.)
\item\label{scatter}
      Generate scattering off of a proton 
      (section~\ref{sec:simulations_app:scatter}), compute the 
      parameters of the new orbit 
      (section~\ref{sec:simulations_app:traject_param}). Generate the 
      column density to be accumulated until the next scattering 
      (section~\ref{sec:simulations_app:generate_path_in_sun}).
      \begin{itemize}
      \item If the particle in on unbound orbit, propagate it to the 
            edge of the Sun, compute the Runge-Lenz vector and go to 
            \ref{finalize}.
      \item If orbit is completely inside the Sun find the next 
            scattering point 
            (section~\ref{sec:simulations_app:motion_in}).
      \item If particle leaves the Sun, rotate the orbit according to the 
            column density required and then, propagate the particle to 
            the next scattering point 
            (section~\ref{sec:simulations_app:motion_out}).
      \end{itemize}
\item Go to \ref{scatter}.
\item\label{finalize}
      Use Runge-Lenz vector to find the $\vec{v}_{\infty}$ or 
      record that the particle is on a non-crossing Sun bound orbit.
      Perform time reflection on initial and final state 
      (section~\ref{sec:simulations_app:motion_out}).

\end{enumerate}


\section{Conserved quantities.}
\label{sec:simulations_app:conserved_quant}

In the considered model, the Sun is a ball of uniform density of mass
$M_{\odot}$ and radius $R_{\odot}$. Thus, the gravitational potential
$U(r)$ is the function of the distance $r$ from the center of the Sun
and is:
\[ U(r)=\left\{\begin{array}{cl}
    -\frac{\alpha}{r} &                            r\geq R_{\odot} \\
   \frac{\alpha}{2R_{\odot}^{3}}r^{2}-\frac{3\alpha}{2R_{\odot}}=
   \frac{\alpha}{2R_{\odot}}(\frac{r^{2}}{R_{\odot}^{2}}-3) & r\leq R_{\odot}
   \end{array}\right. \]
where $\alpha=G_{N}M_{\odot}$ and $G_{N}$ is the gravitational constant. 
The energy and angular momentum are conserved in such a system and are:
\[ {\cal E}=\frac{\vec{v}^{2}}{2}+U(r)=\mbox{const,} \ \ \ \ \ 
                       \vec{\cal J}=\vec{r}\times\vec{v}=\mbox{const} \]

When the particle moves outside of the Sun, its trajectory can be
described by a conical section with the Sun in one of its foci.  The
trajectories of the particles inside the Sun are elliptical only with
the center of the Sun in the center of the ellipse.

There is an additional conserved quantity when the particles is outside 
the Sun. It is the Runge-Lenz vector $\vec{\cal K}$:
\[ \vec{\cal K}=\dot{\vec{r}}\times\vec{\cal J}-\alpha\vec{r}/r
     \ \ \ \ \ |\vec{\cal K}|=\sqrt{\alpha^{2}+2{\cal E}{\cal J}^{2}} \]
The Runge-Lenz vector points along the major axis from focus to
perihelion.


\section{Which orbits cross the Sun.}
\label{sec:simulations_app:cross_Sun}

The energy conservation law outside the Sun has the form:

\[ {\cal E}=\frac{v^{2}_{r}}{2}+
            \frac{{\cal J}^{2}}{2r^{2}}-\frac{\alpha}{r} \]

where $v_{r}$ is the radial velocity of the particle. The particle will 
cross the Sun if its minimum distance to the center of the Sun $r_{min}$ 
is less than $R_{\odot}$. At this point $v_{r}=0$ and:

\[ {\cal E}=\frac{{\cal J}^{2}}{2r^{2}_{min}}-
            \frac{\alpha}{r_{min}} \ \ \ \Rightarrow \ \ \ 
  r_{min}=\frac{\sqrt{\alpha^{2}+2{\cal E}{\cal J}^{2}}-\alpha}
                {2{\cal E}} \]
Thus, only the trajectories for which 
${\cal J}^{2} < 2R_{\odot}({\cal E}R_{\odot}+\alpha)$
will cross the Sun.

A separate remark should be made regarding the unbound orbits which
cross the Sun. Particles on such orbits may never cross the Sun. Indeed,
if a particle is at its annihilation point and if it happens to be on an
unbound orbit (as can be determined from its velocity and position
vectors), the angle between its velocity and the Runge-Lenz vector
should be acute for the particle to pass through the Sun. In other
words, if ${\cal E}\geq 0$ and $\vec{\cal K}\cdot\vec{v}\leq 0$ the
particle will not cross the Sun.


\section{Rotation of a vector $\vec{B}$ around a vector $\vec{L}$
 by an angle $\gamma$.}

Only the proper rotations on angle $\gamma$ are considered where the
rotation is governed by the ``right handed screw'' rule around vector
$\vec{L}$, $|\vec{L}|\neq 0$. The new vector $\vec{B}'$ is obtained from
the original $\vec{B}$ by application of the rotation matrix $A(\gamma,
\vec{L})$.

\[ \vec{B}'=A(\gamma ,\vec{L})\vec{B} \]

The explicit from of the transformation in Cartesian coordinates is:

\[ \vec{B}' =
   \left(\begin{array}{c}
   B'_{x} \\ B'_{y} \\ B'_{z}
   \end{array}\right) =
   \left(\begin{array}{ccc}
   a+cL_{x}^{2} &
   cL_{x}L_{y}-bL_{z} &
   cL_{x}L_{z}+bL_{y} \\

   cL_{x}L_{y}+bL_{z} &
   a+cL_{y}^{2} &
   cL_{y}L_{z}-bL_{x} \\

   cL_{x}L_{z}-bL_{y} &
   cL_{y}L_{z}+bL_{x} &
   a+cL_{z}^{2}
   \end{array}\right)\cdot
   \left(\begin{array}{c}
   B_{x} \\ B_{y} \\ B_{z}
   \end{array}\right)
\]

where

\[ a=\cos\gamma \ \ \ b=\frac{\sin\gamma}{\sqrt{L^{2}}} \ \ \ 
                                        c=\frac{1-\cos\gamma}{L^{2}} \]

If operation of rotation of vector $\vec{\cal K}$ is performed around
vector ${\cal J}$, a special case arises when ${\cal J}=0$. In this
situation the operation of rotation is not defined. However, from the
physics of the situation, it follows that the rotation angle $\gamma$ is
either $0$ or $\pi$. Thus the rotation in this case is very simple:

\[ \begin{array}{ll}
   \mbox{if}\Big(\gamma == \pi \Big) &  \vec{\cal K}' = -\vec{\cal K} \\
   \mbox{else}                       &  \vec{\cal K}' =\; \vec{\cal K}
   \end{array} \]


\section{Some facts about elliptical trajectories.}
\label{sec:simulations_app:traject_param}

In addition to the global coordinate system which is attached to the Sun 
and whose $z$ axis is oriented along the Sun's motion in the Galactic 
disk, there are several auxiliary coordinate systems which are 
convenient to introduce. Both of the auxiliary coordinate systems are 
centered on the Sun's center and one of them is used to analyze the 
particle trajectory inside the Sun, and the other --- outside.

Capital letters for the names of the variables will represent the
trajectory which is outside the Sun and the small letters will represent
the parameters for the trajectories which lie inside the Sun.

\subsection{Equation of the ellipse.}

The equation of ellipse with the coordinate system at its center is
\[ \frac{x^{2}}{a^{2}}+\frac{y^{2}}{b^{2}} = 1 \ \ \ \mbox{or}\ \ \ 
                          r^{2}=\frac{b^{2}}{1-e^{2}\cos^{2}\phi} \]
where $\phi = 0$ is the point of maximal distance from the point on
ellipse to the origin (see 
figure~\ref{fig:simulations_app:ellipse_in_sun}).

The equation of ellipse with the coordinate system at one of its foci
\[ r=\frac{P}{1+E\cos\Phi} \]
where $\Phi = 0$ is the point of minimal distance from the point on
ellipse to the origin (see 
figure~\ref{fig:simulations_app:rungelenz_and_sun}).

The relationship between the semimajor axis $a$ (or $A$) and semiminor 
one $b$ (or $B$) and the eccentricity $e$ (or $E$) and the latus rectum 
$\rho$ (or $P$) is:

\[  \left\{\begin{array}{ll}
     e^{2} = 1-b^{2}/a^{2} &  \rho = b^{2}/a \\
     E^{2} = 1-B^{2}/A^{2} &     P = B^{2}/A
    \end{array}\right. \]

\subsection{Ellipse inside the Sun.}
\label{subsec:simulations_app:ellipse_inside}

\begin{figure}
\centering
\psfrag{rr}{$\vec{r}$}
\psfrag{vv}{$\vec{v}$}
\psfrag{phi1}{$\phi$}
\psfrag{psi}{$\psi$}
\psfrag{xx}{$x$}
\psfrag{yy}{$y$}
\psfrag{angle}{$\widehat{(\vec{r},\vec{v})}$}
\psfrag{JJ}{$\vec{\cal J}^{\bigotimes}$}
\includegraphics[totalheight=3.0in]{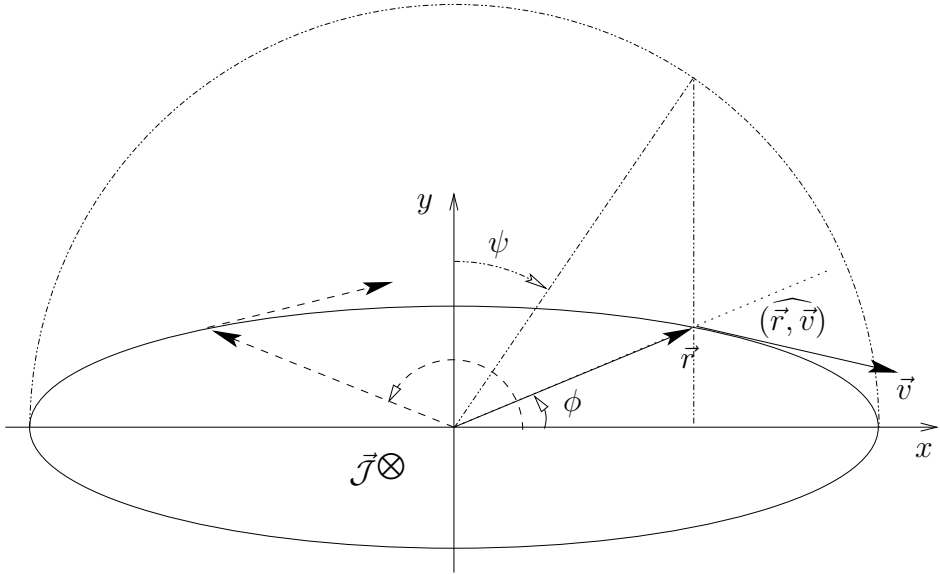}
\caption{Elliptical trajectory inside the Sun.}
\label{fig:simulations_app:ellipse_in_sun}
\end{figure}

The parameters of the ellipse inside the Sun are:

\[ \left\{\begin{array}{l}
   a^{2}=\frac{R_{\odot}^{2}}{\alpha}\left({\cal E'}R_{\odot}+
   \sqrt{{\cal E'}^{2}R_{\odot}^{2}-\alpha{\cal J}^{2}/R_{\odot}}\right) \\
   b^{2}=\frac{R_{\odot}^{2}}{\alpha}\left({\cal E'}R_{\odot}- 
   \sqrt{{\cal E'}^{2}R_{\odot}^{2}-\alpha{\cal J}^{2}/R_{\odot}}\right)
   \end{array}\right| \ \ \ \ 
   {\cal E}' = {\cal E}+\frac{3\alpha}{2R_{\odot}} \]

The trajectory of a particle inside the Sun is an ellipse since the
potential energy varies as distance from the Sun's center squared. 
Moreover, the ellipse is centered on the Sun's center. It is always
possible to choose the coordinate system in such a way that the $OX$
axis is along the semimajor axis of the ellipse and $OY$ --- the
semiminor one. Because same calculations are simpler using one
parameterization of an ellipse and others in another, two different
ellipse parameterizations are used in this work. One is by providing the
polar angle $\phi$ measured counterclockwise from the $OX$ and the
distance to the point from the origin $r$ and the other is by specifying
a phase angle $\psi$ ``{\em measured clockwise from the $OY$ axis}''
only.\footnote{The angle $(\pi/2-\psi)$ is called the eccentric anomaly 
in celestial mechanics.} (See figure~\ref{fig:simulations_app:ellipse_in_sun}.)

The orientation of the coordinate system is chosen by requiring that at
one selected point (usually initial point of propagation) if the angle
between $\vec{r}$ and $\vec{v}$ is acute, the phase angle $\psi$ of the
point should be between zero and $\pi/2$ and between $-\pi/2$ and zero
if the angle is obtuse (see
fig~\ref{fig:simulations_app:ellipse_in_sun}).

The equations of the same ellipse in different parameterizations are:
\begin{equation}
   \left\{\begin{array}{l}
   x=a\sin\psi=\sqrt{\frac{b^{2}}{1-e^{2}\cos^{2}\phi}}\cos\phi \\
   y=b\cos\psi=\sqrt{\frac{b^{2}}{1-e^{2}\cos^{2}\phi}}\sin\phi
   \end{array}\right| \ \Rightarrow\ 
   \left\{\begin{array}{l}
   \tan\phi=\frac{b\cos\psi}{a\sin\psi} \\
   r^{2}=b^{2}+(a^{2}-b^{2})\sin^{2}\psi=
                                      \frac{b^{2}}{1-e^{2}\cos^{2}\phi}
   \end{array}\right|
\label{eq:simulations_app:ellipse_inside}
\end{equation}

The length $s$ of the elliptical arc between the angles $\psi_{1}$ and
$\psi_{2}$ is:
\[ \left\{\begin{array}{l}
   dx=a\cos\psi\,d\psi \\
   dy=-b\sin\psi\,d\psi
   \end{array}\right| \ \Rightarrow\ 
   (ds)^{2}=(dx)^{2}+(dy)^{2}=a^{2}\left[
      1-\frac{a^{2}-b^{2}}{a^{2}}\sin^{2}\psi\right](d\psi)^{2}   \]
\[ ds = a\sqrt{1-e^{2}\sin^{2}\psi}\,d\psi, \ \ e^{2}=1-b^{2}/a^{2} \]

\[ s=a\int_{\psi_{1}}^{\psi_{2}}\sqrt{1-e^{2}\sin^{2}\psi}\,d\psi \]

Thus, the length of an elliptical arc can be expressed in terms of the 
elliptical integral:
\[ EE(\theta,k)=\int_{0}^{\theta}\sqrt{1-k^{2}\sin^{2}t}\,dt,
                             \ \ \ \ k^{2}<1 \ \ \theta\in[0;\pi/2] \]

\subsection{Angle between $\vec{r}$ and $\vec{v}$.}

If current position $\vec{r}$ and velocity $\vec{v}$ are known, it is 
easy to find the angle $\widehat{(\vec{r},\vec{v})}$ between the two:

\[ \cos\widehat{(\vec{r},\vec{v})}=\frac{\vec{r}\cdot\vec{v}}{rv} \]

If velocity $\vec{v}$ is not known, but the the phase angle $\psi$
corresponding to the position $\vec{r}$ on the ellipse is known it is
possible to find the angle between $\vec{r}$ and $\vec{v}$. Inside the
Sun, the energy conservation law is:

\begin{equation}
   {\cal E}=\frac{v^{2}}{2}+
    \frac{\alpha}{2R_{\odot}^{3}}r^{2}-\frac{3\alpha}{2R_{\odot}}
    \ \ \Rightarrow \ \ \ 
   v^{2}=\left(
2{\cal E'}R_{\odot}-\alpha\frac{r^{2}}{R_{\odot}^{2}}\right)/R_{\odot}
\label{eq:simulations_app:energy_conserve_inside}
\end{equation}

\[ \vec{\cal J}=\vec{r}\times\vec{v} \ \ \Rightarrow \ \
   |\vec{\cal J}|=|\vec{r}|\cdot|\vec{v}|\sin\widehat{(\vec{r},\vec{v})} 
  \ \ \Rightarrow \ \ \sin\widehat{(\vec{r},\vec{v})}=\frac{\cal J}{rv}=
   \sqrt{\frac{{\cal J}^{2}}{r^{2}v^{2}}} \]

\begin{equation}
  \widehat{(\vec{r},\vec{v})}=\left\{\begin{array}{clc}
   \arcsin(\sqrt{\frac{{\cal J}^{2}}{r^{2}v^{2}}}), &
     \psi\in [-\pi;-\pi/2]\bigcup [0;\pi/2] &
                              \Leftrightarrow \ \ \tan\psi \geq 0 \\
   \pi -\arcsin(\sqrt{\frac{{\cal J}^{2}}{r^{2}v^{2}}}), &
   \psi\in [-\pi/2;0]\bigcup [\pi/2;\pi]  &
                              \Leftrightarrow \ \ \tan\psi \leq 0
   \end{array}\right.
\label{eq:simulations_app:rv_angle_sin}
\end{equation}

To find $\vec{v}$ it is enough to rotate the vector $\vec{r}$ around
$\vec{\cal J}$ by angle $\widehat{(\vec{r},\vec{v})}$ (see
figure~\ref{fig:simulations_app:ellipse_in_sun}) and rescale 
appropriately:

\[ \vec{v}=\frac{v}{r}
                     \mbox{\tt rotate\_vector}(\vec{r},\vec{\cal J},
                                       \widehat{(\vec{r},\vec{v})}) \]

\subsection{Ellipse outside the Sun.}

\begin{figure}
\centering
\psfrag{KK}{$\vec{\cal K}$}
\psfrag{phi2}{$\Phi$}
\psfrag{JJ}{$\vec{\cal J}^{\bigodot}$}
\includegraphics{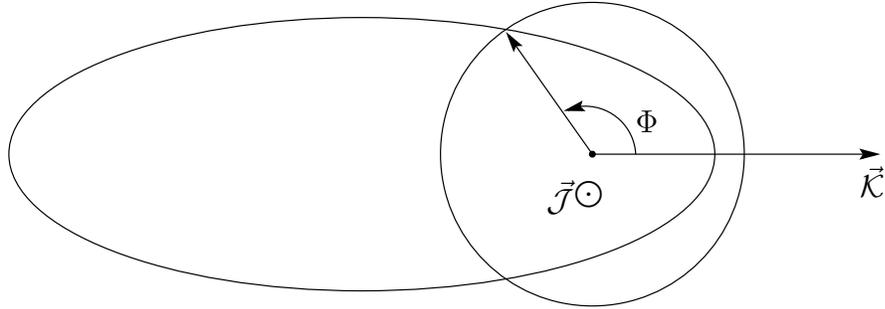}
\caption{Elliptical trajectory outside the Sun.}
\label{fig:simulations_app:rungelenz_and_sun}
\end{figure}

The parameters of the ellipse outside the Sun are:

\[ E^{2}=1+\frac{2{\cal E}{\cal J}^{2}}{\alpha^{2}} \ \ \ \ 
   P={\cal J}^{2}/\alpha \]

The bound trajectory while outside the Sun is an ellipse with the center
of the Sun at one of its foci (``the outside ellipse''or ``the outside
orbit''). It is interesting to find the angle $\Phi$ between the
Runge-Lenz vector (the direction on the perihelion) and the point where
the outside ellipse intersects the Sun (see
figure~\ref{fig:simulations_app:rungelenz_and_sun}). This angle is
easily found from the equation of the outside ellipse:

\[ \frac{P}{R_{\odot}}=1+E\cos\Phi \]
\begin{equation}
    \cos\Phi =\frac{1}{E}\left(P/R_{\odot}-1\right)=
              \frac{{\cal J}^{2}-\alpha R_{\odot}}{E\alpha R_{\odot}}
\label{eq:simulations_app:angle2Runge}
\end{equation}

\subsection{Rotation of outside orbit due to passage through the Sun.}
\label{sec:simulations_app:rotaion_outside_orbit}

\begin{figure}
\centering
\psfrag{ph}{$\phi_{0}$}
\psfrag{Ph}{$\Phi_{0}$}
\psfrag{gg}{$\gamma/2$}
\psfrag{xx}{$x$}
\psfrag{yy}{$y$}
\psfrag{KK}{$\vec{\cal K}$}
\psfrag{JJ}{$\vec{\cal J}^{\bigotimes}$}
\includegraphics[totalheight=2.5in]{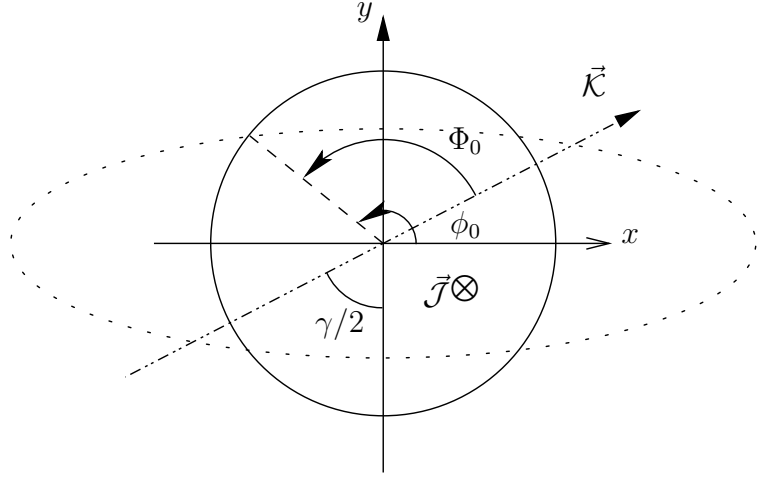}
\caption{Rotation of the outside orbit due to passage through the Sun.}
\label{fig:simulations_app:orbit_rotation}
\end{figure}

The angle of rotation of the orbit in a single pass through the Sun is
defined as the angle of rotation of Runge-Lenz vector due to this
passage.

If $\gamma$ is angle of rotation of the orbit as it passes through the
Sun, and $\Phi_{0}$ is the coordinate describing the point of entrance
of the orbit into the Sun and $\phi_{0}$ is direction to the same point,
but with respect to the orbit which is inside the Sun, we obtain the
relationship for the angle of rotation of the orbit (see
figure~\ref{fig:simulations_app:orbit_rotation}):

\[ \phi_{0} = \Phi_{0} +(\pi/2 -\gamma/2) \ \ \ \Rightarrow\ \ \ 
   \gamma = \pi +2\Phi_{0} -2\phi_{0} \]

From the equation~(\ref{eq:simulations_app:angle2Runge}) $\Phi_{0}$ can 
be obtained:

\[ \Phi_{0}= \arccos\left(
      \frac{{\cal J}^{2}-\alpha R_{\odot}}{E\alpha R_{\odot}}\right) \]

From the equations~(\ref{eq:simulations_app:ellipse_inside})
and~(\ref{eq:simulations_app:rv_angle_sin}) the phase and spatial angles
of the entrance point of the orbit into the Sun are:

\begin{equation}
    \sin\psi_{0}=-\sqrt{\frac{R_{\odot}^{2}-b^{2}}{a^{2}-b^{2}}} 
    \ \ \ \ \     \tan\phi_{0}=\frac{b\cos\psi_{0}}{a\sin\psi_{0}}
\label{eq:simulations_app:entrace_phase_angle}
\end{equation}

The ``$-$'' sign is selected since at the entrance point into the Sun
$(\vec{r}\cdot\vec{v})<0$ always.

Rotation of the Runge-Lenz vector $\vec{\cal K}$ happens on the angle
$(-\gamma)$ around $\vec{\cal J}$ in one passage of the orbit through
the Sun.


\section{Motion inside the Sun {\tt propagate\_in\_sun()}.}
\label{sec:simulations_app:motion_in}

Only the motion which is initialized from inside and while inside the Sun
is treated here. That is, the task considered here is knowing the
requested column density to evolve a particle from its initial position
and velocity inside the Sun to the scattering point inside the Sun or if
the particle exits the Sun --- output the new direction of Runge-Lenz
vector and the remaining column density to be accumulated in subsequent 
passes through the Sun.

There are several situations possible:
\begin{itemize}
\item The orbit is completely inside the Sun. Propagate the particle to 
      its next scattering. Output final $\vec{r}$ and $\vec{v}$.
\item The orbit exits the Sun and column density to be accumulated is 
      large to allow the particle to exit the Sun. Output the final 
      Runge-Lenz vector $\vec{\cal K}$.
\item The orbit exits the Sun geometrically, but the column density to 
      be accumulated is not large enough. Particle scatters before its 
      exist. Output $\vec{r}$ and $\vec{v}$ just before the scattering.
\end{itemize}

\subsection{Particle is inside the Sun.}

An orbit lies completely inside the Sun if the semimajor axis of the
ellipse is not larger than the radius of the Sun: $a\leq R_{\odot}$. The
phase angle $\psi_{1}$ and the spatial angle $\phi_{1}$ corresponding to
the current position $\vec{r}_{1}$ inside the Sun can be found from the
equation of the ellipse in phase coordinates (see
equations~(\ref{eq:simulations_app:rv_angle_sin})
and~(\ref{eq:simulations_app:ellipse_inside})):

\begin{equation}
    \begin{array}{c}
    \sin\psi_{1}=sign(\vec{r}_{1}\cdot\vec{v}_{1})
      \sqrt{\frac{r^{2}_{1}-b^{2}}{a^{2}-b^{2}}},
                                 \ \ \ \ \ \psi_{1}\in [-\pi/2;\pi/2] \\ \\
    \tan\phi_{1}=\frac{b\cos\psi_{1}}{a\sin\psi_{1}}
    \end{array}
\label{eq:simulations_app:psi_1}
\end{equation}

The final angles $\psi_{2}$ and $\phi_{2}$ are such that the column
density accumulated by a particle traveling inside the Sun is equal to
the specified column density $L$. Currently, the equation formulated in 
the  phase angles is considered:

\[ L= \mbox{\tt trajectory\_length}(\psi_{1},\psi_{2}) \]

This equation can be solved 
(see section~\ref{sec:simulations_app:solvepath4psi}) and the 
corresponding spatial angle is:

\[ \tan\phi_{2}=\frac{b\cos\psi_{2}}{a\sin\psi_{2}} \]

Now, the sought for direction of $\vec{r}_{2}$ can be found by rotating
the vector $\vec{r}_{1}$ on angle $-(\phi_{2}-\phi_{1})$ around
$\vec{\cal J}$ and the magnitude of $\vec{r}_{2}$ an be found from the
equation of the ellipse~(\ref{eq:simulations_app:ellipse_inside}):

\[ r_{2}^{2}=b^{2}+(a^{2}-b^{2})\sin^{2}\psi_{2} \ \ \ \ \ 
   \vec{r}_{2}=\mbox{\tt rotate\_vector}
               (\vec{r}_{1},\vec{\cal J},\phi_{1}-\phi_{2})\cdot
                          \sqrt{\frac{r_{2}^{2}}{|\vec{r}_{1}|^{2}}} \]

The velocity vector $\vec{v}_{2}$ at the point $\vec{r}_{2}$ can be
found by rotating the vector $\vec{r}_{2}$ on the angle between
$\vec{r}_{2}$ and $\vec{v}_{2}$ obtained from the
equation~(\ref{eq:simulations_app:rv_angle_sin}). The magnitude of
$\vec{v}_{2}$ is found from the energy conservation 
law~(\ref{eq:simulations_app:energy_conserve_inside}).

In the case of a circular orbit $a\equiv b$ and the previous logic will
fail because the initial phase $\psi_{1}$ is arbitrary and the most
appealing choice is to set $\psi_{1}=0$. The logic of the presented
algorithm is intact.

\subsection{Orbit exits the Sun, but the particle doesn't.}

It is possible to have a situation where a particle has its initial
position inside the Sun, but its orbit leaves the Sun geometrically. It
is also possible, however, that the column density to be accumulated is
small and the particle scatters before it has a chance to exit the Sun.
This is case of motion confined to the interior of the Sun and was
treated above. To make sure that the particle stays inside the Sun, the
column density which can be accumulated until the exit from the Sun
should be greater than the requested column density $L$:

\[ L< \mbox{\tt trajectory\_length}(\psi_{1},\psi_{R_{\odot}}) \]
where $\psi_{1}$ is the initial phase defined in
equation~(\ref{eq:simulations_app:psi_1}) and $\psi_{R_{\odot}}$ is the 
phase of the exit point from the Sun and computed as in 
equation~(\ref{eq:simulations_app:ellipse_inside}):

\begin{equation}
  \sin\psi_{R_{\odot}}=+\sqrt{\frac{R^{2}_{\odot}-b^{2}}{a^{2}-b^{2}}} 
\label{eq:simulations_app:psi_R_def}
\end{equation}

If it is found that the requested column density is greater than maximum 
possible in the configuration, the particle leaves the Sun and this case 
is treated right below.

\subsection{Particle exits the Sun.}

Again, as before, $\phi_{1}$, $\psi_{1}$ describe the initial point 
(equation~(\ref{eq:simulations_app:psi_1})) and 
$\phi_{2}$,$\psi_{2}$ is the final point of propagation ---
the point of orbit exit from the Sun (see 
equations~(\ref{eq:simulations_app:ellipse_inside}) and
(\ref{eq:simulations_app:psi_R_def}), $\psi_{2}\equiv\psi_{R_{\odot}}$):

\begin{equation}
  \sin\psi_{2}=+\sqrt{\frac{R^{2}_{\odot}-b^{2}}{a^{2}-b^{2}}},  \ \ \ 
                     \tan\phi_{2}=\frac{b\cos\psi_{2}}{a\sin\psi_{2}}
\label{eq:simulations_app:sun_exit_phase}
\end{equation}

The accumulated column density inside the Sun between angles $\psi_{1}$
and $\psi_{2}$ should be computed and subtracted from the remaining 
column density. 

The spatial angle between the initial point $\vec{r}_{1}$ and the
Runge-Lenz vector $\vec{\cal K}$ consists of the angle between the
initial point and the exit point from the Sun $(\phi_{1}-\phi_{2})$ and
the angle $\Phi_{2}$ between the exit point from the Sun and the
Runge-Lenz vector which is found from
equation~(\ref{eq:simulations_app:angle2Runge}). Therefore, the vector
$\vec{r}_{1}$ should be rotated by $(\phi_{1}-\phi_{2}-\Phi_{2})$ around
the $\vec{\cal J}$ with {\tt rotate\_vector}$(\vec{r}_{1},\vec{\cal J},
(\phi_{1}-\phi_{2}-\Phi_{2}))$ to find the direction of the Runge-Lenz
vector.


\section[Motion outside the Sun {\tt propagate\_outside\_sun()}.]
{Motion outside the Sun \\ {\tt propagate\_outside\_sun()}.}
\label{sec:simulations_app:motion_out}

The task of this function is given the direction of the Runge-Lenz
vector and the column density {\tt column\_density} to be accumulated
inside the Sun until the next scattering point find the particle
position and velocity at the scattering point. Needless to say, the
function should process only the particles which pass through the Sun.
The other parameters which are expected to be available are the
parameters of the orbit inside the Sun and the phase angle $\psi_{2}$
defined in the equation~(\ref{eq:simulations_app:sun_exit_phase}).

There could be several distinct cases:
\begin{enumerate}
\item The trajectory is an ellipse
\item The trajectory is a hyperbola or parabola (but the particle 
      crosses the Sun)
\end{enumerate}

\subsection{Find the scattering point.}

The parameters of the trajectory do not change between scatterings and
only rotation of the whole orbit is possible as the orbit passes through
the Sun. The angle $\gamma$ of orbit rotation due to a single pass
through the Sun was obtained in
section~\ref{sec:simulations_app:rotaion_outside_orbit}. The number of
required passes through the Sun can be be determined from the required
column density to be accumulated and the column density accumulated in a
single pass through the Sun. This will give the total angle of the orbit
rotation before the scattering.

Due to spherical symmetry of the Sun, the column density accumulated in 
one pass through the Sun has the form of (see 
section~\ref{subsec:simulations_app:ellipse_inside} and 
equation~(\ref{eq:simulations_app:psi_R_def})):

\[ L_{1}= 2\cdot\mbox{\tt trajectory\_length}(0,\psi_{R_{\odot}}) \]

The total angle $n\cdot\gamma$ of rotation of the orbit is:
\[ n\cdot\gamma=\mbox{\tt floor(column\_density/$L_{1}$)}\cdot\gamma \]
\[ \mbox{\tt column\_density}:=\mbox{\tt column\_density}-nL_{1} \]

From figures \ref{fig:simulations_app:rungelenz_and_sun} and
\ref{fig:simulations_app:orbit_rotation}, one sees that the vector
pointing to the entrance point into the Sun can be found by rotating the
Runge-Lenz vector by angle $(-\Phi_{0})$, that is why the total
rotation of $\vec{\cal K}$ to be performed is on angle
$(-n\gamma-\Phi_{0})$ around $\vec{\cal J}$.

The remaining problem is to find the angle between the point of entrance
into the Sun and the scattering point using the remaining column density
to be accumulated. The particle will spend remaining time on an ellipse
inside the Sun, and the phase angle at the scattering point $\psi_{1}$
can be found given the remaining column density and the phase angle
$\psi_{0}$ (equation~(\ref{eq:simulations_app:entrace_phase_angle}))
of particle entrance into the Sun (see 
section~\ref{sec:simulations_app:solvepath4psi}):

\[ \mbox{\tt column\_density}= 
                     \mbox{\tt trajectory\_length}(\psi_{0},\psi_{1})
                     \ \ \ \Rightarrow \ \ \ 
                     \tan\phi_{1}=\frac{b\cos\psi_{1}}{a\sin\psi_{1}} \]

Now, the sought for scattering point $\vec{r}_{1}$ can be found by
rotating the vector $\vec{\cal K}$ on angle
$(-n\gamma-\Phi_{0}+\phi_{0}-\phi_{1})$ around $\vec{\cal J}$ and the
magnitude of $\vec{r}_{1}$ an be found from the equation of the
ellipse~(\ref{eq:simulations_app:ellipse_inside}):

\[ r_{1}^{2}=b^{2}+(a^{2}-b^{2})\sin^{2}\psi_{1} \ \ \ 
   \vec{r}_{1}=\mbox{\tt rotate\_vector}
  (\vec{\cal K},\vec{\cal J},-n\gamma-\Phi_{0}+\phi_{0}-\phi_{1})\cdot
                          \sqrt{\frac{r_{1}^{2}}{|\vec{\cal K}|^{2}}} \]

The velocity vector $\vec{v}_{1}$ at the point $\vec{r}_{1}$ can be
found by rotating the vector $\vec{r}_{1}$ on the angle between
$\vec{r}_{1}$ and $\vec{v}_{1}$ obtained from the
equation~(\ref{eq:simulations_app:rv_angle_sin}) around vector
$\vec{\cal J}$. The magnitude of $\vec{v}_{1}$ is found from the energy
conservation law~(\ref{eq:simulations_app:energy_conserve_inside}).

\subsection{Particle on an unbound orbit crossing the Sun.}

This situation can occur if a particle is generated on a hyperbolic or
parabolic orbit, but it passes through the Sun. If the requested column
density is smaller than can be accumulated in one passage through the
Sun, the above logic will suffice. If however, the requested column
density can not be accumulated in a single pass, the particle will leave
the Solar system, thus, in the previous calculations the angle of orbit
rotation due to passage through the Sun is equal to $\gamma$ from the
section~\ref{sec:simulations_app:rotaion_outside_orbit} only. The output
of the function in this case is the final direction of the Runge-Lenz
vector and is found by rotating the input Runge-Lenz vector on the
angle $(-\gamma)$ around $\vec{\cal J}$.

\subsection{Velocity at infinity.}

\begin{figure}
\centering
\psfrag{J}{$\vec{\cal J}^{\bigotimes}$}
\psfrag{K}{$\vec{\cal K}$}
\psfrag{A}{$\frac{B}{A}$}
\psfrag{VV}{$\vec{V}$}
\includegraphics[totalheight=3in]{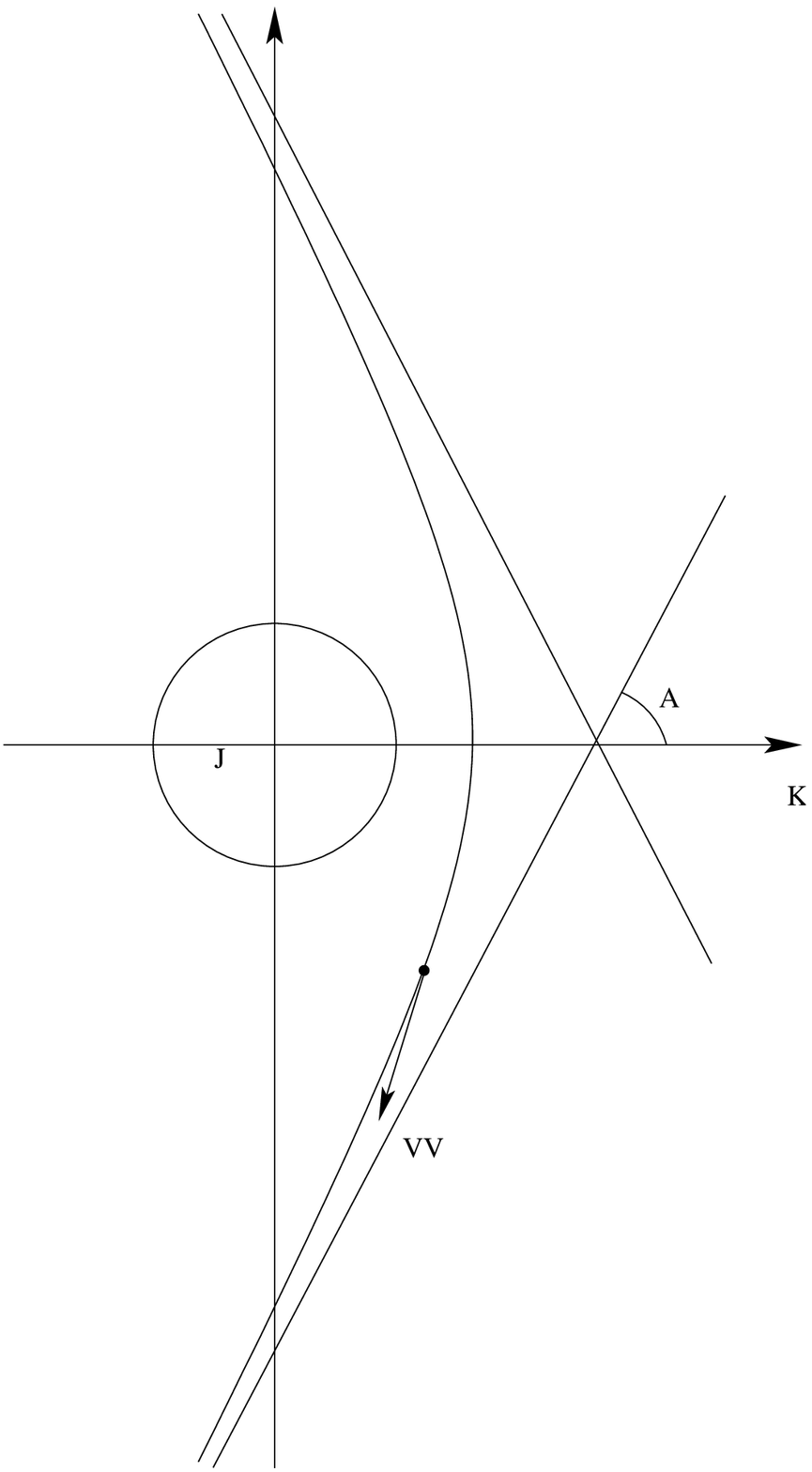}
\caption{Finalize}
\label{fig:simulations_app:finalize}
\end{figure}

This is the final function and its task is to find the velocity of the 
particle at infinity. Since only ``forward'' in time information is 
needed, this function should perform time-reflection on the initial and 
final velocities of the particle.

The time reflection can be performed in the spherical coordinates by
requiring the change in the polar angle $\theta\rightarrow (\pi -\theta)$
and in the azimuth $\phi\rightarrow (\pi +\phi)\mbox{modulo}(2\pi)$.

A particle can exit the Solar system only when it is on a hyperbolic or
a parabolic trajectory when ${\cal E}\geq 0$. The direction of the
velocity at infinity can coincides with the direction of the trajectory
asymptote. Thus, the velocity direction can be found by rotating the
Runge-Lenz vector to point to the asymptote (see
figure~\ref{fig:simulations_app:finalize}) by angle $(\pi-\arctan B/A)$.

\[ \frac{B}{A}=\frac{\cal J}{\sqrt{2\cal E}}\frac{2\cal E}{\alpha}=
   \frac{1}{\alpha}\sqrt{2{\cal EJ}^{2}} \]

and the magnitude of the velocity can be found from the energy 
conservation law:

\[ v_{\infty}=\sqrt{2{\cal E}} \]

The above works for the Solar system escape on a parabolic orbit too,
because in this case the particle starts its fall into the Solar system
in $(\vec{\cal K})$ direction.  Even though $\vec{v}_{\infty}=0$ in this
case, it is necessary to know the direction in which the particle 
started to fall onto the Solar System, that is why $\vec{v}_{\infty}$ 
should be stored in spherical coordinates providing the information 
about the magnitude and the direction of the velocity even if
$|\vec{v}_{\infty}|=0$.

If ${\cal E}<0$ and the particle does not cross the Sun, it stays on
closed bound orbit around the Sun and the velocity at infinity is not
defined.


\section{SolvePath4Psi.}
\label{sec:simulations_app:solvepath4psi}

The task of the function is given initial point on an ellipse inside the
Sun via its phase angle $\psi_{in}$ find the point $\psi_{out}$ such
that the ellipse pathlength from $\psi_{in}$ to $\psi_{out}$ is equal to
the given value of {\tt path\_in}. If such a point is found, its phase 
$\psi_{out}$ is returned. If the point can not be found because the 
particle leaves the Sun before the requested pathlength is accumulated, 
the {\tt path\_in} is decreased by the path traveled in the Sun and 
$\psi_{out}$ is set to the exit point from the Sun.

The length of an ellipse from point $\psi=0$ to $\psi$ is found by the 
elliptical integral:

\[ EE(\psi,e)=sign(\psi)\cdot\int_{0}^{\psi}\sqrt{1-e^{2}\sin^{2}t}dt \]
where $e$ is eccentricity of the ellipse and $|\psi|<\pi/2$.

Thus, elliptical distance between two points on ellipse is 

\[ L(\psi_{in},\psi_{out}) = a\Big(EE(\psi_{out},e) - 
EE(\psi_{in},e)\Big) \]

The problem with this definition is that both initial $\psi_{in}$ and 
final $\psi_{out}$ phases should be less than $\pi/2$, otherwise, the 
definition of the distance on the ellipse must be modified.

The initial phase $\psi_{in}$ is within allowed range by its 
construction.

\subsection{Bracketing the root.}

If the orbit exits the Sun, both $\psi_{in}$ and $\psi_{out}$ are within 
allowed bounds since the phase for the exit point 
$\psi_{out}\leq\psi_{R}\leq\pi/2$.

If the pathlength {\tt path\_in} is larger than the path from
$\psi_{in}$ to $\psi_{R}$ the particle exits the Sun. The path
$L(\psi_{in},\psi_{R})$ is subtracted from {\tt path\_in} and the
remaining {\tt path\_in} and $\psi_{R}$ are returned with a flag that
the particle exited the Sun.

Otherwise, the particle stays inside the Sun even though, its orbit
geometrically exits the Sun and $\psi_{in}<\psi_{out}<\psi_{R}$. The
root bracket is found. If the ellipse is completely inside the Sun, more
things have to be done.

If the total requested pathlength {\tt path\_in} is greater than the 
length of ellipse, particle will make one or several full revolutions 
which are not interesting for us. Pathlength {\tt path\_in} should be 
reduced by the path acquired in full revolutions. After this is done,
$\psi_{out}$ is within: $\psi_{in}<\psi_{out}<2\pi +\psi_{in}$.
The problem is reduced to the one where the particle does not make a 
full revolution in the Sun.

Since it is known that the particle makes less than a full revolution, a
check can be made if $\psi_{out}<\pi/2$ and $\psi_{out}<3\pi/2$ by
comparing {\tt path\_in} with $L(\psi_{in},\pi/2)$ and
$\Big(L(\psi_{in},\pi/2)+L(-\pi/2,\pi/2)\Big)$.

If it is found that the solution is in $\psi_{out}<\pi/2$ the lower
bound on the root $\psi_{l}$ is set to $\psi_{in}$ and the upper
$\psi_{u}$ to $\pi/2$.

If it is found that the solution is in $\pi/2<\psi_{out}<3\pi/2$, the
pathlength $L(\psi_{in},\pi/2)$ is subtracted from {\tt path\_in} and
$\psi_{in}$ is reset to to $-\pi/2$, $\psi_{l}$ is set to $-\pi/2$, and
the upper bound $\psi_{u}$ is set to $\pi/2$. This effectively produces
a rotation of the coordinate system by the angle of $\pi$. Thus, after
the root is found, its value will have to be increased by $\pi$.

If it is found that the solution is in $\psi_{out}\geq 3\pi/2$, the
pathlength
$L(\psi_{in},3\pi/2)=\Big(L(\psi_{in},\pi/2)+L(-\pi/2,\pi/2)\Big)$ is
subtracted from {\tt path\_in} and $\psi_{in}$ becomes the upper bound
on the root $\psi_{u}=\psi_{in}$. The lower bound on the root $\psi_{l}$
is set to $-\pi/2$ and $\psi_{in}$ is reset to to $-\pi/2$. This
effectively produces a rotation of the coordinate system by the angle of
$2\pi$. Thus, after the root is found, its value will have to be
increased by $2\pi$.

When the algorithm arrives to this point, the bracketing of the root is 
finished:

\[ -\pi/2\leq\psi_{l}\leq\psi_{out}\leq\psi_{u}\leq\pi/2\ \  
  \mbox{with possible flag to increase $\psi_{out}$ by $\pi$ or $2\pi$.} \]

\subsection{Circular bracketing.}

The phase angle interval can be reduced further by noticing that the
length of the arc of radius of semimajor axis $a$ is not smaller than
the length of the ellipse within the same boundaries on the phase angle
$\psi$. Likewise, the length of the arc of the semiminor axis radius
$b$ is not greater than the length of the ellipse within the same
boundaries on the phase angle $\psi$. In other words, the solution
$\psi_{out}$ can be bracketed as:

\[ \psi_{1}\leq\psi_{out}\leq\psi_{2}, \ \ \ 
      \psi_{1}=(\psi_{l}+2\pi/a) \ \ \ 
      \psi_{2}=min\Big((\psi_{l}+2\pi/b),\psi_{u}\Big) \]

\subsection{The solution.}

The equation for $\psi_{out}$ which needs to be solved is:

\[ L(\psi_{in},\psi_{out})=\mbox{\tt path\_in} \]

This equation can be solved on the specified interval
$(\psi_{1};\psi_{2})$ using the secant or bisection method until the
error on accumulated pathlength $L$ becomes within allowed range. (It
was found that for $e>0.99$, when the ellipse is close to degeneration
into a line, the bisection method is faster than the secant.)

The solution of this equation should be increased by $\pi$ or $2\pi$ if 
coordinate system rotation was required as described before.


\section{Scattering in the Sun.}
\label{sec:simulations_app:scatter}

\begin{figure}
\centering
\psfrag{tt}{$\theta$}
\psfrag{tt1}{$\theta_{1}$}
\psfrag{pchi}{$\vec{p}\,'_{\chi}$}
\psfrag{pp}{$\vec{p}\,'_{p}$}
\psfrag{nn}{$\vec{n}$}
\includegraphics[totalheight=2.0in]{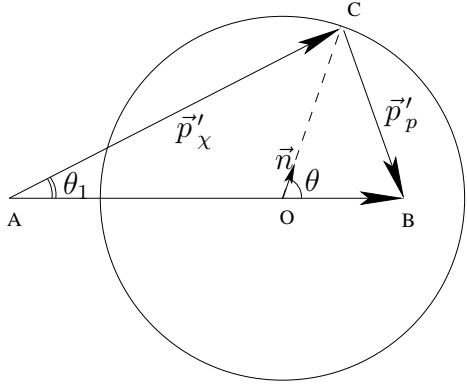}
\caption{Scattering diagram.
 $\vec{AO}=\frac{m_{\chi}}{m_{\chi}+m_{p}}(\vec{p}_{\chi}+\vec{p}_{p})$,
 $\vec{OB}=\frac{m_{p}}{m_{\chi}+m_{p}}(\vec{p}_{\chi}+\vec{p}_{p})$, 
 $\vec{OC}=|\mu\vec{v}|\vec{n}$.}
\label{fig:simulations_apps:collission}
\end{figure}

Let momenta of a neutralino and a scatterer (proton) before scattering
in the laboratory reference frame be $\vec{p}_{\chi}$ and $\vec{p}_{p}$
and after $\vec{p}\,'_{\chi}$ and $\vec{p}\,'_{p}$ accordingly.  It
should be noted that two vectors $\vec{p}_{\chi}$ and
$\vec{p}\,'_{\chi}$ form a plane thus $\vec{p}_{\chi}$ can be found from
$\vec{p}\,'_{\chi}$ by a rotation. The magnitude and the angle of
rotation can be found from the kinematics of the elastic scattering. The
axis of rotation has a random direction in space.

\subsection{Elastic scattering.}

In the center-of-mass frame, the momenta before scattering are:

\[ \left\{ \begin{array}{l}
   \vec{p}_{\chi 0}= \mu\vec{v} \\
   \vec{p}_{p0}    =-\mu\vec{v}
   \end{array}\right| \ \ \ 
   \mu    =\frac{m_{\chi}m_{p}}{m_{\chi}+m_{p}} \ \ \
   \vec{v}=\frac{\vec{p}_{\chi}}{m_{\chi}}-\frac{\vec{p}_{p}}{m_{p}},\ \ 
   \mu v=\frac{1}{\eta +1}|\vec{p}_{\chi}-\eta\vec{p}_{p}| \ \
   \eta =\frac{m_{\chi}}{m_{p}} \]

In the center-of-mass reference frame, the act of elastic scattering can
change the directions of the momenta only \footnote{This is merely a
statement of the energy conservation law}:

\[ \left\{ \begin{array}{l}
   \vec{p}_{\chi 0}^{\ \prime}= \mu v\vec{n}=
       \frac{1}{\eta +1}|\vec{p}_{\chi}-\eta\vec{p}_{p}|\vec{n}  \\ \\
   \vec{p}_{p0}^{\ \prime}    =-\mu v\vec{n}=
      -\frac{1}{\eta +1}|\vec{p}_{\chi}-\eta\vec{p}_{p}|\vec{n}
   \end{array}\right. \]
where $\vec{n}$ is a unit vector along the velocity of the neutralino
after the collision. In laboratory reference frame:
\[ \left\{ \begin{array}{l}
   \vec{p}\,'_{\chi}= \mu v\vec{n}+
          \frac{m_{\chi}}{m_{\chi}+m_{p}}(\vec{p}_{\chi}+\vec{p}_{p})=
     \frac{1}{\eta +1}|\vec{p}_{\chi}-\eta\vec{p}_{p}|\vec{n}+
          \frac{m_{\chi}}{m_{\chi}+m_{p}}(\vec{p}_{\chi}+\vec{p}_{p}) \\ \\
   \vec{p}\,'_{p}   =-\mu v\vec{n}+
          \frac{m_{p}}{m_{\chi}+m_{p}}(\vec{p}_{\chi}+\vec{p}_{p})=
       -\frac{1}{\eta +1}|\vec{p}_{\chi}-\eta\vec{p}_{p}|\vec{n}+
          \frac{m_{p}}{m_{\chi}+m_{p}}(\vec{p}_{\chi}+\vec{p}_{p})
   \end{array}\right. \]

The task is to find the $\vec{p}_{\chi}$ when $\vec{p}\,'_{\chi}$ is
known. If in the laboratory frame the scatter is at rest before the
scattering $\vec{p}_{p}=0$, then (see
figure~\ref{fig:simulations_apps:collission}): $\vec{OB}=\mu\vec{v}$
$\vec{AB}=\vec{p}_{\chi}$ and point $B$ is on the sphere. In this case,
the angle of deflection of neutralino velocity $\theta_{1}$ from its
original direction is related to the scattering angle $\theta$ in the
center-of-mass frame as:

\[ \tan\theta_{1}=\frac{m_{p}\sin\theta}{m_{\chi}+m_{p}\cos\theta}=
                                  \frac{\sin\theta}{\eta +\cos\theta} \]
and
\[ |\vec{p}_{\chi}|= |\vec{p}\,'_{\chi}|
  \frac{m_{\chi}+m_{p}}{\sqrt{m_{\chi}^{2}+m_{p}^{2}+2m_{\chi}m_{p}\cos\theta}}
  \ \Rightarrow\ \ 
   |\vec{v}_{\chi}|= |\vec{v}\,'_{\chi}|
   \frac{\eta +1}{\sqrt{\eta^{2}+2\eta\cos\theta +1}}  \]

\[ |\vec{v}_{\chi}|^{2}=|\vec{v}\,'_{\chi}|^{2}
               \frac{\eta^{2}+2\eta +1}{\eta^{2}+2\eta\cos\theta +1}  \]

The energy loss in a collision is:
\[ \frac{\Big(|\vec{v}_{\chi}|^{2}-|\vec{v}\,'_{\chi}|^{2}\Big)}{2}=
             \frac{2\eta(1-\cos\theta)}{(1+\eta)^{2}}\cdot
                                      \frac{|\vec{v}_{\chi}|^{2}}{2} \]

\subsection{Choosing the axis of rotation.}

Vectors $\vec{v}_{\chi}$ and $\vec{v}\,'_{\chi}$ from a plane which is
defined by some vector $\vec{n}$. A vector $\vec{n}$ perpendicular to 
$\vec{v}\,'_{\chi}$ can be found by solving:

\[ \vec{v}\,'_{\chi}\cdot\vec{n}=0 \ \Rightarrow \
                          v_{x}'n_{x}+v_{y}'n_{y}+v_{z}'n_{z}=0 \]

This equation can be solved by setting trial coordinates for the vector
$\vec{n}=(1,1,1)$ and then modifying one of the coordinates to satisfy
the orthogonality condition:

\begin{tabbing}
margin    \= .....if($v_{z} != 0$)..... \=  \kill \\
 \> if ($v_{z}' \neq 0$)      \> $n_{z} := -(v_{x}' + v_{y}')/v_{z}'$ \\
 \> else if ($v_{y}' \neq 0$) \>  $n_{y} := -v_{x}'/v_{y}'$  \\
 \> else                      \>  $n_{x} := 0$ \\
\end{tabbing}

Then, the obtained vector $\vec{n}$ can be rotated on a random angle
around $\vec{v}\,'_{\chi}$ which will produce a vector perpendicular to
$\vec{v}\,'_{\chi}$ and pointing in a random direction in space. The
obtained vector defines the $(\vec{v}\,'_{\chi},\vec{v}_{\chi})$ plane.


\section{Generate path inside the Sun.}
\label{sec:simulations_app:generate_path_in_sun}

The probability $dp$ that a particle will travel distance $x$ through
matter without scattering and then scatter immediately after that in
the distance $(x,x+dx)$ is:

\[ dp=\frac{1}{\lambda}e^{-x/\lambda}\, dx \]
where $\lambda$ is the mean free path of the particle in matter.

Thus, the pathlength which need to be accumulated in the Sun until next
scattering should be drawn from an exponential distribution with
parameter $\lambda$.

\[ \lambda=\frac{1}{\sigma_{p\chi}n_{p}} \]

$n_{p}$ is the concentration of the scatterers and $\sigma_{p\chi}$ is 
the crossection of the scattering process. In the given model 
$n_{p}=\frac{M_{\odot}}{\frac{4}{3}\pi R_{\odot}^{3}m_{p}}$, $m_{p}$ is 
the mass of scatterers (protons).



\chapter{Comments on the upper limit construction procedure}
\label{chapter:upper_limit}

Very often physicists consider the question of validity of a new theory
to be equivalent to a non-zero value of the parameter(s) of the theory.
Thus, often, the tests of validity of a new theory are designed in such
a way as to measure the value(s) of its parameter(s). If the measured
value is non-zero it is concluded that the new theory is valid with the
obtained value of the parameter. Such a ``physical'' approach seems to
give adequate results but is not correct from a methodological point of
view. Indeed, one can always assume that some theory is correct and
obtain a non-zero value for the parameter of the theory based on the
experiment, but that does not mean that the theory correctly describes
the observed process. Also, if the measurement was ``not successful'',
that is, the experiment could not show that the value of the parameter
is non-zero, it is not possible to decide if the new theory is valid or
not. And more importantly, the new theory might not even have a free 
parameter to be measured.

In the defining work by Neyman~\cite{Neyman} on statistical estimation
the question of a statistical test is separated from the question of the
measurement.  The procedure for parameter estimation (measurement)
demands that it is known that the process, whose parameter is being
measured, exists and the observed data is described by the known
distribution with the parameter being measured. If it is not the case,
the procedure can not guarantee that the constructed confidence interval
will contain the true value of the parameter with requested probability.
This fact if often overlooked.

When a new theory is proposed the experiment should not try to estimate
the value of the parameter of the theory, but, instead, it should be
designed to exploit the differences between the adopted (old) and the
new theories and check if indeed there is evidence to reject the old
one.  It is the difference between the old and the new theories which
provides the ``signal'' in the test. The test of the validity of the new
theory should be designed in the spirit of the proof by contradiction
(also called indirect proof) method. In other words, it should be
assumed that the new theory is wrong and the old one is correct. If the
contradiction between the old theory and observed data is found (i.e.
the ``signal'' is found), the assumption of validity of the old theory
should be rejected and the new theory may be accepted as a valid one. If
the contradiction is not found, nothing can be said regarding the
validity of the new theory and the old theory can not be rejected in
favor of the new one due to lack of evidence.

\section{Sensitivity and upper limit.}

There is a question which an experimenter should answer when designing
the experiment: what is the probability to accept $H_{0}$ due to pure
chance when $H_{1}$ is true i.e. what is the power of the constructed
test. The power of the test depends on the alternative hypothesis and
its parameters as well as on the null hypothesis. If a new theory
provides a large power of the test and is true, the contradiction
between the observed data and the old theory will be found easily by the
constructed test. If, on the other hand, the new theory provides small
power if it is true, the contradiction between the observed data and the
old theory will not be found easily. It is seen that the ``strength'' of
a ``signal'' is defined in terms of the power of the test. The new
theory predicts a ``strong'' detectable ``signal'' if it provides large
power of the test.

Suppose that the alternative hypothesis has the form of
$p_{1}(x;\lambda)$ where $x$ is observed quantity and $\lambda$ is the
parameter of the new theory. For some values of $\lambda$ the power of
the test will be small and for others the power will be big. It is
proposed to define ``sensitivity'' of the test (or experiment) as such
values of the parameter $\lambda$ for which the test has the power of at
least $50\%$. The sensitivity defined this way has several important
properties: it is a detector feature and can be estimated before the
experiment is performed, it is not a random number and does not depend
on the value of the observed quantity.

It is also proposed to state the upper limit when $H_{0}$ is not
rejected as such values of $\lambda$ for which the power of the test is
big (say at least $90\%$). The choice is motivated by the logic that it
is hard to miss such a strong ``signal'' and yet it was not observed. In 
other words the values of $\lambda$ for which there is a big chance to 
not reject the null hypothesis are below the upper limit and the values 
of $\lambda$ for which the chance to reject $H_{0}$ is big can be 
dismissed when $H_{0}$ was not rejected based on the observed data.

\section{Problem with the current approach.}

Currently, the upper limit on the value of the parameter $\lambda$ of a
theory is reported as the upper boundary of the one-sided confidence
interval on the parameter which is obtained assuming that the data came
from a distribution characterized by $p_{1}(x;\lambda)$. This is,
however, true and correct only in the cases when the physical process
originating the observed data exists and according to the theory is
described by $p_{1}(x;\lambda)$. Again, only if there is no question of
existence of the process is it correct to use the procedure for
confidence interval construction for parameter estimation. Note that the
bounds of the confidence interval are random by nature because they are
functions of a random variable and will vary with the data observed.

The application of the same technique to construction of a confidence
interval on $\lambda$ due to a process which is not known to exist when
the hypothesis of absence of this process is not rejected based on the
observed data will lead to meaningless results. The constructed
confidence interval will not have the desired confidence level and more
over, the theory $H_{0}$ which has no parameter $\lambda$ has not been
rejected based on the observed data. It is irrational and illogical to
assume the validity of $H_{1}$ when $H_{0}$ is not rejected.

\section{Example I.}

Let us consider a situation when according to the old adopted theory 
a quantity $X$ is distributed according to Gaussian law with zero mean 
and known standard deviation $\sigma$: 
\[ p_{0}(x)=\frac{1}{\sqrt{2\pi\sigma^{2}}}e^{-x^{2}/2\sigma^{2}} \]

If a new theory is correct, the data $X$ should be distributed according
to the Gaussian distribution with positive mean $\lambda$:

\[ p_{1}(x;m)=\frac{1}{\sqrt{2\pi\sigma^{2}}}
                       e^{-(x-\lambda)^{2}/2\sigma^{2}}\ \ \lambda >0 \]

Let us assume that if $x>3\sigma$ the old theory is rejected and a
confidence interval on the values of $\lambda$ may be constructed. If
$X<3\sigma$ the old theory is not rejected and an upper limit on the
values of $\lambda$ should be set with the confidence of $90\%$ (or
$1.28\sigma$).

Further, let us assume that the outcome of the measurement is $x=0$,
which is the most probable outcome when the old theory is true.
According to the current approach, the upper limit on the $\lambda$
would be $\lambda<1.28\sigma$ (This is a standard one-sided confidence
interval on $\lambda$ with $90\%$ confidence.)

Now, suppose that the new theory is true with the parameter
$\lambda=1.29\sigma$. This signal is above the upper limit. What is the
probability of discovering the signal in this experiment or what is the
power of the test? For a discovery to happen $x>3\sigma$ should be
observed which has a probability of happening of $5\%$ only. Therefore,
the experiment which set a limit of $\lambda<1.28\sigma$ has almost no
capability to discover a stronger signal. This is clearly unsatisfactory
result.

Using the approach proposed here, it is required to state the upper
limit corresponding to $90\%$ power of the test. Since in the observed
data $x<3\sigma$. no discovery is made. The upper limit would be
$\lambda<(3.0+1.28)\sigma=4.28\sigma$. The result is stated as: the
upper limit corresponding to significance $1.35\cdot 10^{-3}$ and the
power $0.9$.

\section{Example II.}

The importance of knowledge of existence of a process before a
measurement can be performed can be illustrated on a somewhat artificial
example. Suppose according to a new theory elephants have wings. An
attempt to measure the length of the wings, for example, resulted in the
limit that their length is between $0$ and $2$ centimeters, for example. 
The approach proposed here would state that no wings were found, but if
they were longer than $5$ centimeters, for example, they would have 
definitely been found.

This illustrates that by following the assumption that the new theory is 
correct without testing it, one is in danger of reporting a limit on an 
absurd parameter while according to the approach proposed it will be 
clearly stated that the new effect was not discovered and how strong the 
effect should have been to be discovered.

\section{Example III.}

Suppose it is known that observed data $X$ comes from a Gaussian 
distribution with unknown mean $\lambda$ and known variance $\sigma$:
\[ p(x;\lambda)=\frac{1}{\sqrt{2\pi\sigma^{2}}}
                             e^{-\frac{(x-\lambda)^{2}}{2\sigma^{2}}} \]

If it is desired to estimate the value of $\lambda$ by a $90\%$ 
one-sided confidence interval, given that the observed data $x$ is 
$x=0$, the interval on the values of $\lambda$ is: 
$-\infty <\lambda <1.28\sigma$.

Suppose, later, an existence of a new process is proposed according to
which the same data $X$ on the same experiment should have the
distribution of:

\[ p(x;\lambda,\mu)=\frac{1}{\sqrt{2\pi\sigma^{2}}}
        e^{-\frac{(x-\lambda-\mu)^{2}}{2\sigma^{2}}}, \ \ \ \mu\geq 0 \]

One is tempted (and according to the current approach this would happen) 
to state that based on the same data the total signal is 
$-\infty <(\lambda +\mu)<1.28\sigma$. This is, of course, incorrect, 
because it is not known if the new process with the parameter $\mu$ 
exists. Instead, one should consider a statistical test with:

\begin{itemize}
\item[$H_{0}$:] $p_{0}(x;\lambda)=\frac{1}{\sqrt{2\pi\sigma^{2}}}
                             e^{-\frac{(x-\lambda)^{2}}{2\sigma^{2}}}$
\item[$H_{1}$:] $p_{1}(x;\lambda,\mu)=\frac{1}{\sqrt{2\pi\sigma^{2}}}
        e^{-\frac{(x-\lambda-\mu)^{2}}{2\sigma^{2}}}, \ \ \ \mu\geq 0$
\end{itemize}

If the value of $\lambda$ is known, the test would be very simple: if
the observed data is greater than some $x_{0}$ the null hypothesis
should be rejected. For a $1.35\cdot 10^{-3}$ significance the critical
region is defined as:

\[ x>x_{0}=\lambda+3\sigma \]

If the observed data is inside of the critical region, a confidence
interval on the value of $\mu$ may be constructed. If the null
hypothesis is not rejected based on the observed data, the upper limit
$\mu_{u}$ corresponding to the power of $0.9$ can be found from the
equation:

\[ \int_{3\sigma+\lambda}^{\infty}
\frac{1}{\sqrt{2\pi\sigma^{2}}}e^{-(x-\lambda-\mu_{u})^{2}/2\sigma^{2}}\,dx=
 \int_{3\sigma-\mu_{u}}^{\infty}
      \frac{1}{\sqrt{2\pi\sigma^{2}}}e^{-y^{2}/2\sigma^{2}}\,dy =0.9 \]

If all the information available on the value of $\lambda$ is in the
form $-\infty <\lambda <\lambda_{u}$ with $100\%$ confidence, one is
forced to construct a conservative test as above assuming that
$\lambda=\lambda_{u}$. This will lead to the critical region
corresponding to the significance $1.35\cdot 10^{-3}$ constructed as:

\[ x>x_{0}=\lambda_{u}+3\sigma \]

Again, as before, if the observed data is inside the critical region, 
the null hypothesis of absence of the secondary process $\mu$ can be 
rejected and the value of $\mu$ may be estimated. If the observed data 
is outside of the critical region, the null hypothesis is not rejected 
and an upper limit on the value of $\mu$ based on power of the test can 
be constructed as above. Note that the information on the value of 
$\lambda_{u}$ should come from an independent experiment otherwise the 
constructed test and upper limit are not correct.

If, however, no information on the value of $\lambda$ is available the
proposed test will not yield any meaningful result as it is not possible
to distinguish data $X$ originating due to process $\lambda$ or $\mu$.

A more general statement can be made. If the null and the alternative
hypotheses on the origin of observed data $X$ come from the same family
of probability distributions i.e:

\[ p_{0}(x;\lambda)=f(x;\lambda)\ \ \ \mbox{and} \ \ \
                             p_{1}(x;\lambda,\mu)=f(x;\lambda +\mu) \]
and if no knowledge regarding the value of $\lambda$ is available, it is
not possible to construct a meaningful statistical test to differentiate
the two.%
\footnote{Indeed, the hypothesis test should be constructed in such a 
way so that the critical region on the values of $X$ does not depend on
$\lambda$. A procedure for constructing such a region was proposed by
Neyman and Pearson~\cite{NeymanPearson} for the case when:

\[ \phi(x;\lambda)=\frac{d\ln p_{0}(x;\lambda)}{d\lambda} \ \ \ \ \ 
   \frac{d\phi(x;\lambda)}{d\lambda}=A(\lambda)+B(\lambda)\phi(x;\lambda) \]

For these conditions to be satisfied the function $p_{0}(x;\lambda)$ 
should come from exponential family of the form:

\[ f(x;\lambda)=e^{Z(\lambda)T(x)+Q(\lambda)+S(x)} \]
where $Z(\lambda)$, $Q(\lambda)$, $T(x)$, $S(x)$ are some functions and 
$Z(\lambda)$, $Q(\lambda)$ are infinitely differentiable functions of 
$\lambda$. The equation of hypersurface $\phi(x;\lambda)=const$ in the 
$X$-space is equivalent to the equation $T(x)=const$.

Since in the considered case both $p_{0}$ and $p_{1}$ are of the same 
type, the likelihood ratio $p_{1}/p_{0}$ is independent of the values of 
$x$ on the hypersurface $\phi(x;\lambda)=const$ and the equation for 
the critical region corresponding to the error of the first kind $\xi$ 
becomes:

\begin{equation}
    \left\{\begin{array}{l}
    \xi\int_{\phi(x;\lambda)=const}p_{0}(x;\lambda)dx = 
          \int_{\phi(x;\lambda)=const;\ p_{1}/p_{0}>q}p_{0}(x;\lambda)dx \\ \\
        \frac{p_{1}(x;\lambda,\mu)}{p_{0}(x;\lambda)}=
         e^{const\cdot (Z(\lambda +\mu)-Z(\lambda))+Q(\lambda +\mu)-Q(\lambda)} >q
   \end{array}\right.
\label{eq:upper_limit:critical}
\end{equation}

Since the integration in~(\ref{eq:upper_limit:critical}) is performed 
over all $x$ for which $p_{1}/p_{0}>q$ is true and because 
$p_{1}/p_{0}>q$ is either always true or always false for all values of 
$X$, the equation~(\ref{eq:upper_limit:critical}) becomes:

\[ \xi\int_{\phi(x;\lambda)=const}p_{0}(x;\lambda)dx =
                     \int_{\phi(x;\lambda)=const}p_{0}(x;\lambda)dx \]

For a given $\xi$ this equation can be satisfied only if 
$p_{0}(x;\lambda)\equiv 0$ or if $\xi=1$. The former condition is not 
interesting and the later one states that the null hypothesis should be 
rejected all the time regardless of the value $x$.

} %
However, if several independent observations $x_{1}$ and $x_{2}$ can
be made such that it is known that data $x_{1}$ originated due to
process with parameter $\lambda$ only and $x_{2}$ is due to possible
$(\lambda+\mu)$ the test constructed in the form:

\[ p_{0}(x_{1},x_{2};\lambda)=f(x_{1};\lambda)f(x_{2};\lambda)
   \ \ \ \mbox{and} \ \ \
 p_{1}(x_{1},x_{2};\lambda,\mu)=f(x_{1};\lambda)f(x_{2};\lambda +\mu) \] 
will have the power to differentiate whether or not the data $x_{2}$ 
came from a new process even if no information on $\lambda$ is available 
before the test.

%
%
%

%
%

\section{When is the new theory valid?}

Any theory of a physical process should be considered ``admissible'' if 
no observed data contradicts predictions based on the theory. 
However, in statistical tests it is only possible to reject a theory in 
favor of some other theory and it is not possible to state with absolute 
certainty whether or not the given theory describes the given process 
correctly.

The question of how much evidence contradicting to validity of the old
theory in favor of the new one should be observed in order to declare
that the new theory correctly describes the given physical process is a
philosophical one. Obviously, the answer to this question is of great
practical importance and will govern the choice of desired significance
and power in formulating and performing the test.


\bibliography{thesis}
\bibliographystyle{plain}

\end{thesisbody}
\end{document}